# How common is LBV S Dor variability at low metallicity?[*]

V. M. Kalari,[1] J. S. Vink,[2] P. L. Dufton[3] and M. Fraser[4]

[1] Departamento de Astronomía, Universidad de Chile, Casilla 36-D Santiago, Chile
    e-mail: venukalari@gmail.com
[2] Armagh Observatory, College Hill, Armagh, BT61 9DG, United Kingdom
[3] Astrophysics Research Centre, School of Mathematics and Physics, Queen's University Belfast, Belfast, BT7 1NN, UK
[4] School of Physics, O'Brien Centre for Science North, University College Dublin, Belfield, Dublin 4, Ireland

## ABSTRACT

It remains unclear whether massive star evolution is facilitated by mass loss through stellar winds only, or whether episodic mass loss during an eruptive luminous blue variable (LBV) phase is also significant. LBVs exhibit unique photometric and spectroscopic variability (termed S Dor variables). This may have tremendous implications for our understanding of the first stars, gravitational wave events, and supernovae. A key question here is whether all evolved massive stars passing through the blue supergiant phase are dormant S Dor variables transforming during a brief period, or whether LBVs are truly unique objects.
By investigating the OGLE light-curves of 64 B supergiants (Bsgs) in the Small Magellanic Cloud (SMC) on a timescale of three years with a cadence of one night, the incidence of S Dor variables amongst the Bsgs population is investigated. From our sample, we find just one Bsg, AzV 261, that displays the photometric behaviour characteristic of S Doradus variables. We obtain and study a new VLT X-shooter spectrum of AzV 261 in order to investigate whether the object has changed its effective temperature over the last decade. We do not find any effective temperature variations indicating that the object is unlikely to be a LBV S Dor variable. As there is only *one* previous bona-fide S Dor variable known to be present in the SMC (R 40), we find the maximum duration of the LBV phase in the SMC to be at most ~few $10^3$yrs, or more likely that canonical Bsgs and S Dor LBVs are intrinsically different objects. We discuss the implications for massive star evolution in low metallicity environments, characteristic of the early Universe.

**Key words.** Stars: variables: S Doradus, Stars: evolution, Magellanic Clouds

## 1. Introduction

Massive stars are important producers of radiative and kinetic energy in the Milky Way, whilst massive stars in low metallicity environments may provide us with key insights into their ionizing input in the early Universe. However, the evolution and fate of the most massive stars in the Universe, especially at low metal content, is highly uncertain (Langer 2012). A key question is whether the bulk of the mass loss is lost in stellar winds (Chiosi & Maeder 1986), or in Luminous Blue Variable (LBV) type eruptions or binary interactions (Smith & Tombleson 2015), and how mass loss varies with metal content (Vink et al. 2001; Puls et al. 2008).

If low-metallicity massive stars would indeed loose substantial amounts of mass in the LBV phase, this could be pivotal for our understanding of pristine Pop III stars (Vink & de Koter 2005; Smith & Owocki 2006), as it is usually assumed these objects keep their initial masses intact (Woosley & Heger 2015). The mass-loss process at low metallicity is key for understanding seed black hole masses in the early Universe, the "heavy" black holes recently identified from the gravitational wave event GW 150914 (Abbott et al. 2016), and supernovae and supernovae impostors at low metallicity (Trundle et al. 2008).

### 1.1. Photometric variability of Bsgs and S Dor variables

Physically, LBVs can be understood as variables within the evolved massive blue supergiant zoo. B supergiants (Bsgs) are evolved massive stars ($>15\,M_\odot$) which are thought to be in a transitional stage from the main sequence towards the evolved Wolf-Rayet stage (although as noted in Vink et al. 2010 it is unclear if Bsgs are hydrogen-burning main sequence stars with a large overshooting parameter, or helium-core burning stars). Most Bsgs display intrinsic photometric variability that are characterized by their amplitude and timescales as–

1. $\alpha$-Cyg variables: display (likely periodic) micro-variations of $\lesssim 0.1^m$ on timescales of days-months, now commonly thought to be due to non-radial gravity mode pulsations (Lefever et al. 2007)
2. S Doradus variables (or Luminous Blue Variables; LBVs): display aperiodic variations of $\gtrsim 0.1^m$ (van Genderen 2001) on timescales of years-decades encompassing a broad variety of light curves in amplitude and time, with a clear distinction made in colour-brightness changes. Although no commonly accepted paradigm exists for the cause of the variability it must consist of radius inflation to explain the observed changes in optical brightness and spectral type, rather than dust obscuration events. Almost all S Dor variables also display $\alpha$-Cyg variations, whilst the opposite is not the case.
3. $\eta$ Car variables (or eruptive LBVs; or supernova impostors): show changes of $>1-2^m$ within days indicative of an eruptive

event. These events are extremely rare (around one per century in our Galaxy), with the most famous example being of $\eta$ Car (de Vaucouleurs & Eggen 1952).

Classically, most S Dor variables have been found serendipitously either through identification of their peculiar variability (Humphreys et al. 2016), or more recently, from their compact shells by infrared imaging (Gvaramadze et al. 2012). LBVs are rare, $\sim 20$ bona-fide LBVs are currently known in the Milky Way and the $1/2 Z_\odot$ Large Magellanic Cloud (LMC), and only one bona-fide member, R 40 in the $1/5 Z_\odot$ Small Magellanic Cloud (SMC), whilst it is not clear whether the SMC star HD 5980 is an LBV or Wolf-Rayet star (Humphreys et al. 2016; Smith & Tombleson 2015; Vink 2012; Humphreys & Davidson 1994).

Additionally, while $\alpha$-Cyg variables are considered to form a large fraction (if not most) of all Bsgs (Lefever et al. 2007), it remains unclear how common S Dor and eruptive LBV variability are among Bsgs. I.e. do all Bsgs display S Dor variability, albeit at smaller amplitudes or varying baselines?

### 1.2. Evolutionary status of S Dor variables

For many decades given their extreme rarity (even for massive stars), LBVs were thought to represent a brief evolutionary phase ($\sim 10^4$ years; Lamers et al. 1998) during a massive star's life between the hydrogen burning main sequence, and the core helium burning Wolf-Rayet phase. As LBVs lie close to the Humphreys-Davidson instability limit (Humphreys & Davidson 1994), they are linked to an instability which allows massive stars to lose mass, preventing them from becoming red supergiants (RSGs). Within this period of extreme mass loss the brightness increases in line with an inflation of the envelope, resulting in a temperature decrease (Gräfener et al. 2012; Vink 2015). This picture is corroborated by their light-curves, which show S Dor-type decadal variations (0.5–2 mag) interspersed by micro variations (of order 0.1-0.2 mag).

Although, S Doradus LBVs have traditionally been suggested to be transitional objects (Humphreys & Davidson 1994; Lamers et al. 1998), in the last decade, they have been proposed as direct core-collapse supernovae SNe progenitors (Kotak & Vink 2006; Gal-Yam & Leonard 2009; Groh et al. 2013; Smith 2014). Additionally, searches for LBVs in nearby galaxies (e.g. Massey et al. 2007) have found a vast number of LBV candidates without the distinctive photometric variability, and have suggested out that there are large number of Bsgs that are not eruptive LBVs but probably dormant S Dor LBV variables. This would suggest a much larger lifetime for the LBV phase $\sim 10^5$ yr, comprising a significant fraction of the He-burning phase. However, a number of the LBV candidates found by Massey et al. (2007) have been shown to be sgB[e] stars (a small subset of Bsgs whose evolutionary status remains unclear; see Miroshnichenko 2006). It has been suggested sgB[e] and LBVs do not evolve into one another (see Humphreys et al. 2017b).

In order to properly unravel the evolutionary status of the LBVs at low metallicity, we need to increase the number of confirmed LBVs in low metallicity galaxies. The key question is whether S Dor LBVs are truly special Bsg objects, or if *all* Bsgs might actually be *dormant* LBVs. In the latter case the total number of LBVs might be an order of magnitude larger than the number of bona-fide LBVs would suggest (see Massey et al. 2007) and hence their lifetimes would increase to $10^5$ years or more, with relevant consequences for their evolutionary state.

In this study, we present a systematic search for LBVs using long-baseline (3 years) medium-cadence (nightly) light curves of known Bsgs in the $1/5 Z_\odot$ local group dwarf galaxy Small Magellanic Cloud (SMC). The data allow us to quantify well the incidence of S Dor variability among known B Sgs. The main aim of the study is to quantify how common is the incidence of S Dor variability among the Bsgs population, to establish if most Bsgs are in fact dormant S Dor LBVs, or if S Dor LBVs are intrinsically different to Bsgs.

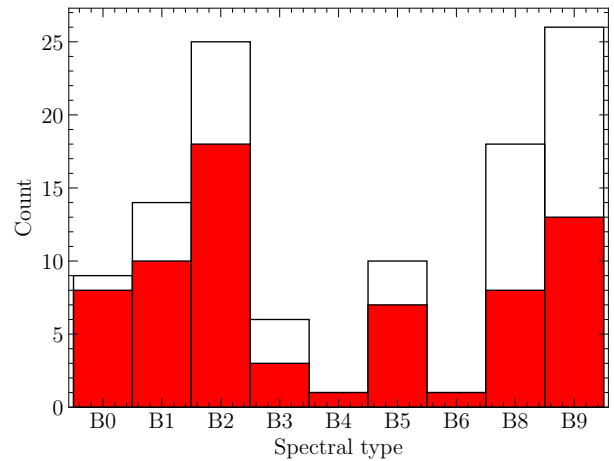

Fig. 1: Recovered spectral types of Bsgs in our sample (red) and the total number of Bsgs in the SMC (solid line) within the OGLE II survey area.

This paper is organized as follows: Section 2 describes the analysis of the light curves of known B Sgs. Section 3 describes the identification of the sole potential S Dor variable, AzV 261. Section 4 presents a discussion of the results. Our conclusions are given in Section 5.

## 2. Analysis of periodicity of B Sgs in the SMC

### 2.1. Bsgs sample

Bsgs having multi-epoch photometric data obtained during the second phase of the OGLE II survey were selected for analysis. The OGLE II survey obtained multi-epoch $I$-band imaging of the SMC. The photometry covers a period of 3 years spanning approximately from June 1997 to November 2000, at a cadence of approximately every night. The data cover only the periods (from approximately 8 months in each year) across the baseline in which the SMC is visible from the Warsaw telescope situated at the Las Campanas Observatory in Chile. The median seeing of the imaging is around $1''$, with a resulting median random photometric uncertainty around $\sim 0.007$ mag. Bad data points are identified and removed according to difference imaging criterion. The reader is referred to the OGLE II survey publication (Szymański 2005; Udalski et al. 1997) for further details of the data.

The stellar classifications given by Bonanos et al. (2010) were used to compile a list of 179 known SMC Bsgs. In a case of multiple varying classifications, we chose the most recent classification based on spectroscopy. A cross-match with a tolerance of $1''$ (the median seeing of the photometry) was performed with the OGLE II $I$-band imaging. Over the total field of view of the OGLE II survey, there are 110 known Bsgs, of which 69 Bsgs had multi-epoch $I$-band photometry (20 Bsgs did have OGLE II photometry, but not light curves, while 21 did not have OGLE II photometry even though they fall within the OGLE II field of view). The recovered 69 Bsgs cover the entire B spectral type, and are split based on their luminosity classification into 16 BIa, 24 BIab, and 24 BIb Bsgs, and 5 with no luminosity classification. The 5 Bsgs without luminosity classification are discarded from the final catalogue. The total number of identified Bsgs, and the total known in the OGLE II survey area are shown as a fraction in Fig. 1 as a function of spectral type. The recovery fraction is ~70% until spectral type B6. For binomial statistics, the recovered sample size can predict simple binomial properties at 99% confidence level with a 10% margin of error. Although the selection criteria are unbiased, OGLE II photometry saturates around $V$ ~12. Hence our final sample is not uniformly spread in either spectral type or brightness, but does contain a sufficiently representative sample of Bsgs for our analysis. A list of the Bsgs, along with their photometric properties are given in Table 1.

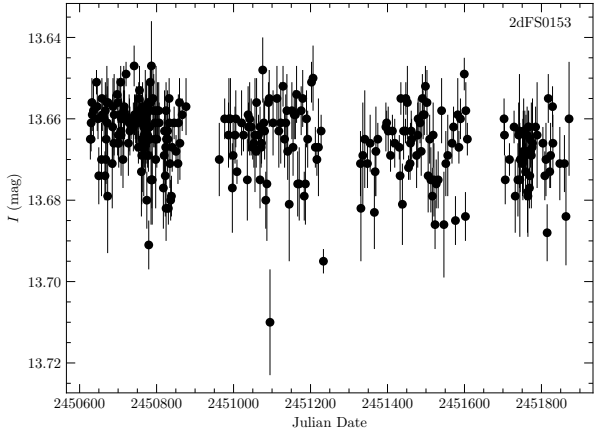

Fig. 2: Representative light curve of the Bsg 2dFS0153 from the OGLE survey, where the measured OGLE $I$-band magnitude is plotted against the Julian Date. Light curves of all 64 Bsgs in our sample are shown in Fig. A1.

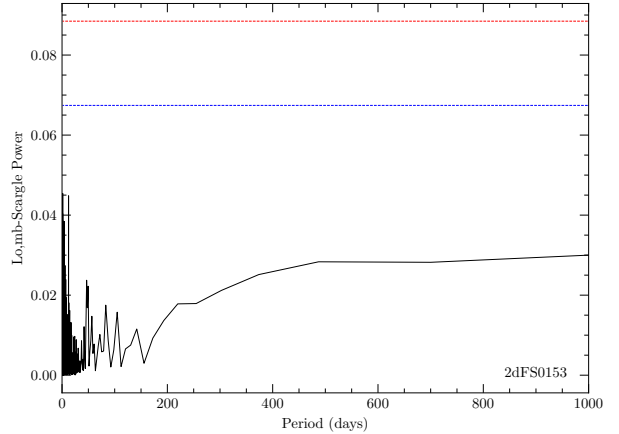

Fig. 3: Representative Lomb Scargle Periodogram of Bsg 2dFS0153. The period in days is plotted against the Lomb Scargle power. The dashed red and blue lines are the 99% and 95% significance levels identified using the bootstrap Monte Carlo simulations. Periodic variables are those whose peak in the power spectrum is greater than the 99% significance level. Lomb Scargle Periodograms of the remaining 63 Bsgs are displayed in Fig. A2.

### 2.2. Analysis of time series data

S Dor variables present multiple types of photometric variability (van Genderen 2001), presumably due to an inflating radius and decreasing photospheric temperature. The photometric variability is between 0.1 (below which most micro-variations are generally caused by the $\alpha$-Cyg $g$-mode pulsations) – 2.5 mag (above which most variations are due to eruptive events, falling into the eruptive phase of LBVs), on timescales of years to decades. Light curves within this broad magnitude and time span show diverse characteristics, yet few distinctive features excluding them from most other variables.

Primarily, the light curve varies on timescales greater than a year (not considering $\alpha$-Cyg variations), with a systematic trend in brightness. Additionally, a decrease in brightness is accompanied by a bluer colour and vice-versa. While the former rules out short period variability, the latter precludes variability due to dust obscuration events, and is considered a genuine increase in stellar radius. Considering the periodicity of S Dor variables, there are multiple cycles/periodicities present in the light curve with pronounced beats, although overall, the light curve increases/decreases aperiodically.

In this section, we search the light curves of the Bsgs to identify stars with multiple periodicities, and/or irregular photometric variability above 0.1m. To place this into context, we also compare our light curves with those of known S Dor variables in the Magellanic Clouds. The light curves of the Bsgs are presented in Fig. 2 and A1. We discuss our results below.

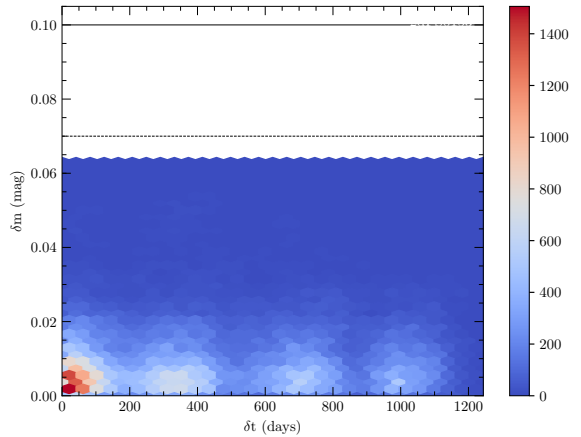

Fig. 4: $\delta$m-$\delta$t two-dimensional histogram of the Bsg 2dFS1053 from the OGLE II data, where the colour gradient marks the density of the points (with details in the colourbar). The dashed line marks the $10\sigma$ level of the random photometric uncertainty, while the solid line differentiates variations above $0.1^m$ level. The $\delta$m-$\delta$t two-dimensional histograms of the remaining Bsgs are included in Fig. A3.

### 2.2.1. Searching for periodic variability

In Fig. 2 and A1, we plot the light curves of 64 Bsgs and search for photometric variability at periodic, and aperiodic timescales which may have been caused due to outbursts. Identification for periodic variability is comparatively straightforward. To do so, we employed the Lomb-Scargle periodogram, which is an oft-used statistical tool to identify periodic variations in unevenly spaced observations. A power spectrum for each Bsg light curve was constructed. The period range was 0.5− 1000 d, with a frequency grid spacing of 0.0037 d$^{-1}$. The significance of the identified peaks was tested by a Monte Carlo simulation (e.g., see Vander Plas 2017; Frescura et al. 2007). To do so, 1000 simulated light curves were generated using a bootstrapping technique, wherein random data pairs are replaced from the original light curves. The bootstrapping technique gives us a significance level for the periods detected. Each Bsg showing a peak in the power spectrum greater than at the 99% level can be identified as a periodic variable following the methods outlined in Vander Plas (2017), and shown in Fig. 3 and A2.

Table 1: Properties of Bsgs having photometry in the OGLE II survey. Only the first ten lines are shown here to display the form and content, but the full table is available in the online version of this journal, or electronically via remote ftp.

| Name | SpT | Right Ascension J2000 | Declination J2000 | $I$ (mag) | $I_{err}$ (mag) | Period (days) | Significance (>95%) | Aperiodic (>0.1 mag) |
|---|---|---|---|---|---|---|---|---|
| 2dFS0153 | B9Ib | 0:36:37.00 | −73:34:48.7 | 13.665 | 0.006 | 12.35 | No | No |
| 2dFS0240 | B9Ib | 0:38:44.83 | −73:35:51.4 | 14.309 | 0.007 | 9.07 | No | No |
| AzV356 | B0Iw | 1:04:35.47 | −72:04:43.2 | 13.649 | 0.007 | 1245.78 | No | No |
| AzV376 | B1-2Ib | 1:05:07.30 | −72:24:36.7 | 13.419 | 0.007 | 1245.78 | No | No |
| AzV351 | B1-2Ibe | 1:04:21.38 | −72:09:14.6 | 13.412 | 0.007 | 414.99 | No | No |
| AzV394 | B1-5Iab | 1:06:01.52 | −72:23:28.1 | 13.567 | 0.007 | 1245.78 | No | No |
| AzV329 | B1.5I | 1:03:22.99 | −72:02:35.5 | 14.028 | 0.007 | 1245.78 | No | No |
| AzV340 | B1Ia | 1:03:49.43 | −72:07:29.4 | 12.702 | 0.006 | 1245.78 | No | No |
| AzV339 | B2.5Iab | 1:03:44.18 | −72:52:25.8 | 12.721 | 0.007 | 1245.78 | No | No |
| AzV342 | B2.5Iab | 1:03:56.07 | −72:02:47.4 | 13.099 | 0.007 | 1245.78 | No | No |

When searching for periodicity, we attempt to differentiate between short and intermediate period variables. Bsgs having short periodic variations (essentially with periods less than 2 days), are generally α-Cyg variables (unless the amplitude of their variations exceeds a few tenths of magnitude). Nearly all Bsgs show periods around 1-2 days with varying amplitudes around 0.05-0.1 mag. However, given that our sampling frequency is around the same, we cannot conclusively identify such variations as being strictly periodic due to the star. Note that periodic variations around 0.5 and 1d are in most cases almost certainly caused by sampling below the sampling frequency, representing the nightly half cadence and cadence levels. Therefore, although in many cases multiple frequencies in this time regime are found, we cannot isolate them as being strictly stellar (and thereby α-Cyg) with our data. Hence, any periodic variables below 1d are discarded from the final sample. In total, we identified 5 periodic variable Bsgs in our sample. Split based on luminosity class, they are 1 out 24 BIabs, 1 out of 16 BIas, and 3 out of 24 BIbs. In all instances, the period is greater than the Nyquist frequency at 3σ level (1% level as due to random variations). We did not attempt to smooth out any long term variability in the light curve. Only in the case of AzV 261, multiple periods above the 3σ level were found.

### 2.2.2. Searching for aperiodic variability

Searching for aperiodic variability is not straightforward compared to periodic variability as there are no commonly accepted metrics for massive stars (e.g. see Labadie-Bartz et al. 2017; Findeisen et al. 2015). We employed the δm-δt plots which are suggested by Mahabal et al. (2011); Findeisen et al. (2015) as a better indicator of variability in a series of tests on simulated light curves, although the correlation with any period is low. The δm-δt plots are only used to identify variability, and further checks were done by eye as at times some features may not be compared directly, as suggested by Labadie-Bartz et al. (2017).

The δm-δt plot presents the incidence of variability at a particular timescale in a light-curve. It is defined as the sum of all observations at magnitude $m$ and time $t$ of a light curve, by computing the differences,

$$\Delta m_{ij} = |m_i - m_j| \ (i > j)$$
$$\Delta t_{ij} = |t_i - t_j| \ (i > j)$$
(1)

and $i > j$ such that each pair is only considered once. By presenting the differences between magnitude and time as a density histogram, we can directly observe how much variability is observed on what time scales (see Fig. 4 and A3). Note that the δm-δt is closely related to the structure function used to characterize variability in galaxies. Overall, different timescales are sampled to varying levels in the plots, however we are only interested in aperiodic/micro-variations greater than 0.1m on timescales greater than a few days. To aid identification we compute the δm-δt plots of know S Dor variables S Dor, R 40 and R 110 in the Magellanic Clouds [1], and compare these with our Bsgs sample. From the δm-δt plots of know S Dor variables shown in Fig. 5, we identify only one BIb star as displaying aperiodic variability above the significance threshold: AzV 261.

## 3. AzV 261: A candidate S Dor variable

Searching systematically through archival light curves from the OGLE database, we found that AzV 261 in the SMC is an excellent S Dor candidate. In Fig. 6a, we show the $I$-band light-curve of the star spanning a decade (1998-2009). The $(V - I)$ colour behaviour is also shown (the additional time-series data come from OGLE III data of Kourniotis et al. 2014). From Fig. 6, AzV 261 firstly displays micro-variations of ∼0.2 mag on timescales of days. In addition, the brightness increased by 0.53 mag from Nov. 13 2000 to Oct. 20 2008 (the OGLE-III data are until 2009). Using the Period04 software (Lenz & Berger 2005), we attempted to characterize further the periodicity and micro-variations. In the Lomb-Scargle periodogram, AzV 261 shows four significant peaks (see Fig. A2 and Table 1). However, this light curve has a long-term trend of increasing brightness. We fit a fourth order-polynomial to remove this trend, and attempt using Period04 to fit periods to replicate the light curve of AzV 261. However, even with three periods, the residuals of the light curve are significant (see Fig. 6b) indicating that the observed possibly semi-periodic variations are microvaritaions (see van Genderen 2001; Lamers et al. 1998). Based on its light-curve, AzV 261 is the single likely LBV candidate in our sample.

---

[1] Data are from the American Association of Variable Star Observers International Database, at http://www.aavso.org

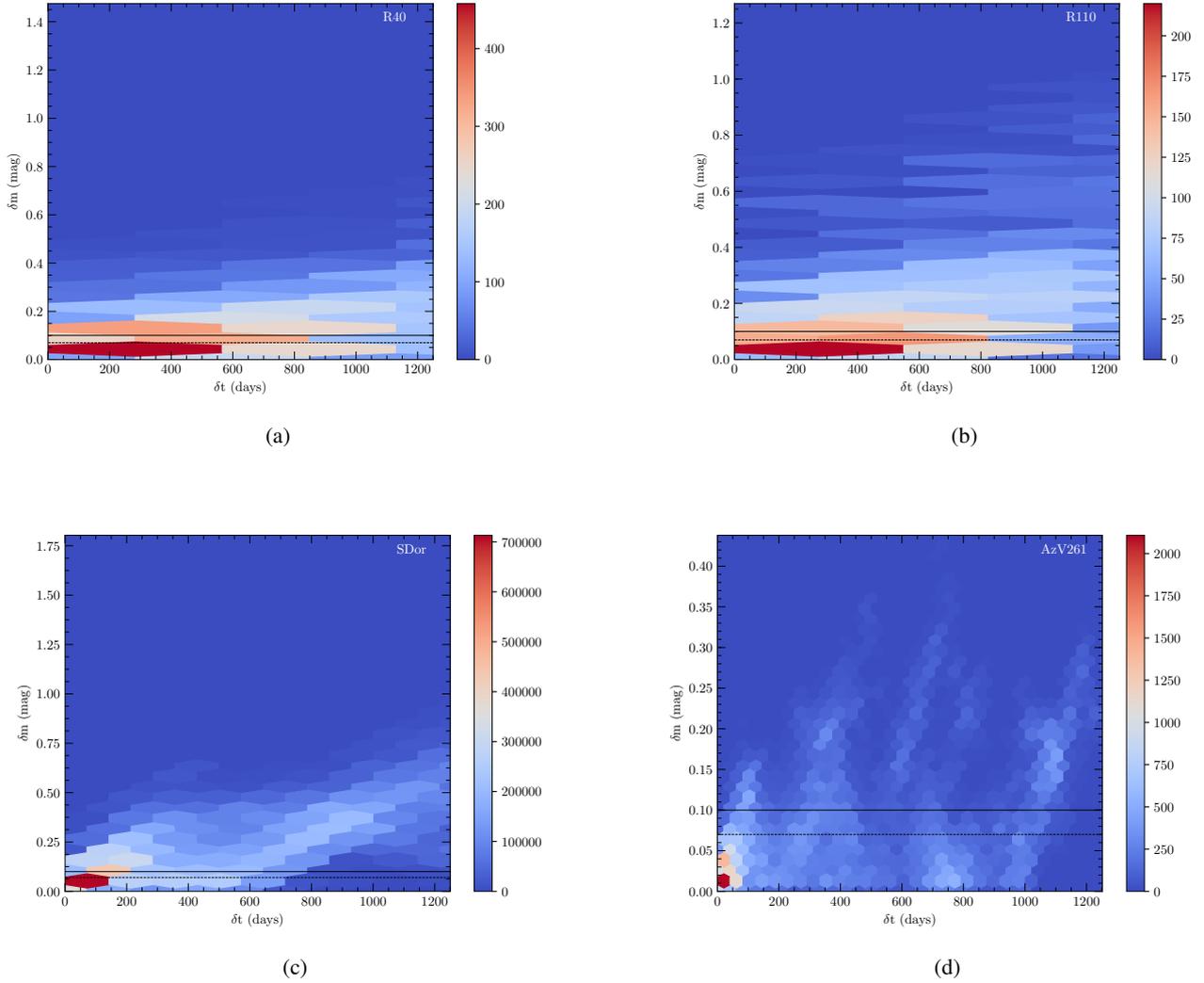

(a)

(b)

(c)

(d)

Fig. 5: δm-δt two-dimensional histogram of known S Dor variables R40, R110, S Dor on (a), (b) and (c) respectively. The candidate S Dor variable in our Bsgs sample, AzV 261 is shown in (d). Symbols are same as Fig. 4.

In addition, the colour information from Fig. 6c strongly supports the LBV candidacy: the $(V-I)$ colour becomes redder with time, corresponding with an actual temperature decrease. This is consistent with the picture of a brightness increase as the temperature decreases due to radius inflation, while the object's bolometric luminosity stays roughly constant. Therefore, AzV 261 is a prime S Dor-type LBV candidate, the only one among 64 known Bsgs identified by our systematic search. Its light curve displays micro-variations on the order of 0.2 mag over timescales of few days to months, with an overall increase over eight years of 0.53 mag in brightness. The increase in brightness corresponds to a redder colour and therefore a cooler temperature. All these features are unique to S Dor variables (van Genderen 2001; Lamers et al. 1998). Although the amplitude in overall change in brightness is on the lower end of most confirmed LBVs (which are on the order 0.5–0.1 mag). Overall, the light curve characteristics of AzV 261 point towards a S Dor classification.

In the literature, spectroscopic classification of the object taken when its magnitude was fainter (in 1999) from Evans et al. (2004) suggests a B2Ib(e) classification. Evans et al. (2004) used the 2dF multi-fibre spectrograph mounted on the Anglo-Australian 3.9m telescope to obtain low resolution spectra over the 3900-4800 Å region with a spectral element resolution $R \sim 1500$ (for spectral classification), along with Hα spectra ($R \sim 2500$) in 1999 September 30. The blue spectrum is shown in Fig.7, while the Hα spectrum is plotted in Fig.12.

In addition, in Massey et al. (1995), the authors obtained slit spectra in 1992 December 13 using the Cerro Tololo Interamerican Observatory 4.m telescope in the blue-air Schmidt camera. The final data cover the 3900–4800 Å range, at a resolution of 3Å pixel$^{-1}$. Based on these spectra the stellar classification is noted by the authors of that paper as O8.5Ie, however a re-classification of the stellar spectrum (show in in Fig.7) points towards an early B2 supergiant classification (priv. communication P. Massey), agreeing with the classification by Evans et al. (2004). Overall, on the one hand, photometric variability suggests S Dor candidacy for AzV 261, however, on the other hand, the spectral features themselves do not really agree with this notion. So, in order to find out, we must take and analyze further spectra to find out if the effective temperature changes between epochs.

### 3.1. Spectral analysis of AzV 261

#### 3.1.1. Medium resolution spectroscopy

We have obtained medium resolution spectroscopy with the X-shooter slit spectrograph mounted on the UT2 telescope of the Very Large Telescope (VLT) in Paranal, Chile (see Vernet et al. 2011 for details). The spectra were obtained on 2017 December 12. The X-shooter spectrograph simultaneously observes the 300–2500 nm wavelength range in medium resolution ($R \sim 7000$) slit spectra using three arms for ultraviolet (UVB; 300-560 nm), visual (VIS; 560-1020 nm) and infrared (1020-2500 nm) wavelength ranges. For the purpose of this study, slit nodding was not used and hence the infrared observations are not usable due to sky contamination, and were discarded. The spectra were observed in moderate seeing conditions ($\sim 1''$) using a slit of of 0.8 and 0.9$''$ in the UVB and VIS arm respectively. The achieved $R$ is 6200 and 7500 in the UVB and VIS arm respectively. Spectra were reduced using the X-shooter pipeline, with sky subtraction performed using the optimal extraction routine. The final sky-subtracted spectra are given in Fig. 7.

#### 3.1.2. Radial and rotational velocities

Initial stellar radial velocity estimates were obtained from the Doppler shifts of the He ɪ spectrum; the strong Balmer lines were not considered as they had significant emission in their cores. The observed He ɪ lines were fitted using a simple Gaussian plus a low order polynomial to represent the continuum. As discussed below, there is evidence for some stellar rotational broadening, whilst the diffuse lines (due to $^1P$-$^1D$ and $^3P$-$^3D$ transitions) exhibited significant intrinsic broadening. However visual inspection showed that the fits appeared to identify the line centre. For transitions between triplet levels, the fine structure multiplet transitions covered a wavelength range corresponding to $\sim$12-15 km s$^{-1}$ and in these cases the midpoint wavelength was adopted. Several lines were not considered as they were either very weak (e.g. 4169Å) or had emission in their wings (e.g. 5876 and 6678Å). The radial velocity estimates are summarized in Table 2 and have a mean value of 167±11 km s$^{-1}$.

It was also possible to identify features due to the ions, C ɪɪ, Mg ɪɪ, O ɪɪ and Si ɪɪɪ. However these lines were weak and often blended; additionally several appeared to be affected by emission near to the line cores (see Figs. 9-11). Radial velocity estimates were found from four metal lines using the same methodology as for the He ɪ spectrum and these are summarized in Table 2. They lead to a radial velocity estimate of 168±5 km s$^{-1}$, in excellent agreement with that found from the He ɪ spectrum.

The intrinsically narrow metal line spectra showed a significant amount of broadening, which could be due to either rotation or macro-turbulence. Also the cores of the He ɪ lines were not sharp again implying additional broadening mechanisms. We have attempted to estimate the contribution of rotational broadening using a Fourier Transform (FT) methodology (Simón-Díaz & Herrero 2007), which has been widely used for early-type stars (Dufton et al. 2006; Lefever et al. 2007; Markova & Puls 2008; Simón-Díaz et al. 2010; Fraser et al. 2010; Dufton et al. 2013; Simón-Díaz & Herrero 2014; Simón-Díaz et al. 2017). It relies on the convolution theorem (Gray 2005), viz. that the Fourier transform of convolved functions is proportional to the product of their individual Fourier Transforms. Then the first minimum in the Fourier transform for an observed spectral line, is assumed to be the first zero in the Fourier transform of the rotational broadening profile. Further details on the implementation of this methodology can be found in for example Simón-Díaz & Herrero (2007) and Dufton et al. (2013), whilst the projected rotational estimates, $v_e \sin i$, are listed in Table 2.

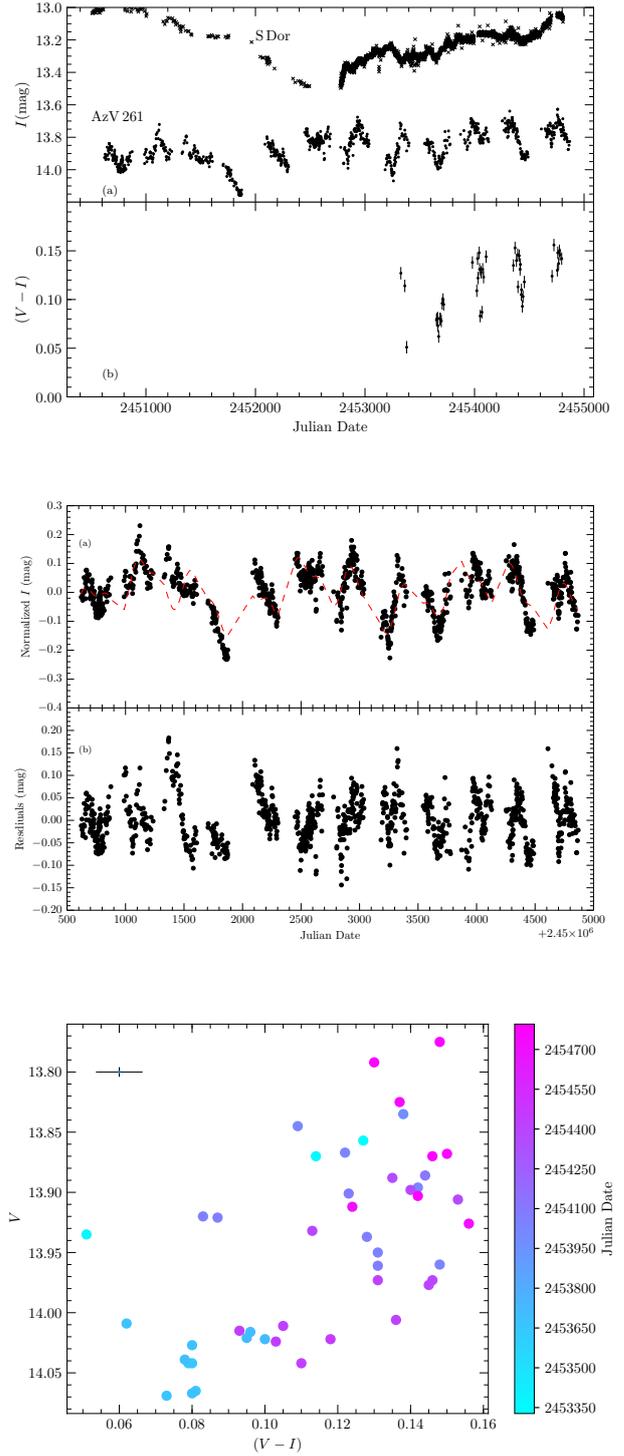

Fig. 6: (Top): *I*-band light curve of AzV 261 spanning from 1998 to 2009. Also shown is the scaled historical light curve of the classic LBV star S Dor for comparison (as crosses). (b) ($V - I$) colour plotted against the Julian date for AzV 261. (Middle): (a) The light curve with the overall trend removed by a $4^{th}$ degree polynomial, with the best-fit predicted light curve (the three peaks are given in Table 1). (b) The residuals between the fit period and the observed light curve.(Bottom): $V$ vs. ($V - I$) colour-magnitude diagram of AzV 261 over time, with time given by the colourbar. The behaviour of a brightness increase corresponding to a temperature decrease is seen here.

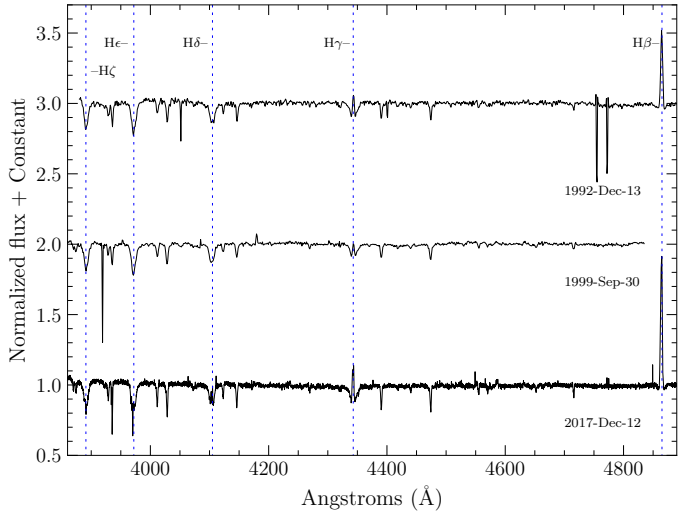

Fig. 7: Normalized X-shooter spectra of AzV 261 in the MK classification wavelength range, with key spectral lines marked. Also shown are the archival spectra from Massey et al. (1995) taken in 1992, and from Evans et al. (2004) taken in 1999. A constant is applied for visualization purposes.

Table 2: Radial velocities ($v_r$) and projected rotational velocity ($v_e \sin i$) estimates in km s$^{-1}$ from the He I and metal line spectra. The adopted laboratory wavelengths are also listed – see text for more details.

| Species | Wavelength | $v_r$ | $v_e \sin i$ | |
|---------|-----------|-------|-----|-----|
| | | | FT | PF |
| He I | 3926.54 | 165 | 116 | 101 |
| He I | 4009.26 | 191 | 88 | 122 |
| He I | 4026.27 | 157 | 97 | 113 |
| He I | 4120.90 | 159 | 78 | 91 |
| He I | 4143.75 | 169 | 103 | 105 |
| He I | 4387.93 | 169 | 76 | 93 |
| He I | 4437.55 | 179 | – | – |
| He I | 4471.58 | 158 | 88 | 88 |
| He I | 4713.24 | 175 | 96 | 60 |
| He I | 4921.93 | 151 | – | – |
| He I | 5047.74 | 168 | – | – |
| | | | | |
| C II | 4267.15 | 166 | 86 | 87 |
| Si III | 4552.62 | 162 | 106 | 97 |
| Si III | 4567.84 | 174 | – | – |
| Si III | 4574.76 | 172 | – | – |

Projected rotational velocities were also estimated from the He I spectrum by fitting the profiles of the observed lines, using a Gaussian function (representing the instrumental profile) with a rotational broadening function. Particularly for the diffuse helium lines, the intrinsic broadening may be significant, whilst macro-turbulence and micro-turbulence (Simón-Díaz & Herrero 2014; Simón-Díaz et al. 2010, 2017) may also be significant. Hence these estimates (also listed in Table 2) should be considered as upper limits. This is confirmed by the PF estimates for the He I lines being on average 16% larger than their FT equivalents with the latter providing a useful check on the FT estimates.

Convincing minima in the FT of the metal lines were only found in two cases with the $v_e \sin i$ estimates being consistent with those from the He I spectrum. Additionally no convincing minima could be found for the He I lines at 4437 and 5047Å, whilst the line at 4921Å was affected by emission in its red wing. Combining all the available FT estimates, yields a mean value, $v_e \sin i = 93 \pm 13$ km s$^{-1}$.

### 3.1.3. Atmospheric modelling

We have employed model-atmosphere grids calculated with the TLUSTY and SYNSPEC codes (Hubeny 1988; Hubeny & Lanz 1995; Hubeny et al. 1998; Lanz & Hubeny 2007). They cover a range of effective temperature, $10\,000\,\mathrm{K} \leq T_{eff} \leq 35\,000\,\mathrm{K}$ in steps of either 1 500 K or 2 000 K. Logarithmic gravities (in cm s$^{-2}$) range from 4.5 dex down to the Eddington limit in steps of 0.25 dex, and microturbulences range from 0-30 km s$^{-1}$ in steps of 5 km s$^{-1}$. As discussed in Ryans et al. (2003) and Dufton et al. (2005), equivalent widths and line profiles interpolated within these grids are in good agreement with those calculated explicitly at the relevant atmospheric parameters.

These non-LTE codes adopt the 'classical' stationary model atmosphere assumptions, i.e. plane-parallel geometry, hydrostatic equilibrium and the optical spectrum is unaffected by winds. The grids assumed a normal helium to hydrogen ratio (0.1 by number of atoms) and have been calculated for a range of metallicities with that for an SMC metallicity being used here. However tests using the Galactic (7.5 dex) or SMC (6.9 dex) grids led to only small changes in the atmospheric parameters (primarily to compensate for the different amounts of line blanketing). For further information on the TLUSTY grids see Ryans et al. (2003) and Dufton et al. (2005). Obviously a hydrostatic code like TLUSTY cannot be used to derive wind parameters, and can also not be employed to obtain meaningful fits to wind affected hydrogen lines such as Hα. For such modelling, an alternative code such as CMFGEN or FASTWIND (e.g. Massey et al. 2013) would be needed. However in the first instance, we are mostly interested in deriving reliable effective temperatures, and comparisons between TLUSTY and FASTWIND have generally shown good agreement between both codes for effective temperature estimation (e.g. Trundle et al. 2007).

Initial estimates for the effective temperature and surface gravity could be obtained from fitting the neutral helium and hydrogen spectrum. There is a loci of possible solutions (see, for example, Dufton et al. 1999 for further details), which range from an effective temperature and surface gravity (in cm s$^{-2}$) of $T_{eff} \approx 16\,000$ K and $\log g \approx 2.40$ dex to $T_{eff} \approx 23\,000$ K and $\log g \approx 3.05$. For lower effective temperatures, the theoretical He I becomes too weak, whilst for higher values, the He II spectra would be observable.

This range can be constrained using the observed metal lines and assuming that AzV261 has SMC metal abundances similar to those found in H II regions (see, for example, Kurt & Dufour 1998; Garnett 1999; Carlos Reyes et al. 2015; Toribio

San Cipriano et al. 2017) and for other hot stars (see, for example, Venn 1999; Rolleston et al. 2003; Hunter et al. 2007; Trundle et al. 2007), as summarized in Dufton et al. (2005). In Fig. 9-11, observed and theoretical spectra are presented for the wavelength region 3900-4700Å. The wavelength region 4200-4300Å is omitted as the only feature observable was the C II 4267Å doublet, which is subject to complex non-LTE effects (see, Sigut 1996; Nieva & Przybilla 2006). The theoretical spectra have been convolved with both a Gaussian function to allow for instrumental broadening and a rotational broadening function with $v_e \sin i$=95km s$^{-1}$. A micro-turbulence of 10 km s$^{-1}$ has been adopted which is similar to that found in analyses of Magellanic Cloud targets (Hunter et al. 2007; Trundle et al. 2007; McEvoy et al. 2015) with similar effective temperatures and gravities to the locus of estimates found for AzV 261. Spectra for three sets of atmospheric parameters ($T_{\mathrm{eff}}$, $\log g$) are shown, viz. (16 000, 2.45), (18 000, 2.65), (18 000, 2.80).

Overall the best agreement between observation and theory appears to be for the model with an effective temperature of 18 000 K. Below we discuss the spectra of the different ionic species in more detail:

**H I, He I:** As discussed above all three models provide a satisfactory fit to the helium spectrum and to the wings of the Balmer lines. The theoretical spectra for the He I line at 3926Å are too weak, possibly due to the adoption of a semi-classical quasi-static approximation for the line broadening (Dimitrijevic & Sahal-Brechot 1984). Additionally the theoretical profiles for the He I line at 4713Å appear to be too strong, although the observed profiles may exhibit emission in its wings. The lack of significant He II absorption is all the theoretical spectra (at, e.g. 4541 and 4686Å) is also consistent with the observations.

**C II:** For the gravities considered here, the C II spectrum has a maximum strength at $T_{\mathrm{eff}} \simeq$20 000 K. As such it shows little variation in the theoretical spectrum, with the the doublet at 4267Å being in good agreement with observation. As discussed above, the doublet at 4267Å is difficult to model and indeed all our theoretical spectra are stronger than observed, which is consistent with previous investigations using this model grid (Trundle et al. 2007; Hunter et al. 2007).

**N II:** The theoretical spectra adopt a nitrogen abundance of 7.1 dex, which is higher than that found in H II regions (for example, 6.55 dex, Carlos Reyes et al. 2015) or the lowest stellar abundances (for example, 6.66 dex, Rolleston et al. 2003). It was chosen to be representative of the typical nitrogen enrichment observed in SMC B-type stars (Trundle et al. 2007; Hunter et al. 2007). and as such, it provides no constraint on the atmospheric parameters. However the reasonable agreement between theory and observation for the N II line at 3995Å, argues against any large nitrogen enhancement.

**O II:** There is a rich O II spectrum with some of the stronger features identified in Fig. 9-11; additionally most of the unidentified absorption features are also due to this ion. The theoretical line strengths are sensitive to the adopted effective temperature and the model with $T_{\mathrm{eff}}$ =18 000 K provides the best overall agreement with observation.

**Mg II:** The doublet at 4481 Å is to be too strong in all three theoretical spectrum. However the observed profile appears to be affected by emission in both wings making any comparison

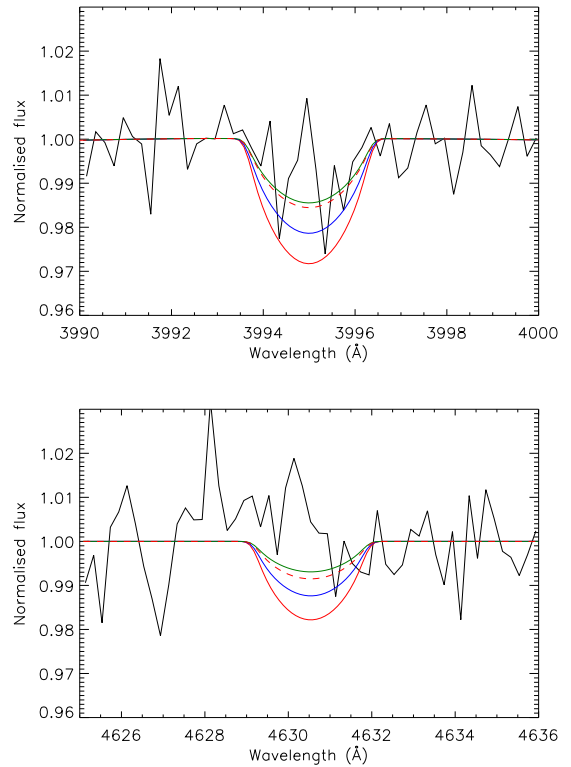

Fig. 8: Observed and theoretical profiles for the N II lines at 3995 and 4630Å. The theoretical profiles have been convolved with a rotational broadening function and are for atmospheric parameters of $T_{\mathrm{eff}}$ =18 000 K and $\log g$ = 2.65. The adopted nitrogen abundances are 6.7 (green), 6.9 (blue) and 7.1 (red) dex. Also shown is a theoretical plot for $T_{\mathrm{eff}}$ =16 000 K and $\log g$ = 2.45 and a nitrogen abundance of 7.1 dex (dotted red)

problematic.

**Si II, Si III:** The Si II doublet at 4128-4130 Å appears to be very weak in the observed spectrum. The spectra for the two hotter models appear consistent with the observations. The Si III triplet transitions between 4552 and 4574 Å are best fitted by the theoretical spectrum with $T_{\mathrm{eff}}$ = 18 000 K, although the other spectra also provide acceptable fits.

In summary the model with $T_{\mathrm{eff}}$ =18 000 K provides the best overall fit to the spectrum and, we estimate a *stochastic* error of ±2 000 K. This error would translate into an error of ~0.2 dex in the gravity. Combining this in quadrature with a fitting uncertainty of ~0.2 dex leads to an error in the gravity of ±0.3 dex. Hence our best estimate of the atmospheric parameters and their errors are $T_{\mathrm{eff}}$ =18 000±2 000 K and $\log g$ =2.65±0.3 dex.

As discussed above, the strength of the N II spectrum is close to or below the limit for detection. As nitrogen is important for constraining the evolutionary status of early-type stars, we have attempted to set an upper limit on its abundance. Several N II lines can be seen in B-type spectra, especially for targets showing significant nitrogen enrichments (see, for example Dufton et al. 2018). In Fig. 8, we show the two intrinsically strongest features at 3995 Å and 4630 Å that are not significantly affected by blending (Dufton et al. 2018), together with theoretical profiles. The latter are for our adopted atmospheric parameters and have been convolved with a rotational broadening functions and are for nitrogen abundances between 6.7 and 7.1 dex. Both observational and theoretical profiles were normalised in an identical fashion by a least-squares fitting of a linear polynomial to continuum regions at ±3–5 Å from the line centre.

There is marginal evidence for an observed N ɪɪ line at 3995 Å and this would imply a nitrogen abundance of ~6.7 dex. Alternatively, a conservative upper limit for the nitrogen abundance would appear to be 7.1 dex. The N ɪɪ at 4630 Å is not seen again implying a similar upper limit on the nitrogen abundance. Adopting a baseline nitrogen abundance of 6.6 dex (Carlos Reyes et al. 2015; Rolleston et al. 2003; Trundle et al. 2007; Hunter et al. 2007) would then imply a nitrogen enhancement of ≤0.5 dex Uncertainties due to possible errors in the atmospheric parameters are dominated by that due to the effective temperature. This is illustrated in Fig. 8, where a theoretical profile is shown for an effective temperature of 16 000 K (log $g$ = 2.45 dex). Adopting this lower effective temperature would increase the abundance limits by approximately 0.4 dex.

The estimation of the atmospheric parameters depended particularly on the assumption of baseline abundances for helium, oxygen and silicon. It is unlikely that the silicon abundance has been effected by nucleosynthetic process since the star formed. However the other two elements could be affected by mixing of material processed by the CNO bi-cycle to the surface. Adopting a nitrogen enhancement of 0.5 dex, the SMC models of (Brott et al. 2011) imply changes in the helium and oxygen abundances of less than <0.01 and <0.04 dex for masses between 15 M$_\odot$ and 60 M$_\odot$. Increasing the nitrogen enhancement to 0.9 dex (that is allowing for possible errors in the effective temperature) would lead to changes of <0.02 and <0.09 dex. This would imply that our use of a normal SMC abundance should not be a major source of error.

### 3.1.4. Hα line profile

The Hα line strength and profile in Bsgs is a key indicator of mass loss via stellar winds. The rate of mass loss is still under debate, with differences in measured mass-loss rates varying by up to factors of 10 from different indicators (e.g. the Hα line versus ultraviolet P Cygni lines). This is because clumping in the winds significantly affects the profile shapes of emission lines, including Hα. In addition, stellar pulsations in Bsgs may be linked to periods of enhanced mass loss.

The Hα emission line strength (Fig.12) changes from −22.4±1.2Å in the 1999 Evans et al. spectra to −38.5±0.3Å in 2017 X-shooter spectra. It is clear that AzV 261 displays a variable stellar wind, although the model fits to observations may not result in accurate mass-loss rates due to the choice of models (e.g. TLUSTY). If the sky subtraction for the multi-fibre spectroscopy of Evans et al. (2004) is poor, the measured equivalent width would be higher from their spectrum. In fact their equivalent width estimate is lower implying that uncertainties in the sky subtraction cannot explain the difference. We do not fit more sophisticated model atmospheres as we are merely interested in observing a change of spectral type, rather than wind variability, in order to characterize if AzV 261 is a S Dor variable. Regarding the absence of P Cyg absorption, this actually suggests that the star is hotter than the Bi-stability jump of 21 000 K, as absorption appears at and below the jump (Petrov et al. 2014).

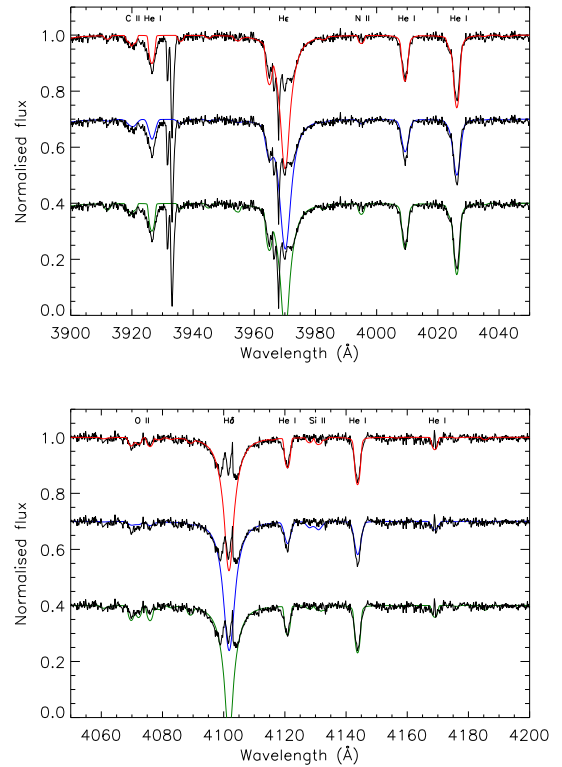

Fig. 9: Observed and theoretical profiles for selected wavelength regions. The theoretical profiles have been convolved with a rotational broadening function and are for atmospheric parameters of $T_{\rm eff}$ =18 000 K and log $g$ = 2.65 (red); $T_{\rm eff}$ =16 000 K and log $g$ = 2.45 (blue) and $T_{\rm eff}$ =20 000 K and log $g$ = 2.80 (green). Selected absorption lines are marked whilst the narrow features at 3933 and 3968 Å are due to interstellar Ca ɪɪ

### 3.1.5. Circumstellar dust

Most S Dor LBV variables show evidence for excess free-free emission (Humphreys et al. 2017a,b; de Wit et al. 2014). In contrast with evolved supergiant B[e] (sgB[e]), and warm hypergiant stars, they do not display evidence for warm dust. Compared to the other Hα emitting objects in the Bsgs zoo, S Dor variables therefore have excess free-free emission resembling that of classical Be (cBe) stars (which are main sequence fast rotators).

We can therefore use the infrared colour-colour diagrams to distinguish AzV 261 from other objects in the Bsgs zoo. In Fig.13a, we plot the near-infrared ($J - H$) vs. ($H - Ks$) colour-colour diagram showing the position of known S Dor variables, sgB[e], warm hypergiants, cBe stars (taken from Humphreys et al. 2017a) and Bsgs in our sample. Also shown is the main-sequence locus at SMC metallicity taken from Kalari et al. (2018). In addition, we also show the data in the mid-infrared (from Humphreys et al. 2017a) for the same stars in the [3.6μm]-[4.5μm] vs. [5.8μm]-[8.0μm] mid-infrared colour-colour diagram. From these diagrams, AzV 261 exhibits modest free-free emission, at levels similar to other know S Dor LBVs, at the higher end of those exhibited by cBe stars. In Fig. 14, we plot the spectral energy distribution of AzV 261, along with the TLUSTY model atmosphere for the best-fit stellar parameters. The best-fit model for the infrared excess is a blackbody of $T_b$=2550 K. The position of AzV 261 in infrared colour-colour

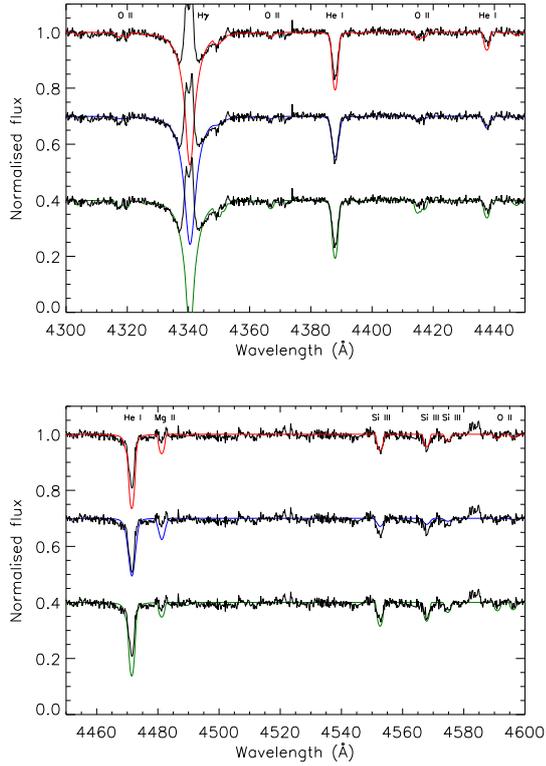

Fig. 10: Observed and theoretical profiles for selected wavelength regions. See caption to Fig. 9 for details.

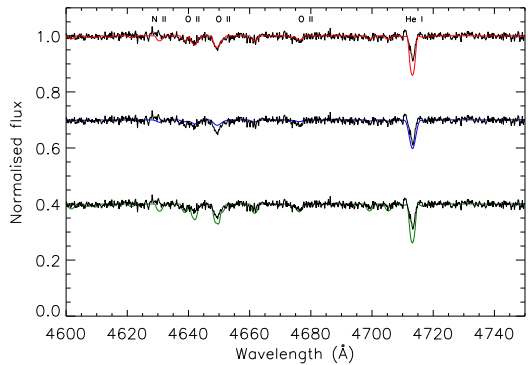

Fig. 11: Observed and theoretical profiles for selected wavelength regions. See caption to Fig. 9 for details.

diagrams, and its spectral energy distribution point towards free-free emission.

Since the spectroscopic analysis of the surface gravity precludes the possibility that AzV 261 is a cBe star, it is most likely a S Dor variable based on its photometric variability and infrared colours. However, the absence of spectroscopic variability is puzzling. Based on the absence of spectral non-variability AzV 261 is not a LBV S Dor variable.

### 3.1.6. AzV 261 within the Bsgs zoo

Based on the observed properties of AzV 261, we attempt to classify AzV 261 within the Bsgs zoo (assuming also that it may be a S Dor LBV). Bsgs are diverse, but we attempt to classify

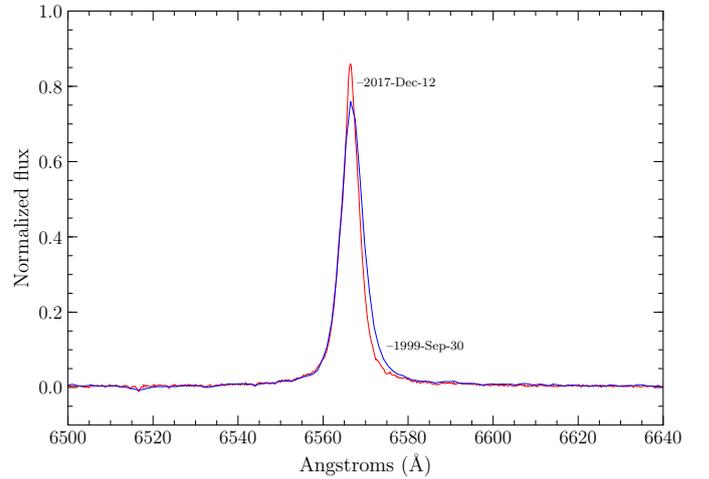

Fig. 12: Normalized Hα line profile from the X shooter spectra taken in 2017 Dec (red) and archival spectra from Evans et al. (2004) taken in 1999 Sep shown as solid blue line. The emission line equivalent width increased by −16.1Å within this period. Note that if the sky subtraction from the Evans et al. (2004) fibre spectra is poor, this difference will only be magnified.

AzV 261 between the main types, namely (i) Bsgs (ii) sgB[e] (iii) S Dor LBV variables. In addition, we include the (iv) cBe classification although note that cBe stars are main-sequence stars. The main classifying properties of AzV 261 are thus: (i) the star displays a supergiant luminosity classification (ii) does not display effective temperature variations (iii) displays aperiodic brightness increase/decrease inter-sped with periodic microvariations (iv) single peaked Hα line emission (v) infrared excess resembling free-free emission due to cold circumstellar disc/ ejecta.

In Table 3, we list the properties of the different Bsg subtypes, together with those for AzV 261. In general these agree best with an S Dor classification apart from the lack of characteristic temperature variability seen in most S Dor LBV variables, which remains puzzling. Finally, we note that AzV 261 displays an extremely low luminosity for a S Dor LBV variable (of logarithm of luminosity log $L = 4.9\pm0.1$ $L_\odot$), which implies a stellar mass of $15.9\pm^1_3 M_\odot$ according to the models of Brott et al. (2011). This places AzV 261 on the low mass end of known S Dor LBV variables.

In Fig. 15, we plot the position of AzV 261 and the Bsgs studied here in the Hertzsprung-Russell diagram with respect to stellar tracks from Brott et al. (2011), along with known S Dor variables in the SMC, LMC, and the Milky Way. The effective temperatures and bolometric corrections (to calculate the luminosities of the Bsgs) were estimated using the spectral type-effective temperature scale of Trundle et al. (2007), and computed following the procedure described in McEvoy et al. (2015). The data for the luminosities and effective temperatures of the literature S Dor variables here were taken from Smith et al. (2018, subd.), who present revised distances and luminosities for known galactic S Dor variables using recent GAIA DR2 parallaxes. Their results identify a population of low luminosity confirmed S Dor variables in the galaxy which do not have extragalactic counterparts (for e.g. W 243 and HD 168607). The authors of that paper suggest a systematic search for new S Dor variables in external galaxies might well uncover new low-luminosity LBVs, such as shown here. The existence of these low-luminosity S Dor LBV variables in the Milky Way, and also the SMC suggest the existence of a different instability mechanism, with important implications for SNe (see Smith et al. 2018 subd.). However, we consider AzV 261 unlikely to be a member of this group since it does not show the characteristic temperature variations seen in other LBVs.

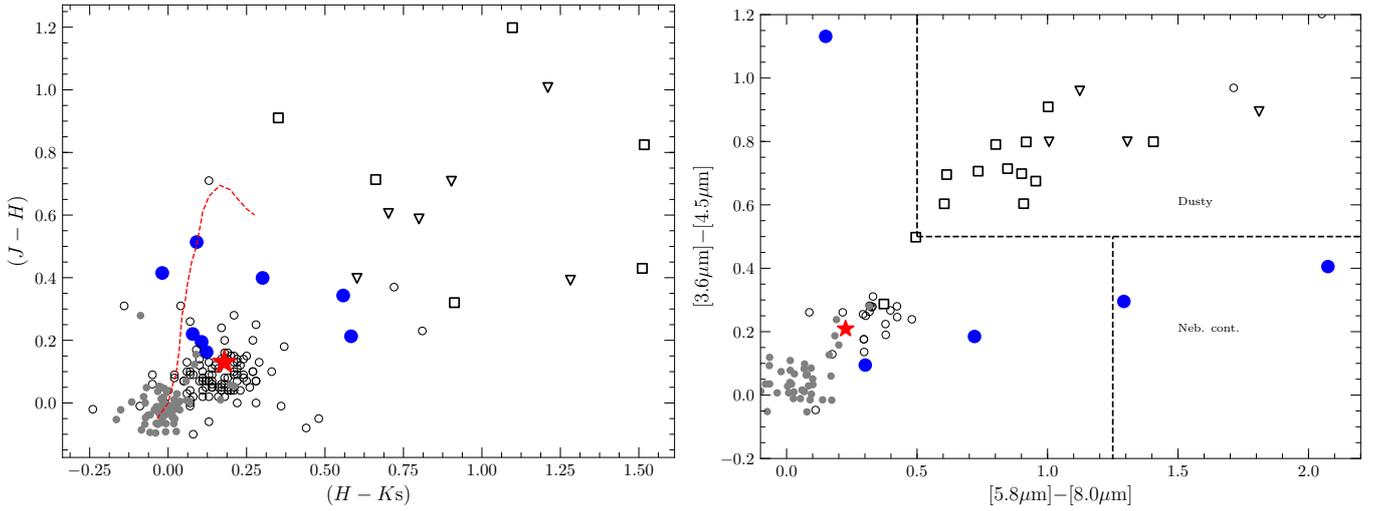

Fig. 13: (Left) Near-infrared $(J − H)$ vs. $(H − Ks)$ colour-colour diagram of known sgB[e] (open squares), warm hypergiants (open triangles), cBe stars (open circles), known S Dor variables (blue dots), BSgs (small dots), and AzV 261 (red asterisk). Data taken from Humphreys et al. (2017a). The dashed red line is the main-sequence locus at SMC metallicity. (Right) Mid-infrared $[3.6\mu m]$-$[4.5\mu m]$ vs. $[5.8\mu m]$-$[8.0\mu m]$ colour-colour diagram. Symbols are same as in (Left). The position of AzV 261 in both diagrams suggests the presence of free-free emission. Note that the LBVs redward of colours $[5.8\mu m]$-$[8.0\mu m]>1.0$ are contaminated by PAH emission from nebulosity, with the regions of nebular contamination, and dusty stars delianted following Humphreys et al. (2017a).

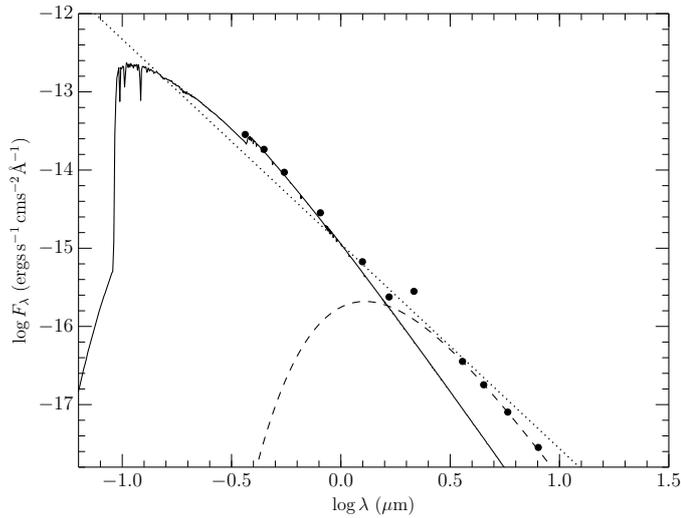

Fig. 14: Spectral energy distribution of AzV 261 covering photometry from the literature from $U$-$8.0\mu m$. Overplotted as a solid line is the TLUSTY model atmosphere for the best-fit stellar parmeters of $T_{\mathrm{eff}}$=18000 K, log $g$=2.5. The infrared-excess, corresponding to possibly free-free emission of cold dust is best fit using a blackbody model having $T_b$=2550 K (shown as the dashed line). The dotted line is of constant frequency flux density normalized to the 3.6$\mu$m photometry to make the free-free contribution clearer.

## 4. Discussion: S Dor variables among the Bsg population

The key question we wish to address was whether Bsgs are fundamentally dormant LBVs, or if they represent separate stellar populations, which would make bona-fide S Dor LBVs "special objects".

Although individual LBVs and LBV candidates have been discovered at metallicities potentially even lower than that of the SMC (Drissen et al. 2001; Izotov & Thuan 2008; Pustilnik et al. 2008; Herrero et al. 2010; Bomans & Weis 2011), in this systematic study of the SMC we found just one S Dor candidate – out of a total of 64 Bsgs.

If we assumed that all Bsgs pass through the S Dor phase in their lifetime, we can set an upper limit to the duration of the S Dor phase. Assuming that the mean duration between the end of the H-burning lifetime of a massive star and the onset of the Wolf-Rayet phase is ∼10⁵ yr, and that all Bsgs become S Dor variables, we expect that the duration of the S Dor phase at SMC metallicities is ∼ 10³ yr. This is much lower the number suggested if Bsgs were truly dormant S Dor variables (Massey 2006; Smith 2015) where their predicted evolutionary lifetime is ∼10⁵ yr; but also lower than that determined in the standard picture of massive star evolution ∼ 10⁴ yr (Humphreys & Davidson 1994; Conti 1976).

The low incidence of LBVs in the SMC seems to suggest that Bsgs and S Dor LBVs are truly different from one another, and that not all Bsgs become S Dor variables. This may suggest that the search for Hα P Cygni profiles – as suggested by Massey et al. (2007) – might not provide the answer to the total duration of the LBV phase in massive star evolution. However it might still be possible that related searches at high spectral resolution, allowing the presence of multiple-troughed P-Cygni absorption components to be revealed, may become an efficient tool to detect extra-galactic LBVs, as suggested by Groh & Vink (2011).

Additional evidence for an intrinsic difference between Bsgs and S Dor variables comes from the fact that there is a complete absence of rapidly rotating Bsgs with $v_e \sin i \gtrsim 200$ km s⁻¹ (see

Table 3: Properties of AzV 261 among the Bsg zoo and cBe stars. The properties of AzV 261 are shaded in blue.

| | Bsgs | sgB[e] | S Dor LBV | cBe |
|---|---|---|---|---|
| Luminosity | I-III | I-III | I-III | IV-V |
| Light curve | Quasi-periodic | Irregular | Periodic microvariations +Irregular increase/decrease | Periodic |
| Spectral type changes | Stable | Stable | Changes aperiodic | Stable |
| Hα emission | Variable emission | Yes+ forbidden line emission | Variable emission | Double-peaked variable |
| Infrared colours | None | Hot dust+disc | Cold dust/free-free emission | Free-free emission |

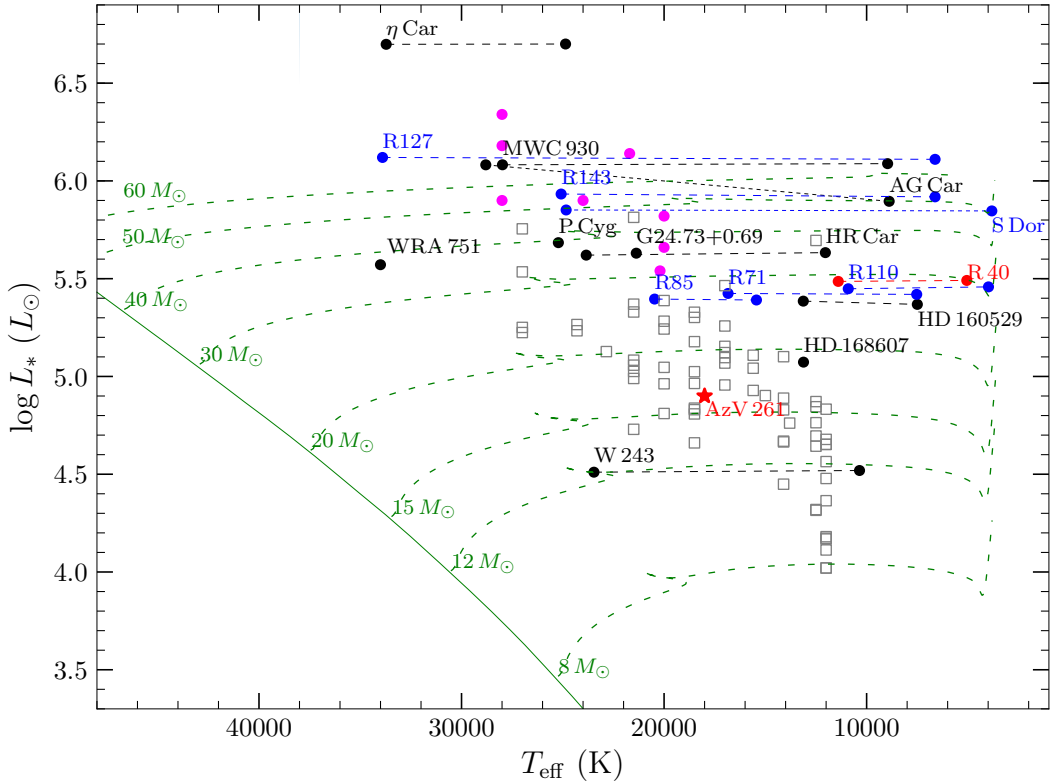

Fig. 15: Position of AzV 261 on the Hertzsprung-Russell diagram marked by the red asterisk, along the Bsgs studied here as gray squares. The positions of confirmed LBVs from the literature in the galaxy (black circles), LMC (blue circles), and the SMC (red circle) are also shown. Cyan circles marks the position of known LBVs in M31 and M33 (from Humphreys et al. 2017b). The dashed green lines are the stellar tracks from Brott et al. (2011) with the corresponding initial stellar mass indicated, and the solid line is the zero-age main sequence isochrone.

Vink et al. 2010) whilst several S Dor variables have been found to be rotating rather rapidly (Groh et al. 2006; 2009). For this reason Justham et al. (2014) considered stellar merging to be a likely evolutionary path to produce LBV supernovae.

The low number of confirmed LBVs in the SMC represents a puzzle, as one would seemingly require enhanced LBV mass loss if canonical stationary wind mass loss from normal stars is reduced at the lower metal content of the SMC (Mokiem et al. 2007). This raises the question of whether the known population of evolved Wolf-Rayet stars in the SMC have been produced by rotationally-induced chemical homogeneous evolution (Brott et al. 2011; but see Vink & Harries 2017), or by binary interactions (Schootemeijer & Langer 2018; but see Shenar et al. 2016).

Finally, Vink (2018, subd.) recently performed radiative transfer computations with a dynamic version of the Monte Carlo method to study the bi-stability jump of LBVs as a function of metallicity. He found the bi-stability jump to be smaller at low metallicity. These results may be related to the low number of S Dor variables in the SMC that we find here. Metal-poor galaxies are generally lower-mass and less luminous than their metal-rich counterparts, resulting in a smaller absolute population of massive stars (although recent evidence for a top-heavy stellar initial mass function at low metallicities compared to a Salpeter mass function, suggests that the relative fraction of massive stars may be higher, e.g. Kalari et al. 2018; Schneider et al. 2018). Therefore, short-lived phases such as LBVs might be-come harder to observe serendipitously, hence the systematic approach such as presented here carries important weight in defining if the incidence and characteristics of LBV S Dor variables do indeed vary by metallicity, and how this influences our view of stellar evolution in the early Universe, and of how LBVs evolve.

## 5. Conclusions

In this paper, we presented a systematic search for S Dor variables among the Bsgs population in the $1/5\,Z_\odot$ SMC. We conducted our search using light curves spread across three years (with a cadence of one night) of 64 Bsgs from a total known Bsgs population of 110 in the SMC within our field of view.

Our motivation was to identify whether all Bsgs display dormant LBV behaviour at differing timescales or amplitudes, or whether S Dor LBV variables are truly unique objects. From examining the light curves of the 64 SMC Bsgs, only one Bsg, AzV 261 demonstrated S Dor-like photometric variability. AzV 261 did not display variation in the spectral type. Therefore, AzV 261 is not a LBV S Dor variable. Our findings imply that the duration of the LBV phase is $\sim 10^3$ yr, or more likely S Dor LBVs and Bsgs are likely intrinsically different objects.

*Acknowledgements.* The authors thank the referee, R. Humphreys for insightful comments. We are grateful to C. J. Evans, and P. Massey for kindly providing their spectra of AzV 261. The authors also thank L. Wyrzykowski and the OGLE consortium for providing their data. V. M. K. also thanks G. Blanc, K. Boutsia, and N. Morrell for collecting a spectrum from Las Campanas. The authors thank ESO for providing Directors Discretionary time for performing the X-shooter observations. V.M.K. acknowledges support from the FONDECYT-CONICYT grant No. 3160117. MF is supported by a Royal Society - Science Foundation Ireland University Research Fellowship. We acknowledge with thanks the variable star observations from the AAVSO International Database contributed by observers worldwide and used in this research.

**Appendix A: Light curves of Bsgs having data from the OGLE II survey.**

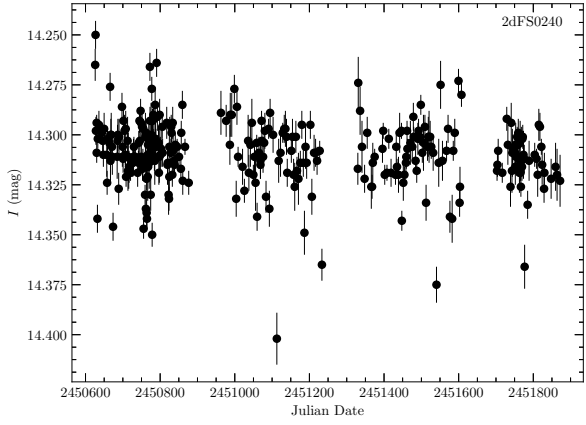

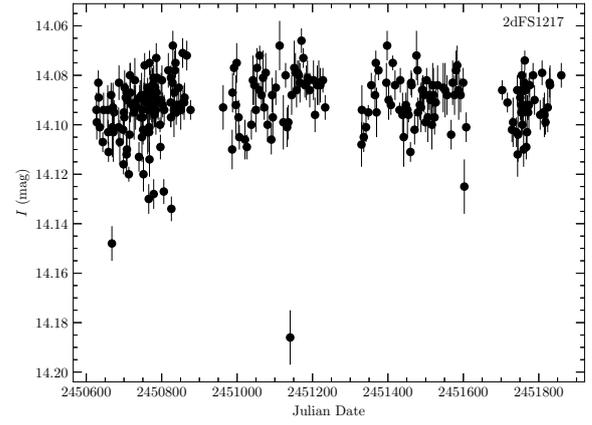

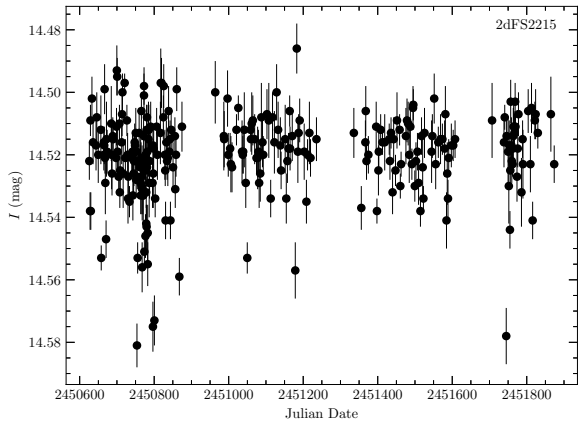

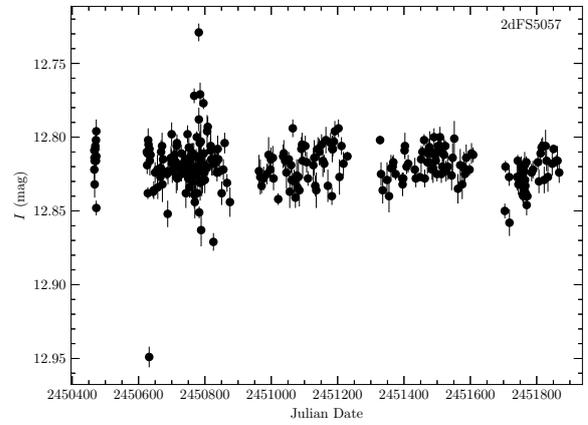

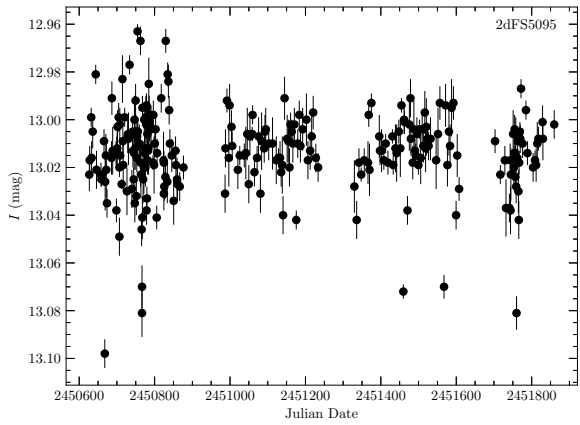

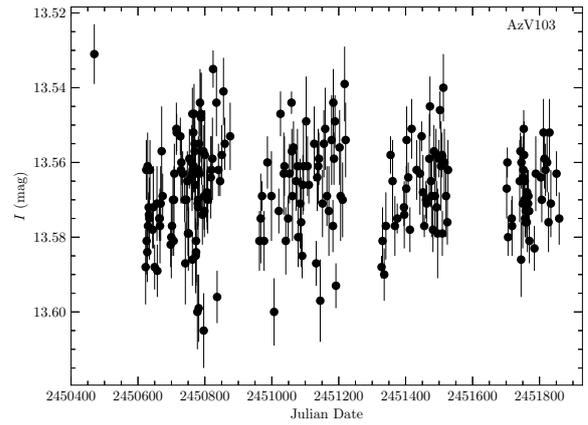

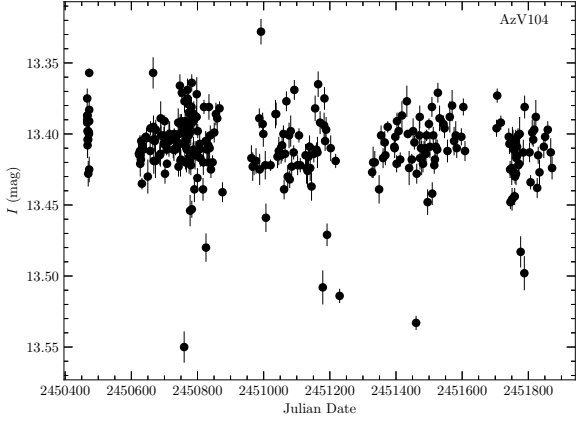
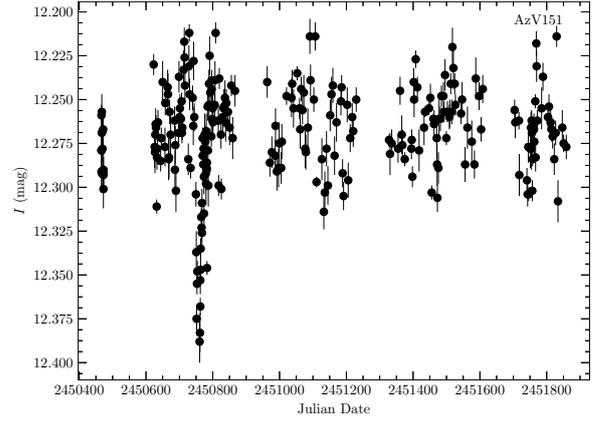

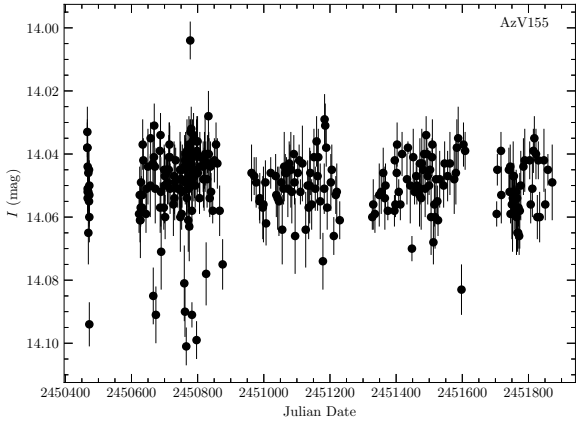
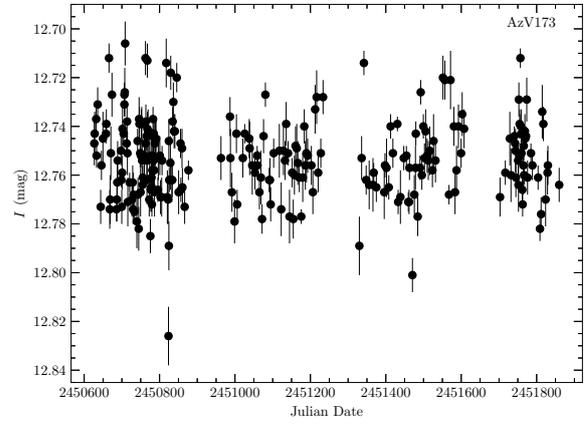

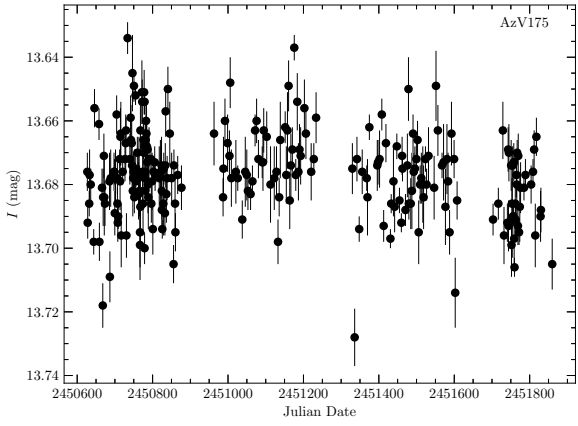
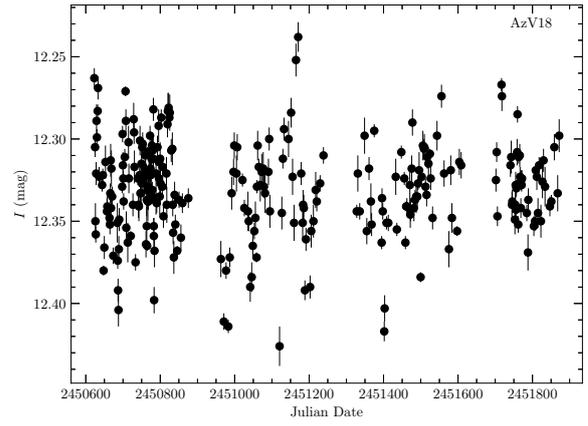

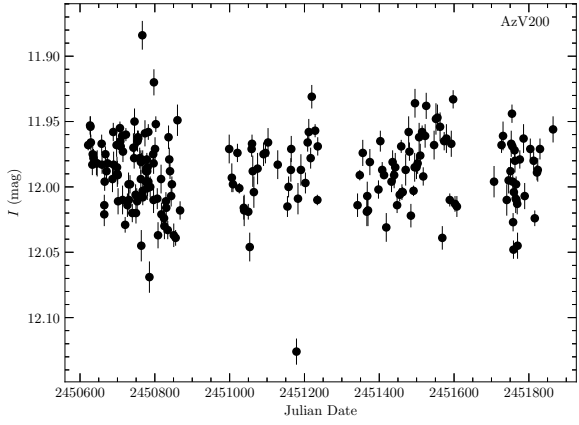
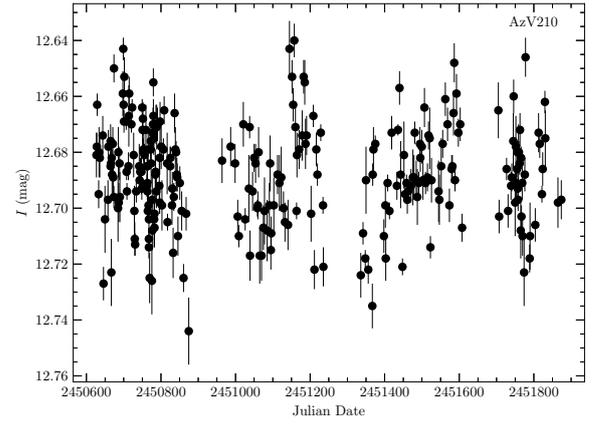
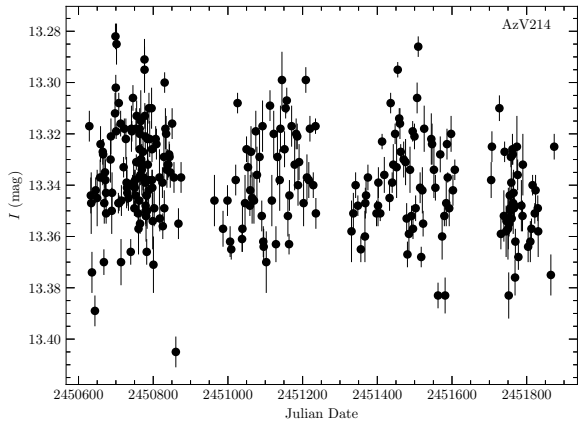
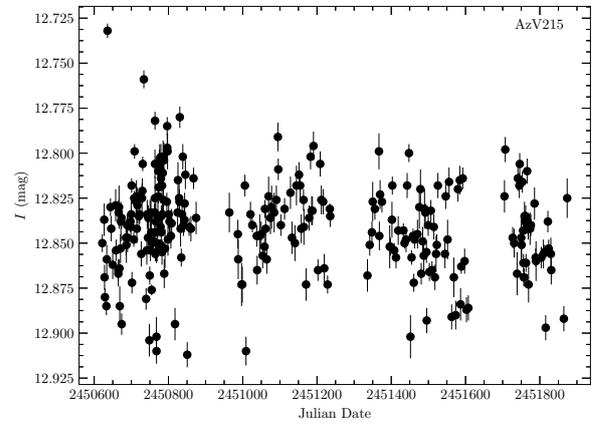
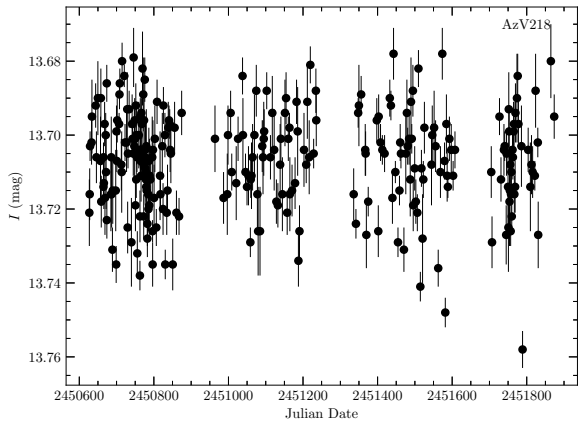
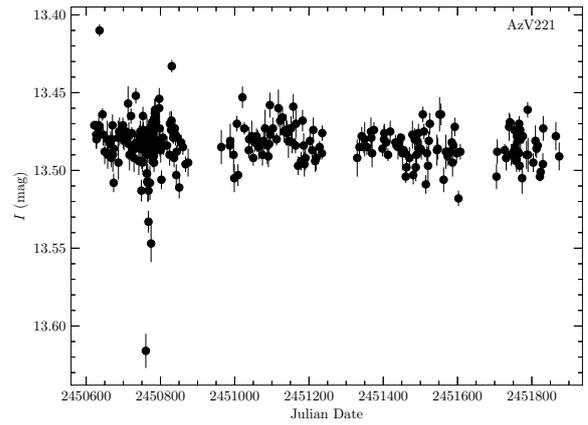

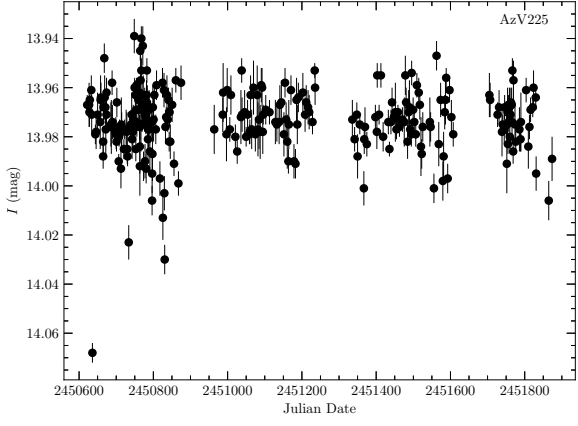
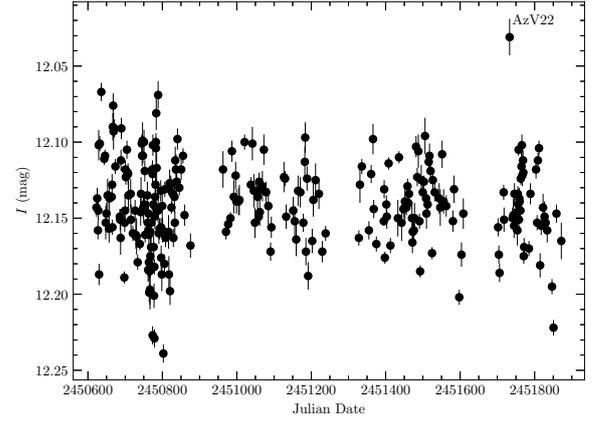

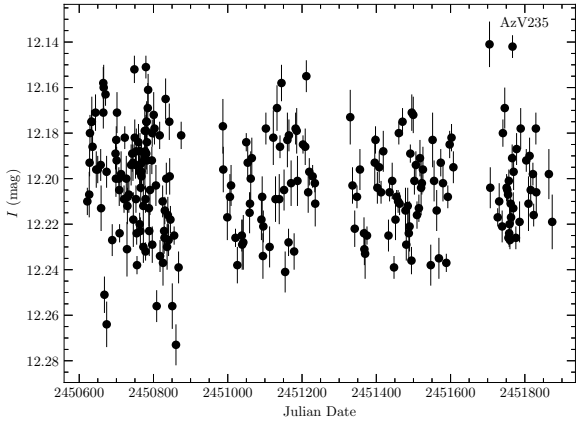
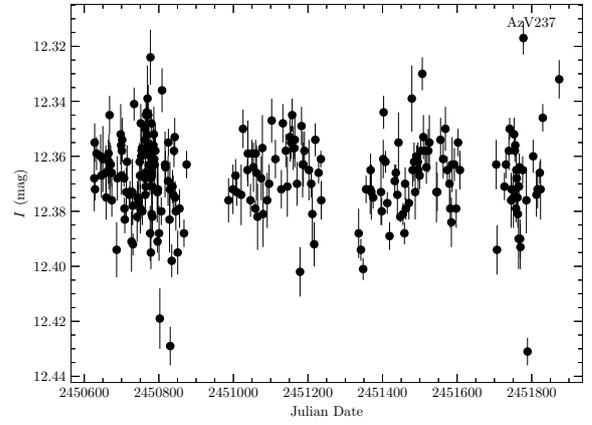

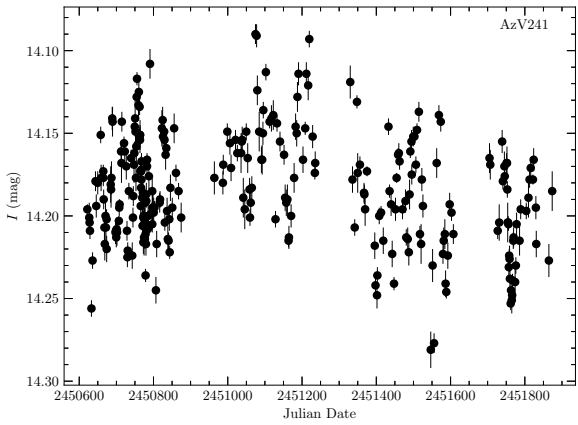
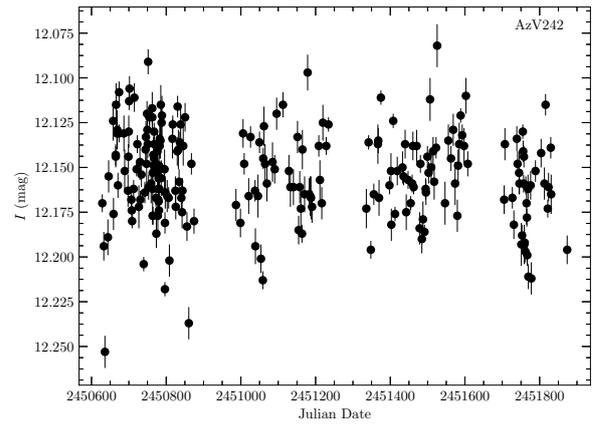

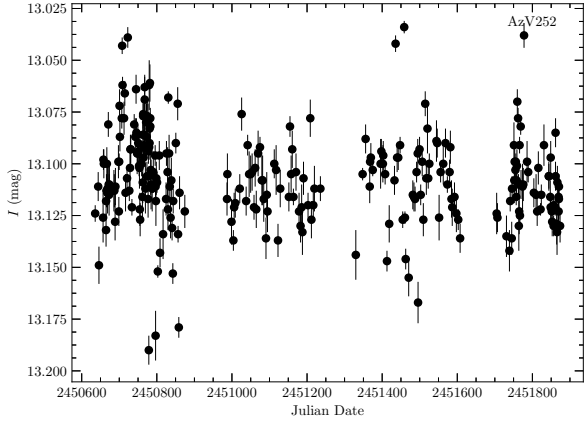
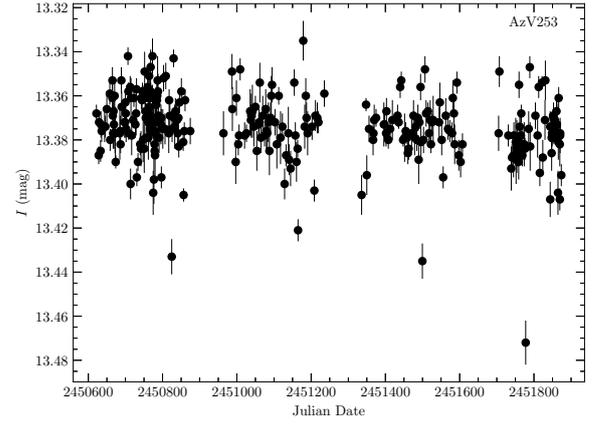

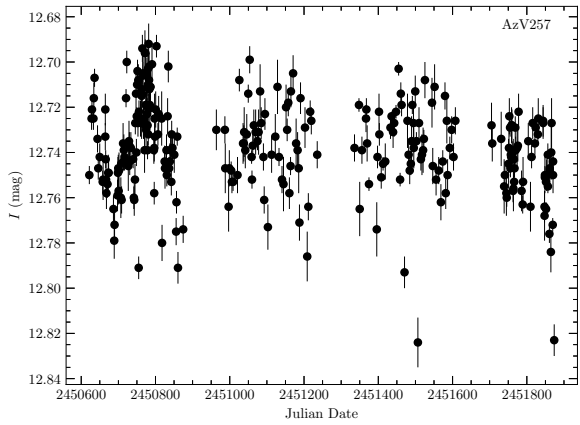
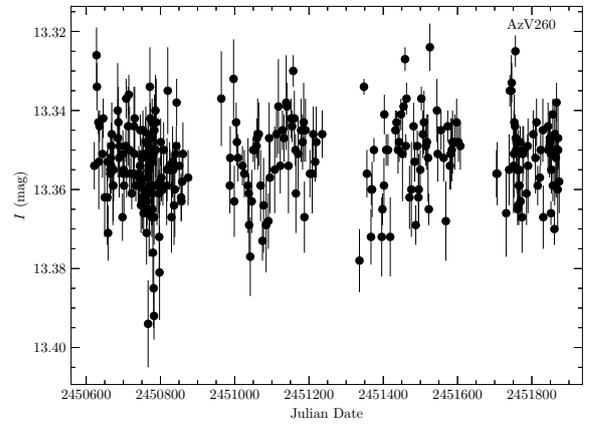

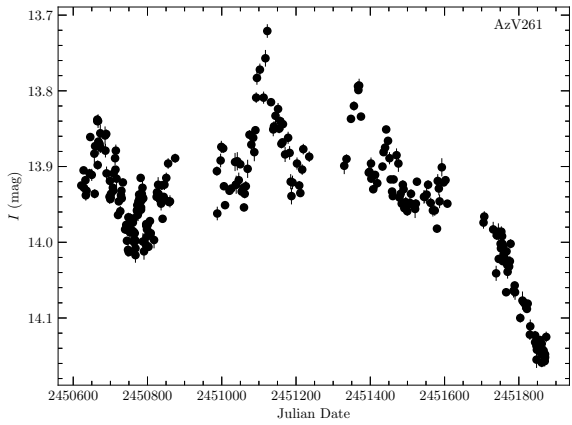
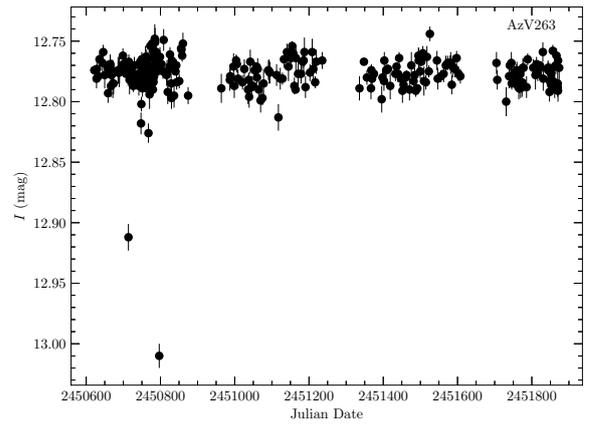

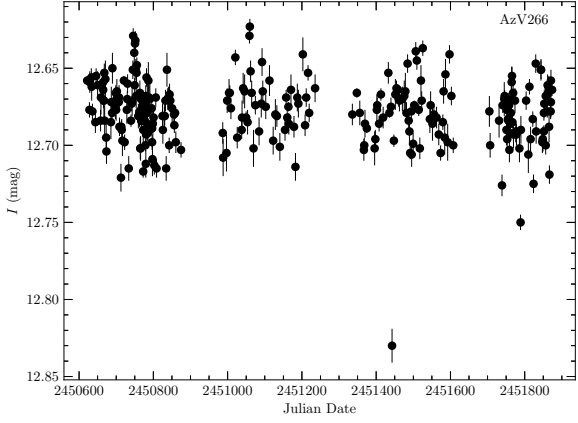

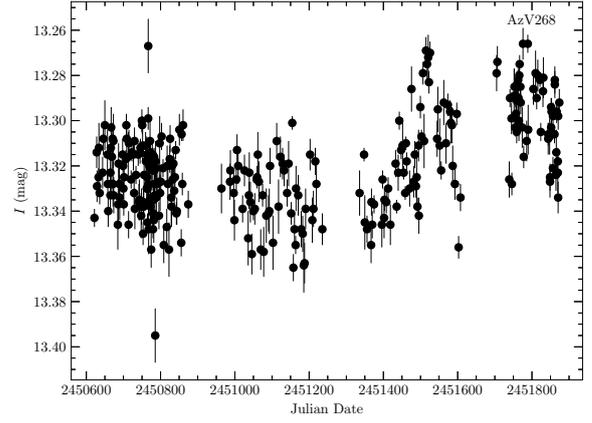

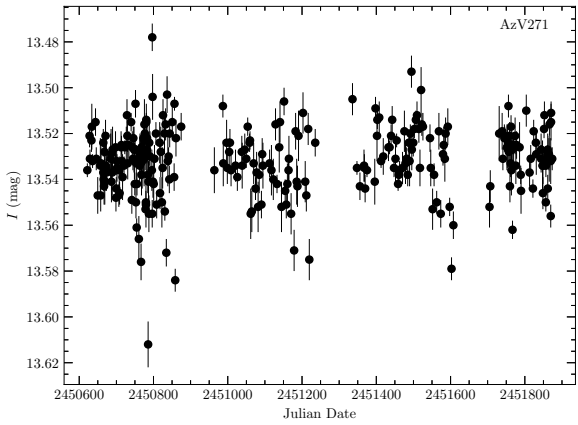

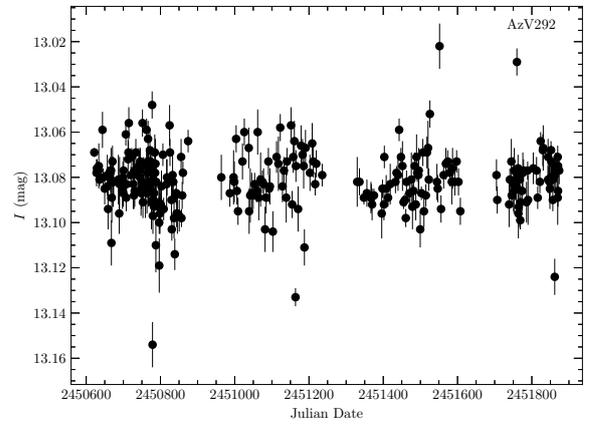

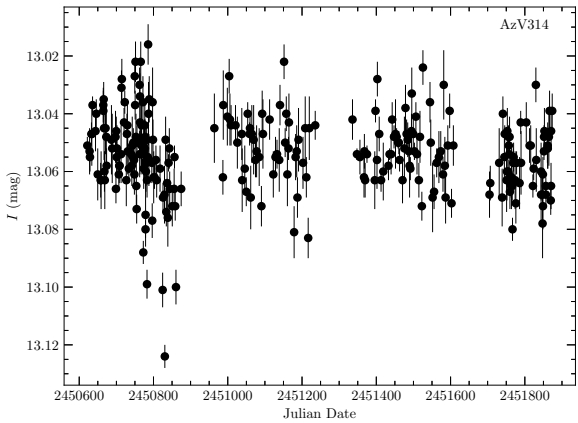

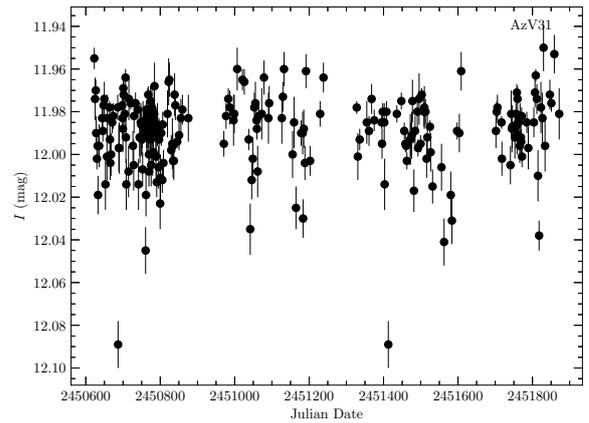

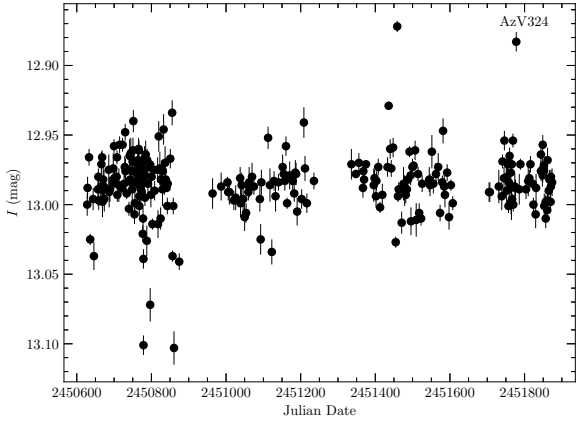

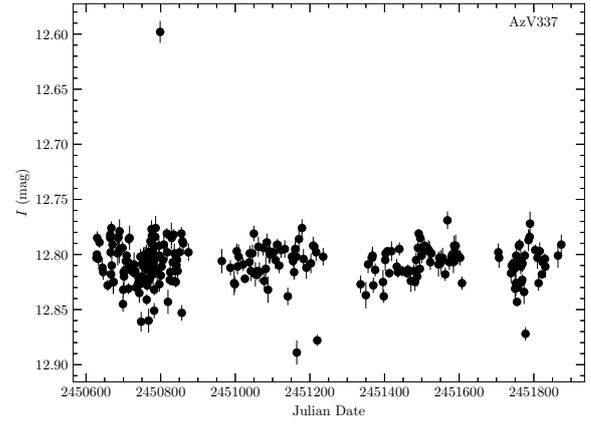

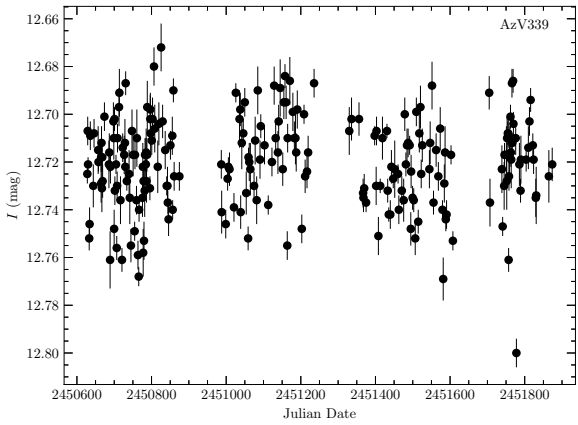

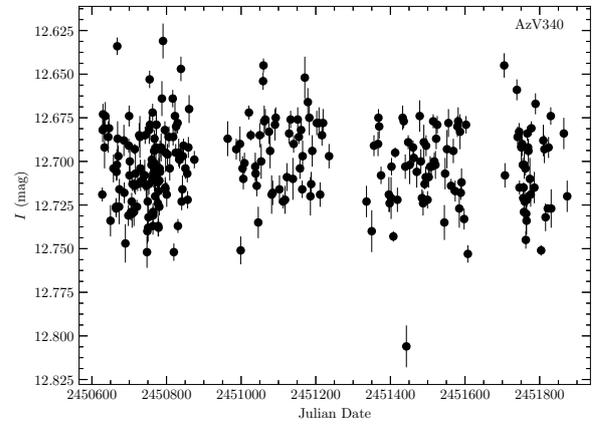

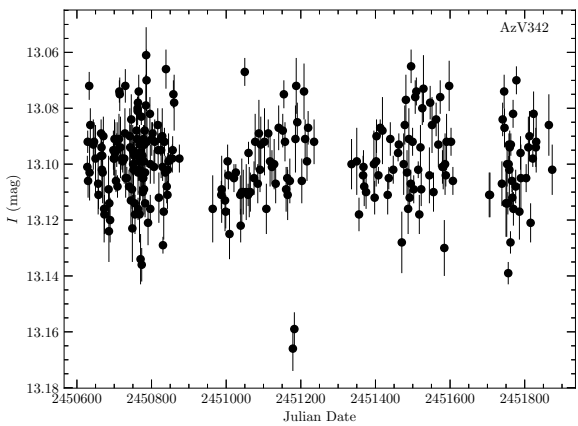

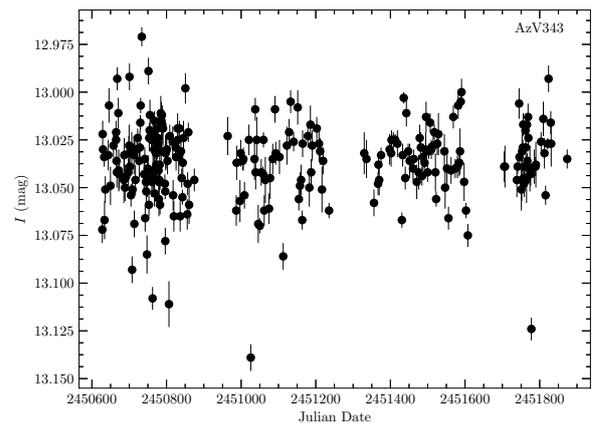

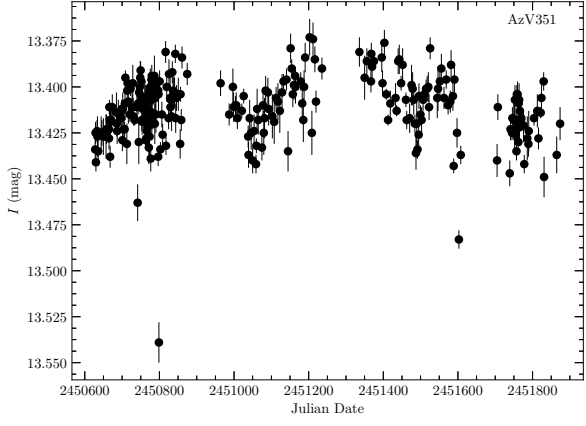
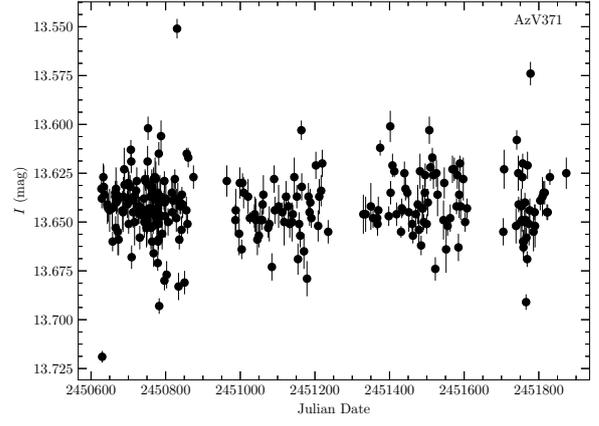

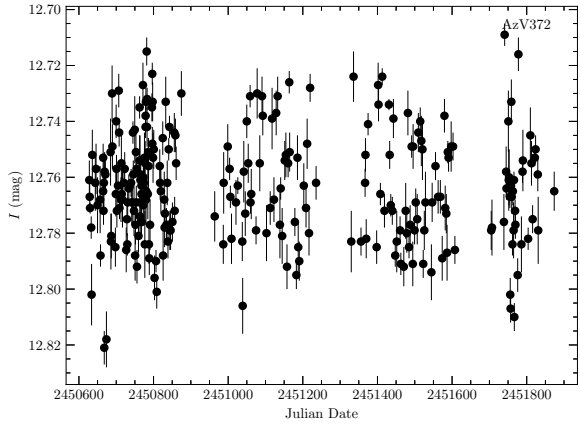
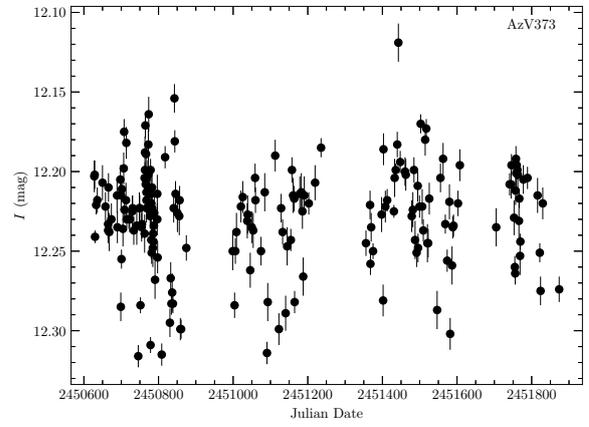

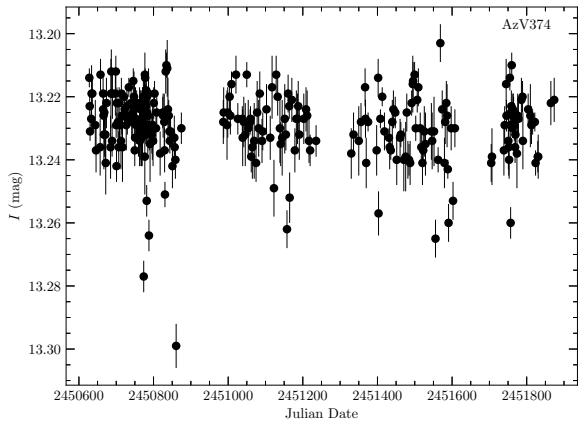
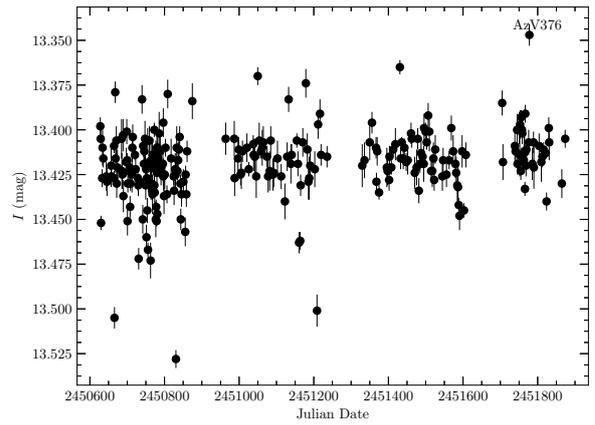

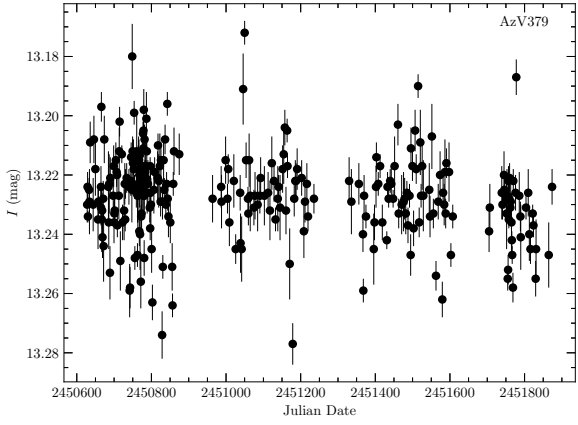

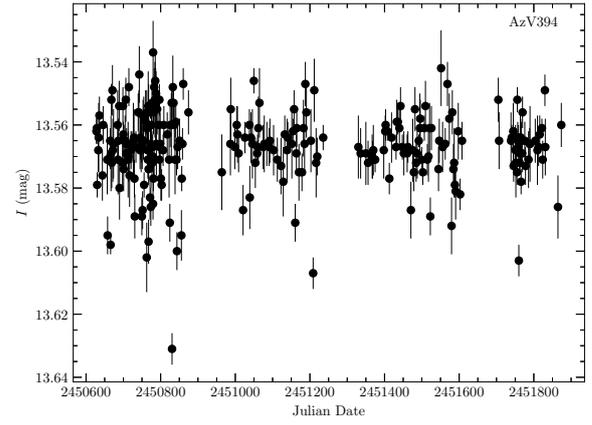

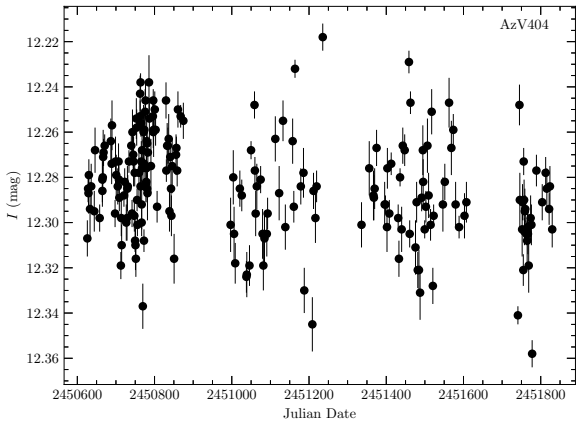

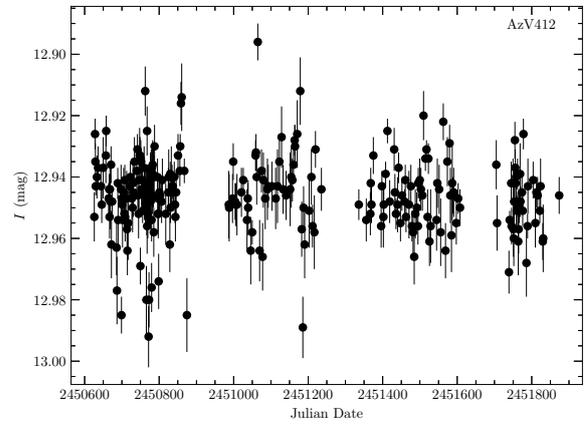

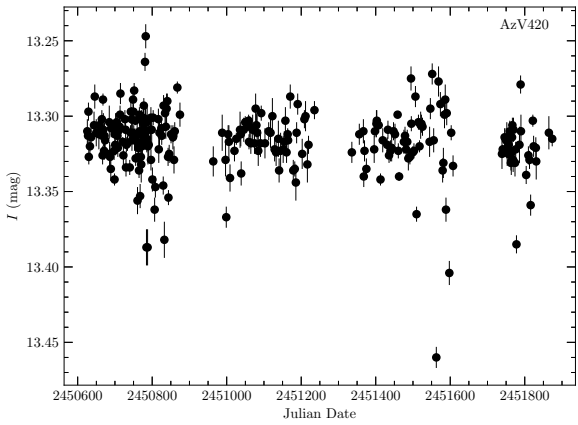

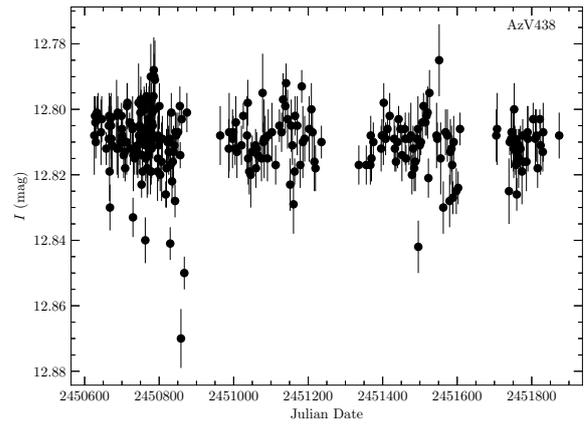

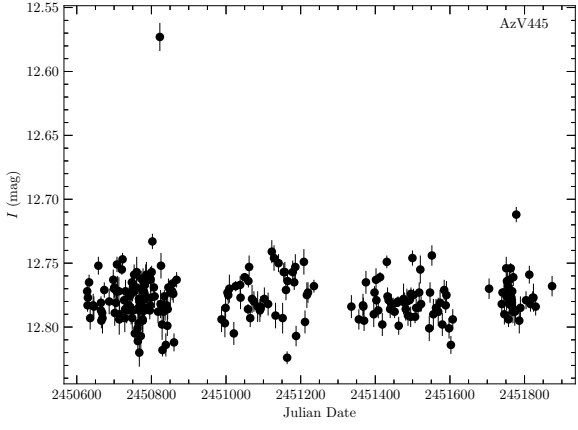
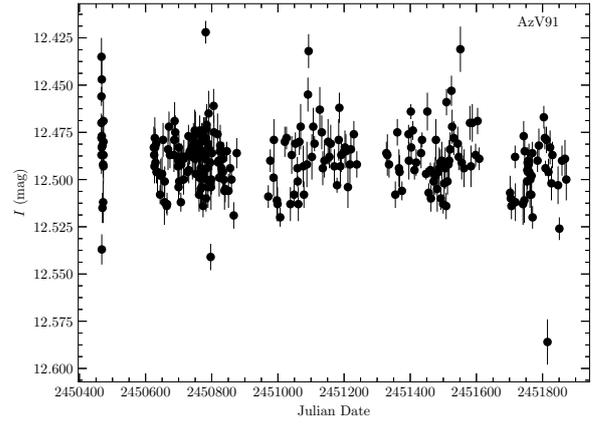

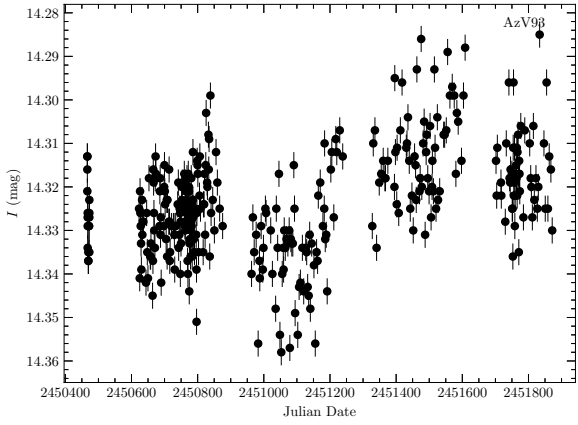
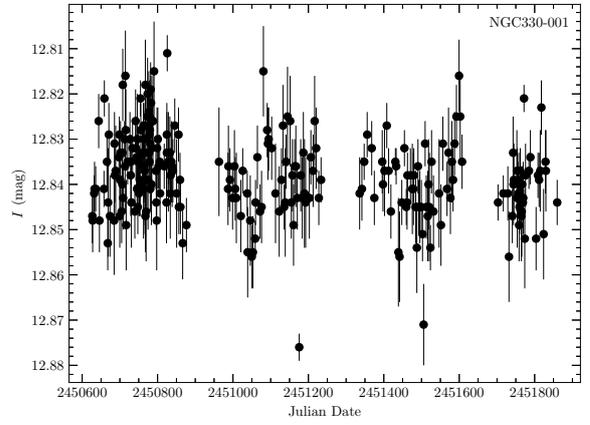

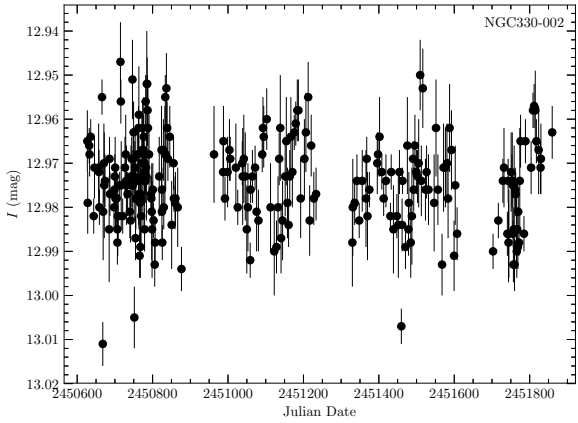
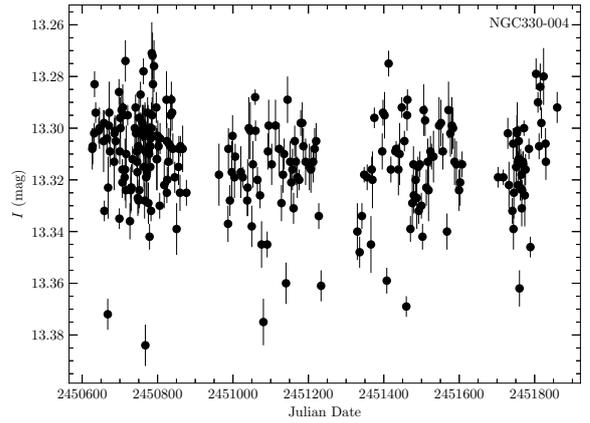

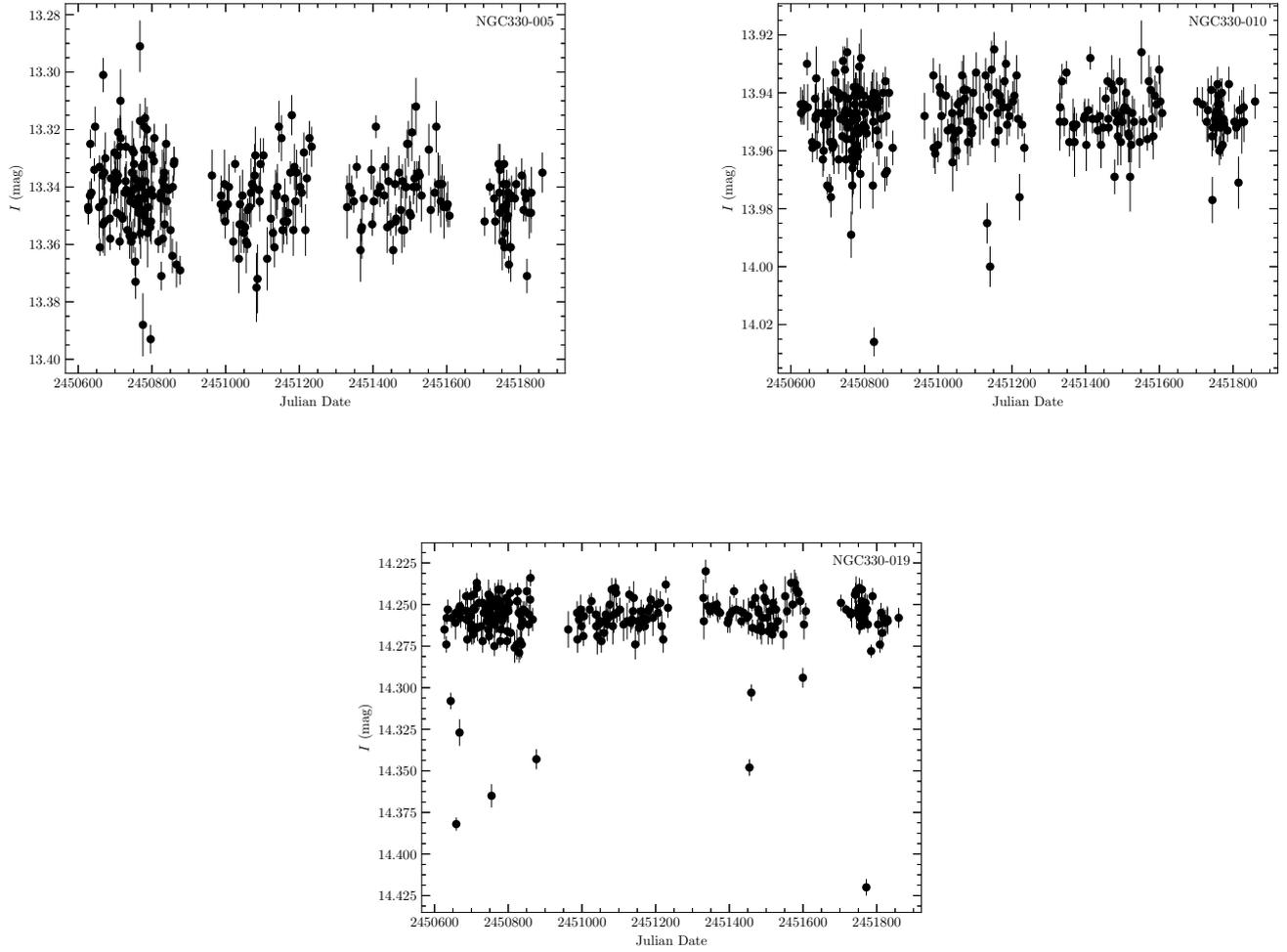

Fig. A1: Light curves of 63 Bsgs having data from the OGLE II survey. The Julian date is plotted against the measured *I*-band magnitude.

# Appendix B: Lomb Scargle periodgram of Bsgs having OGLE II data

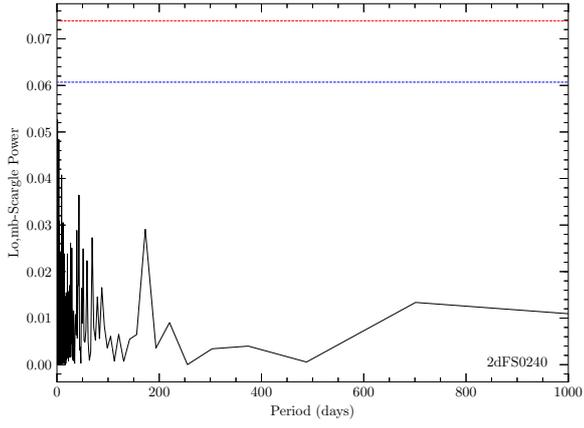
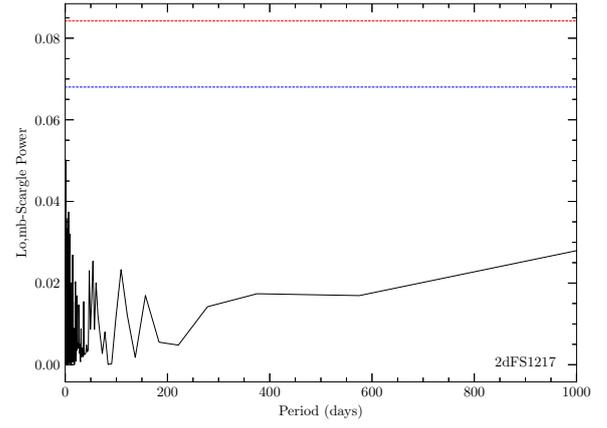

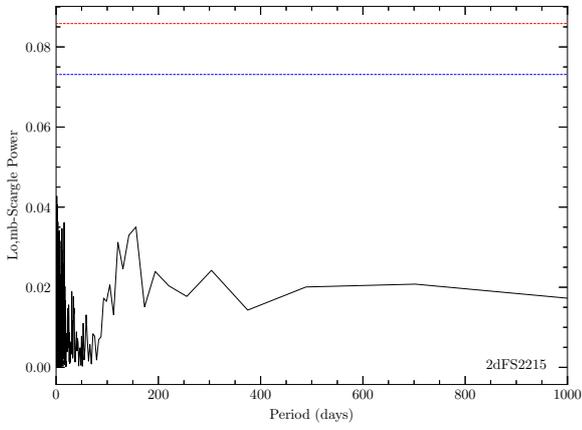
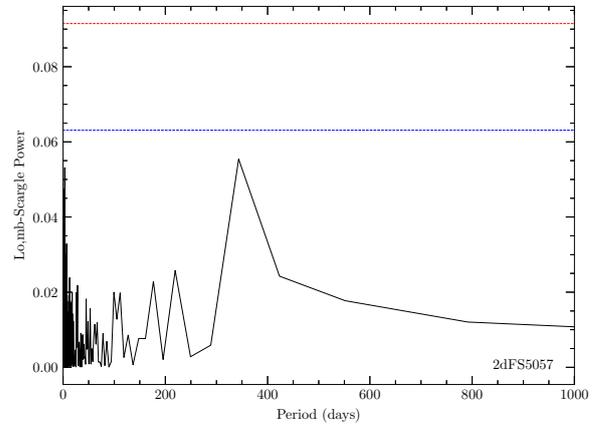

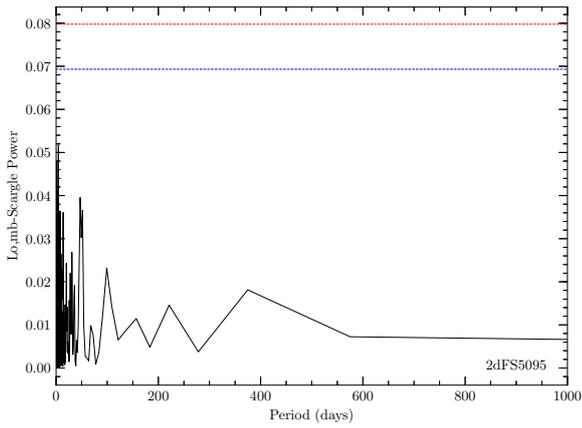
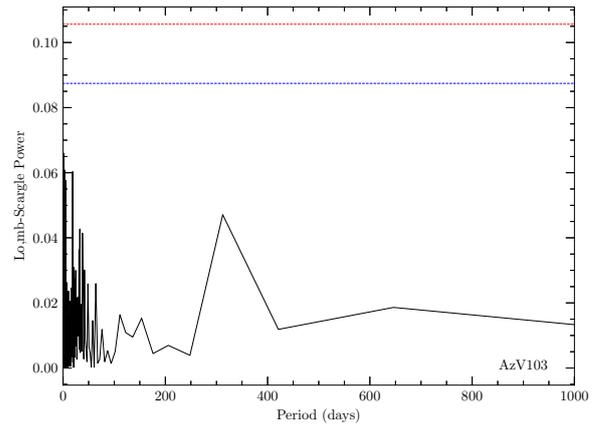

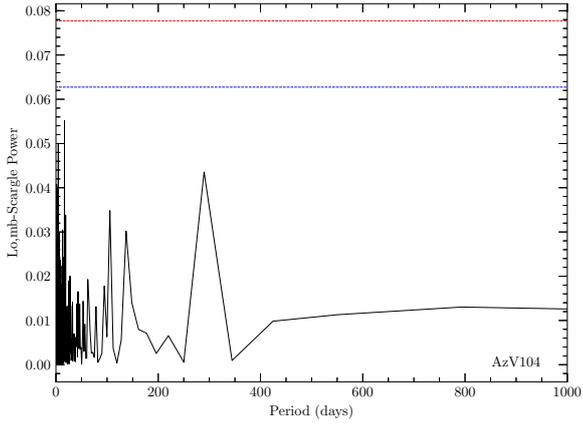
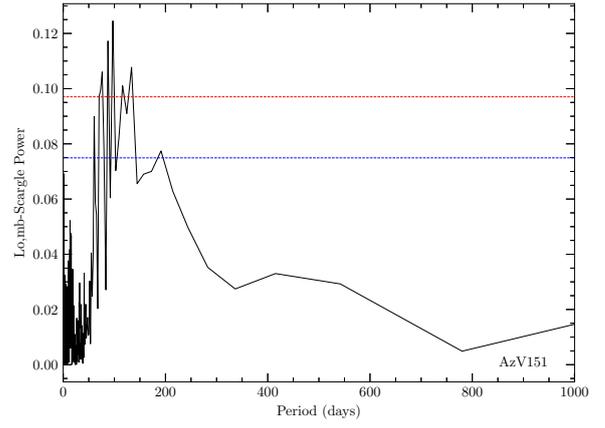

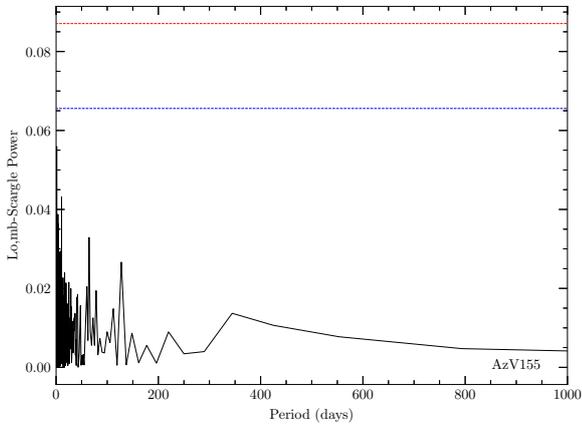
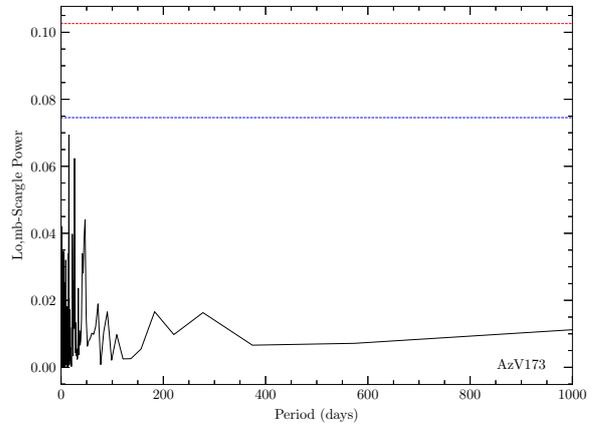

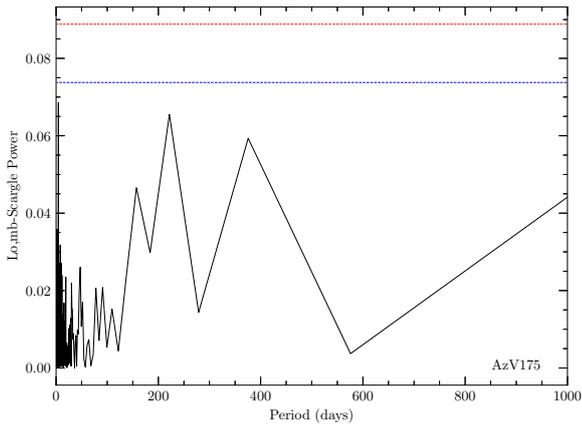
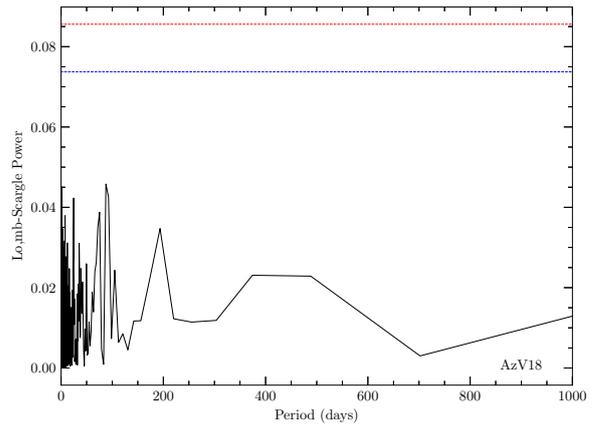

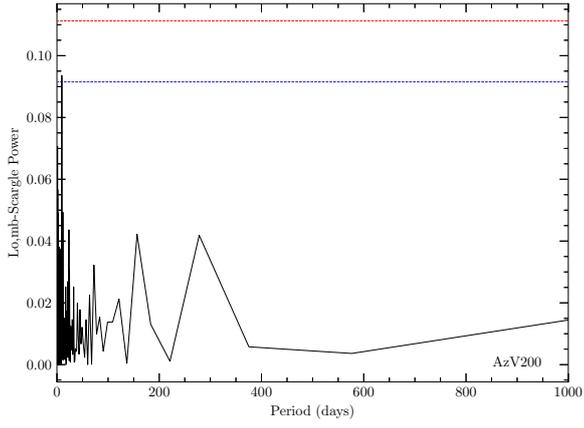
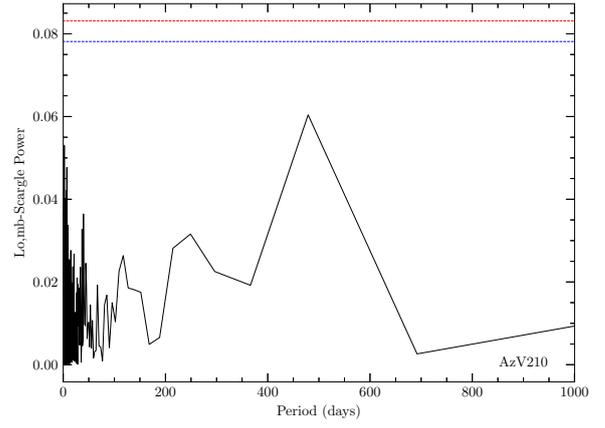

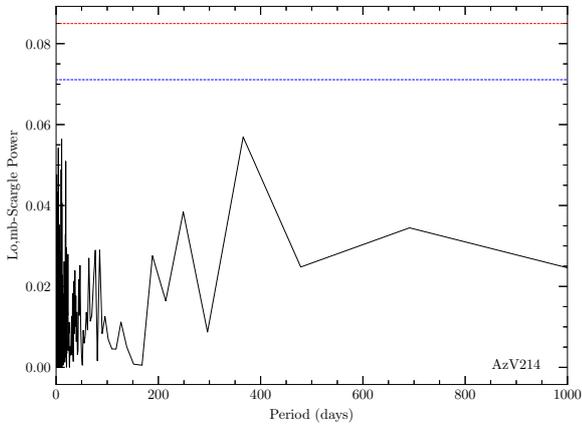
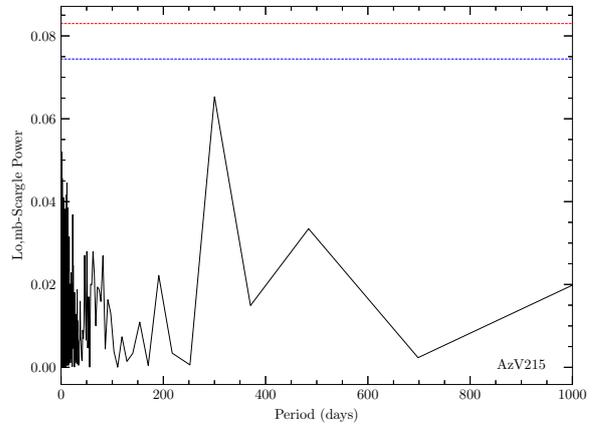

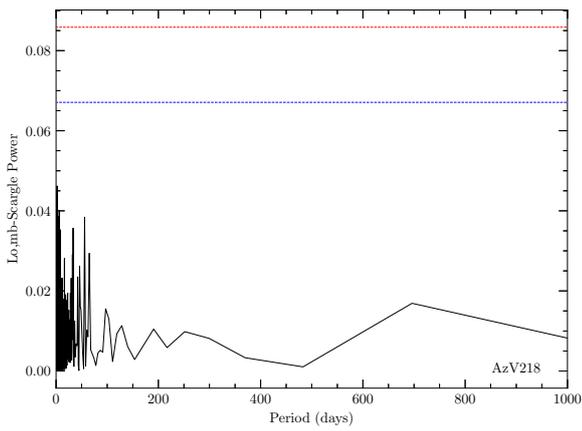
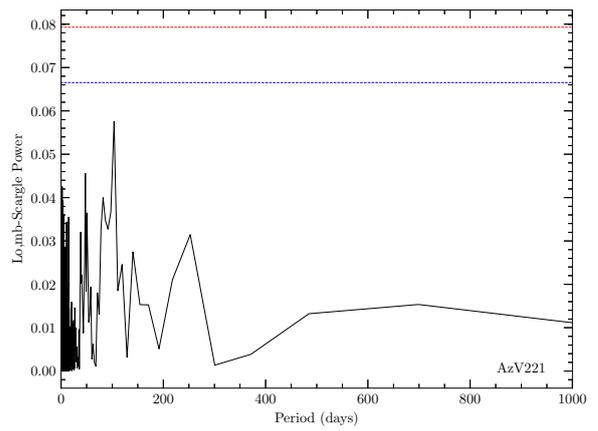

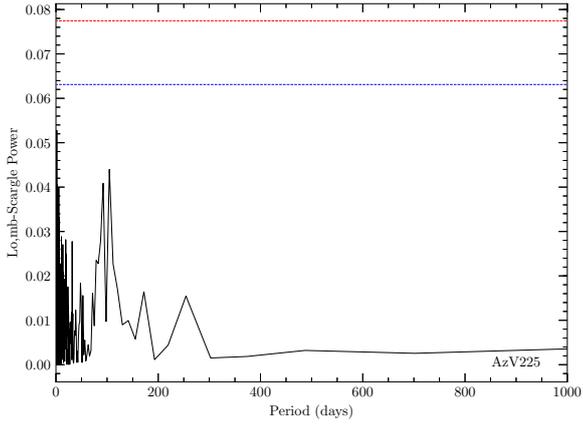
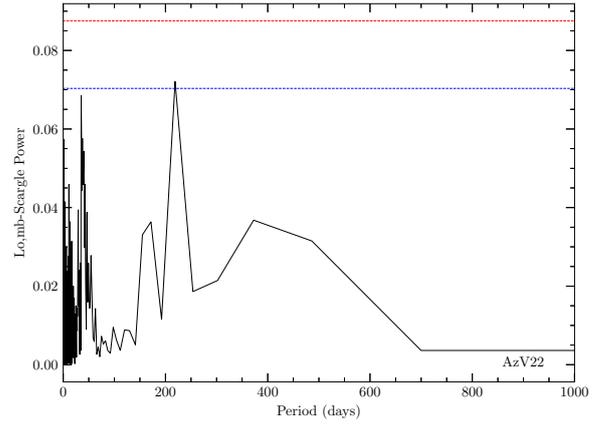
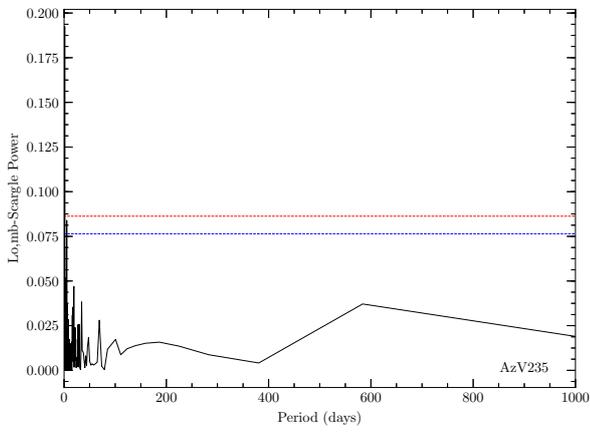
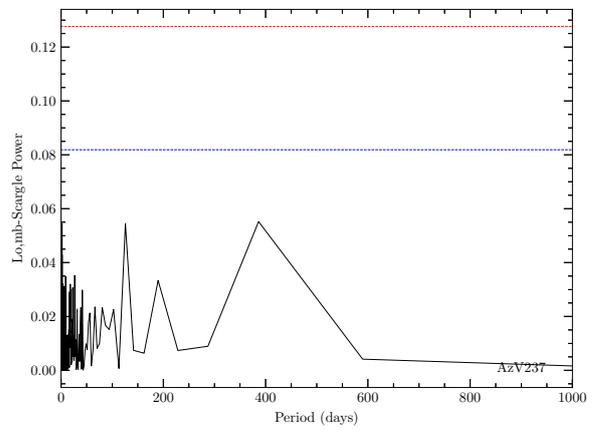
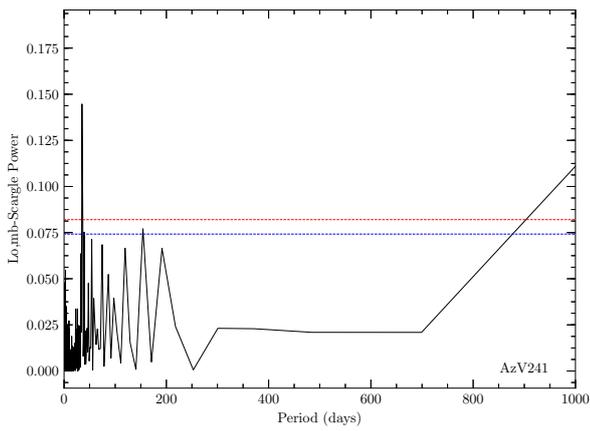
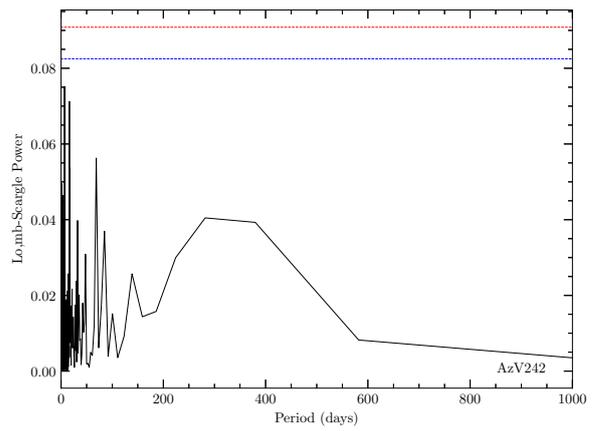

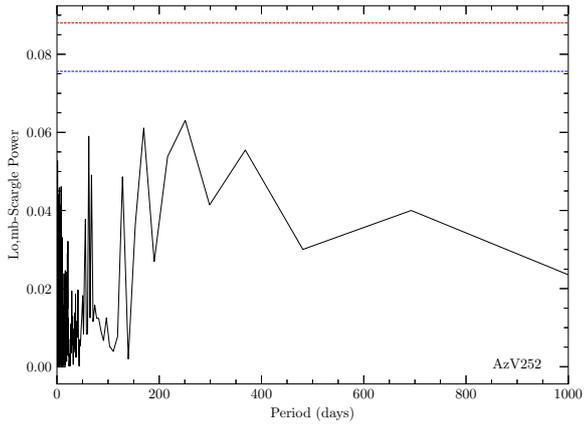

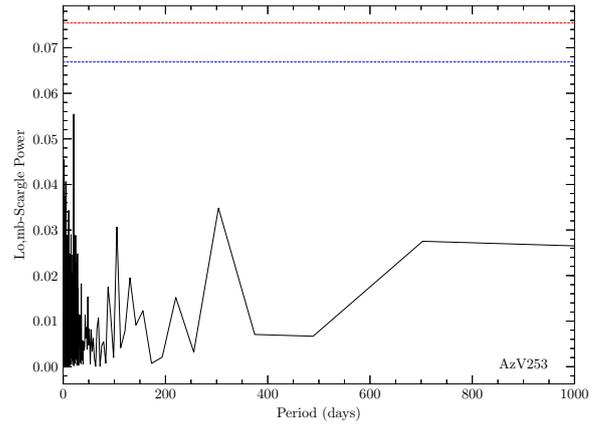

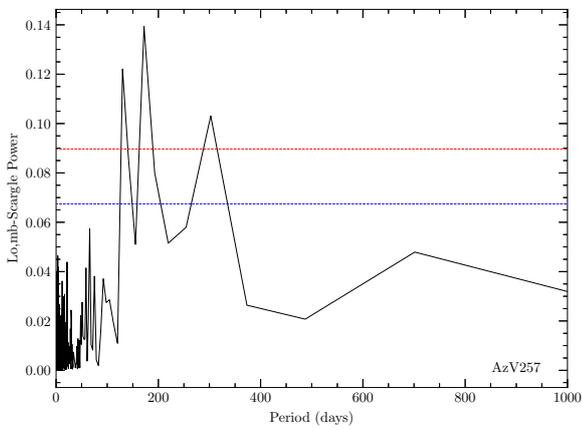

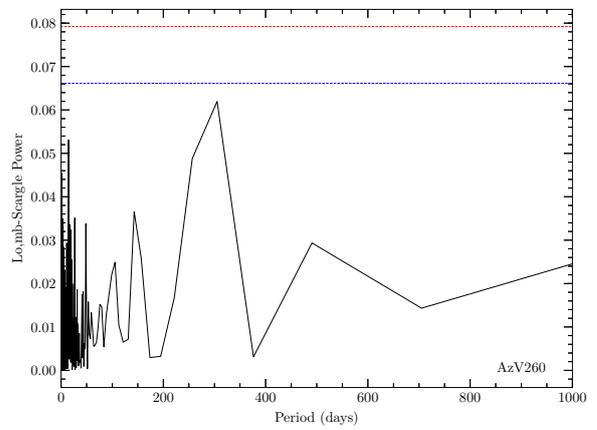

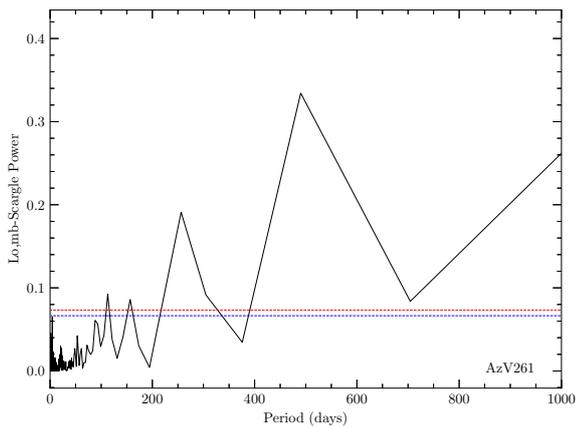

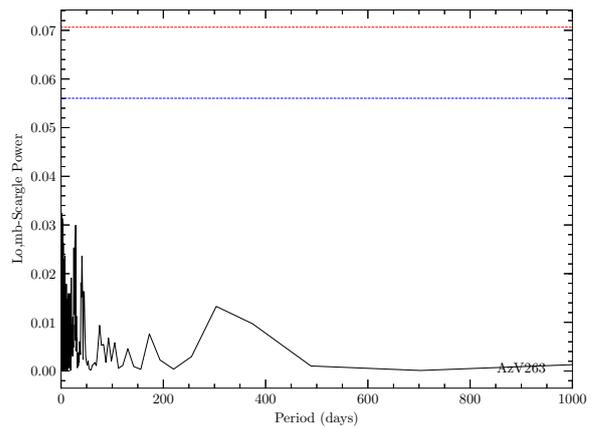

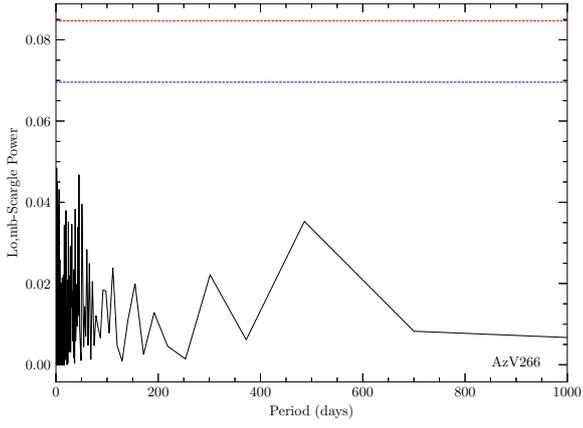
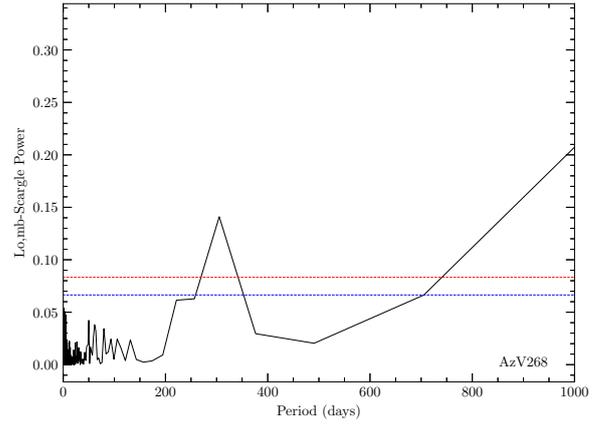

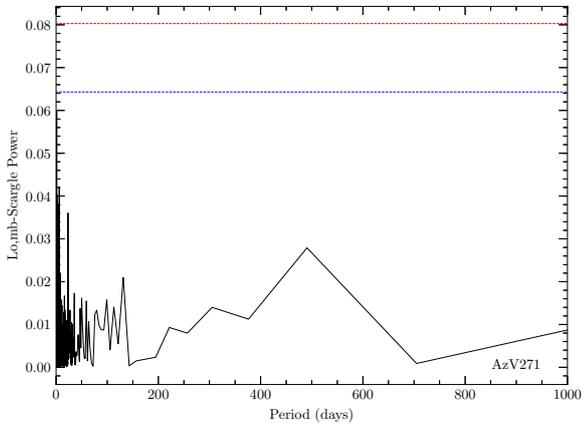
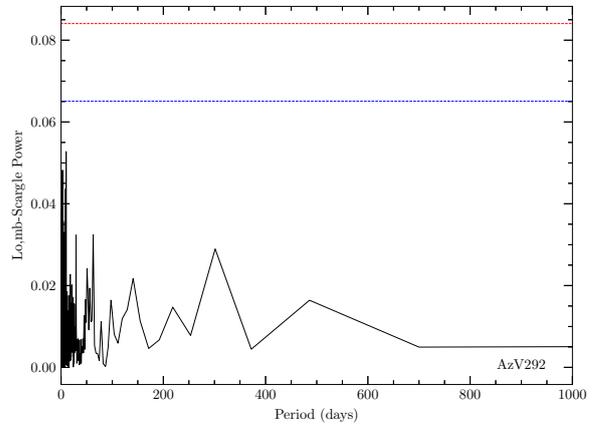

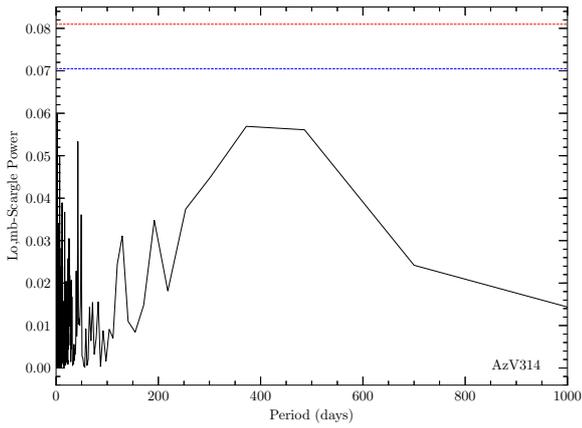
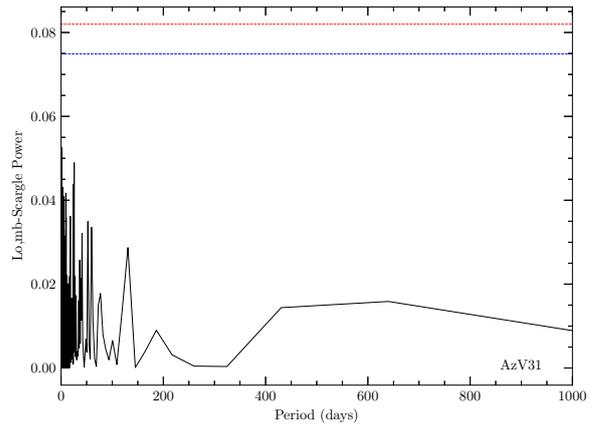

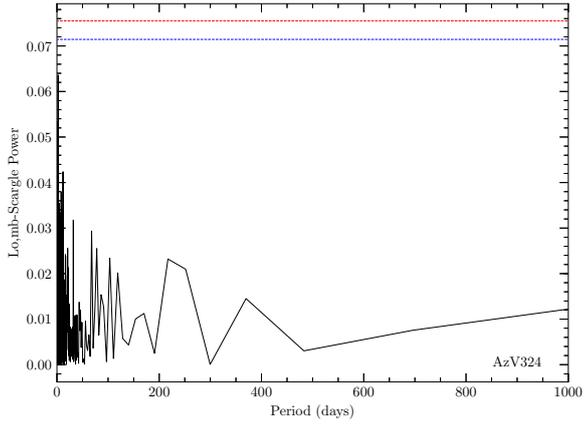
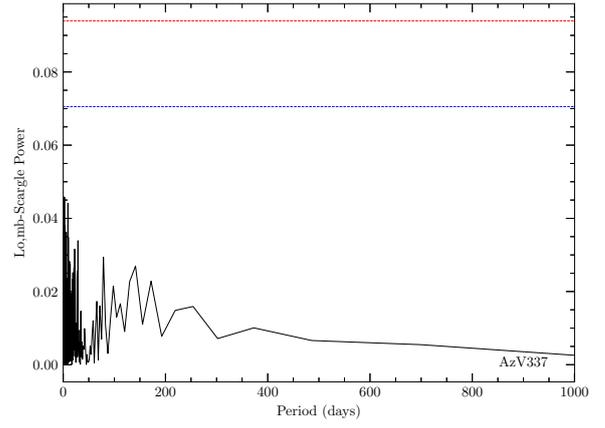
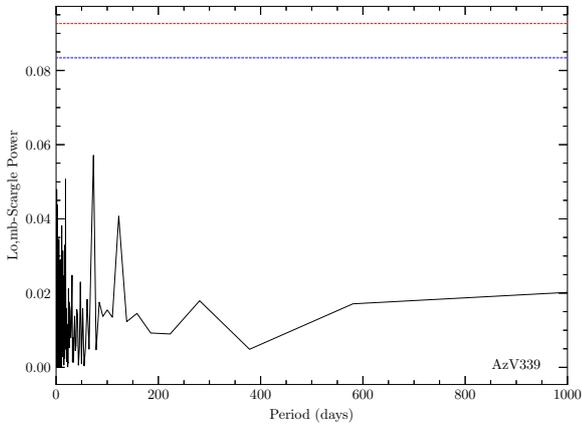
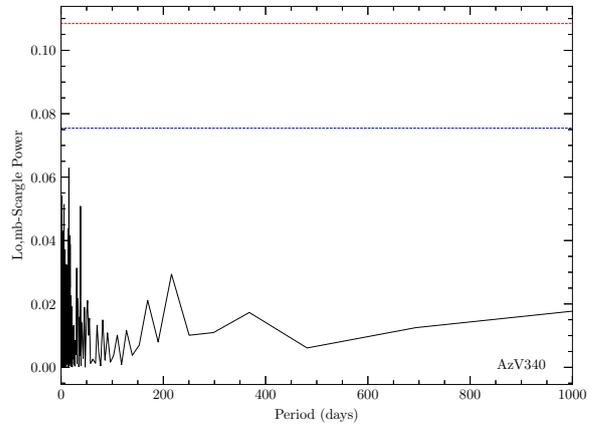
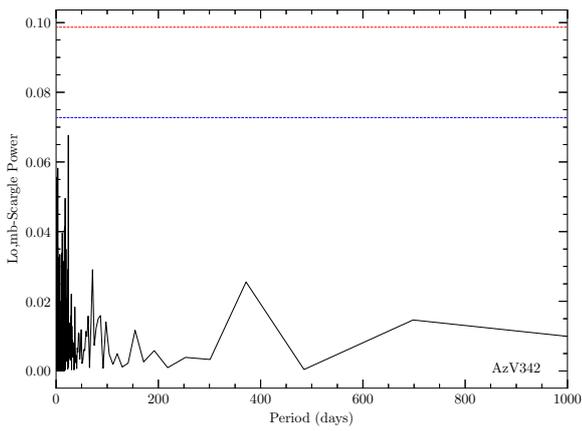
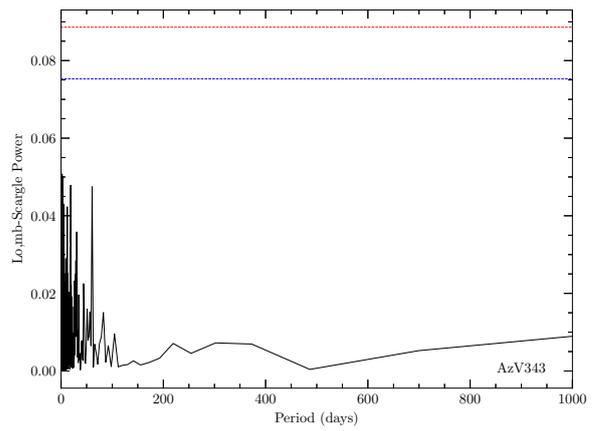

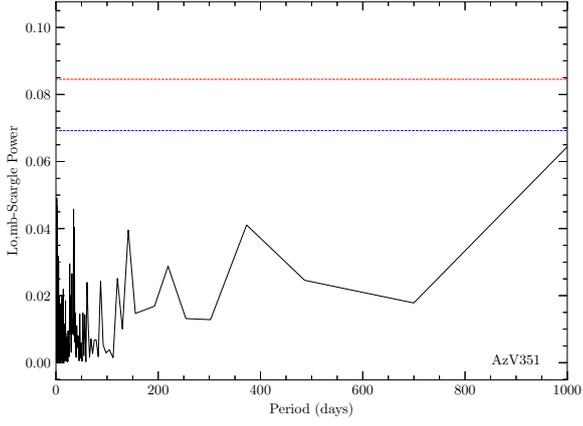

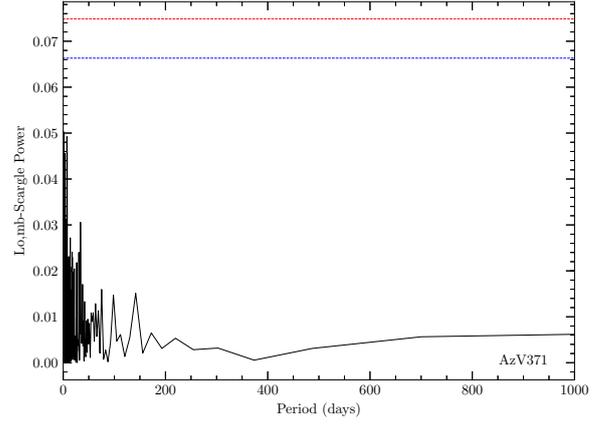

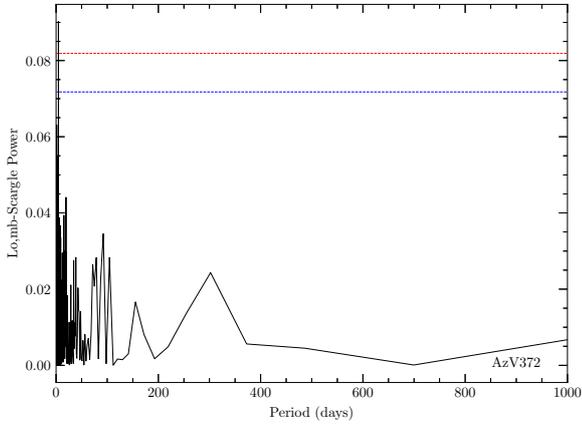

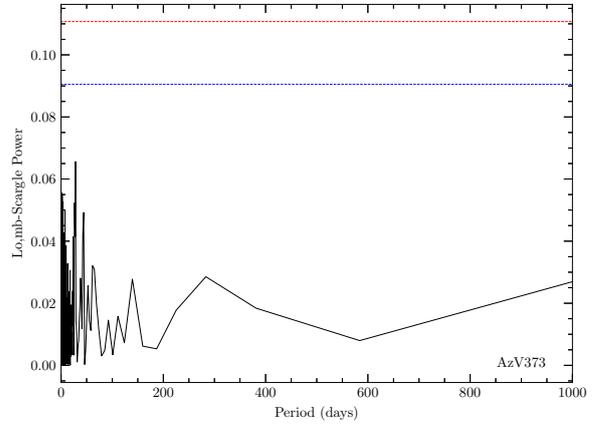

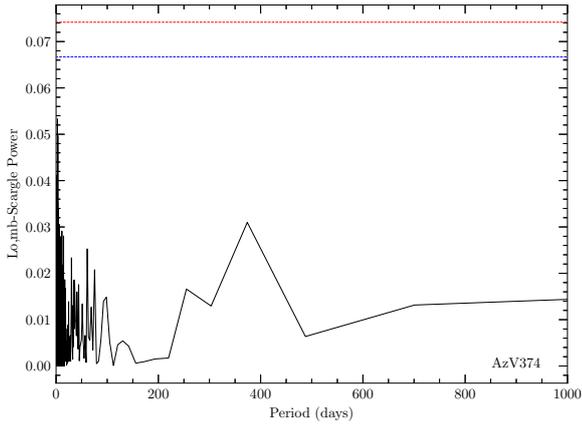

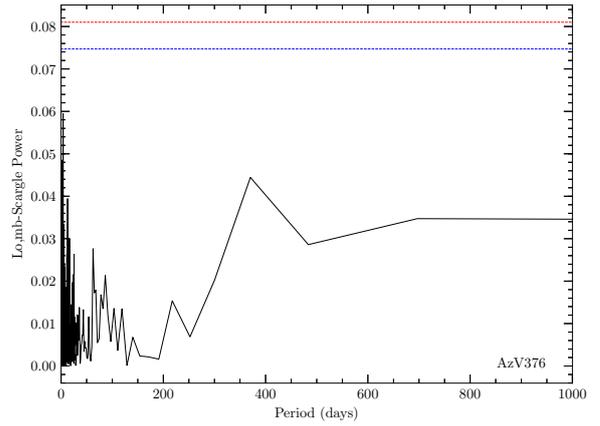

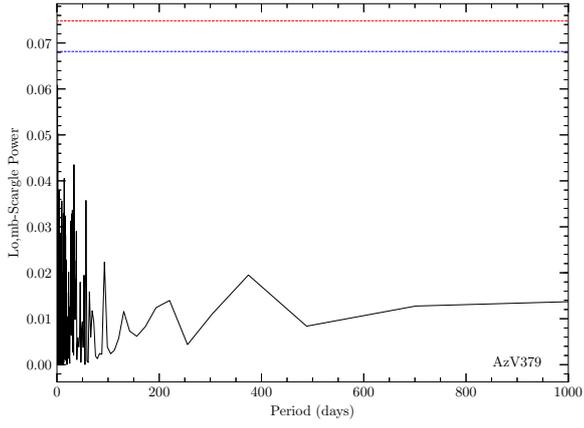
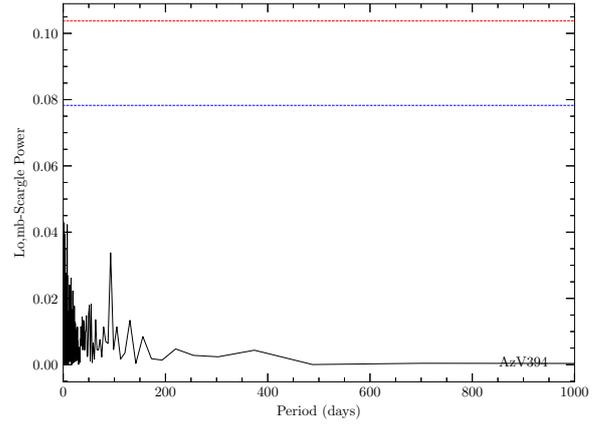

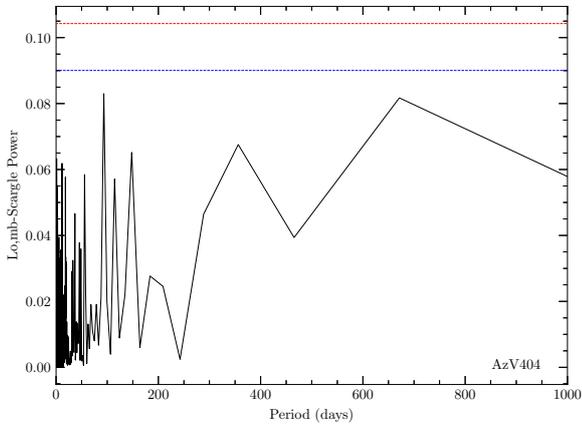
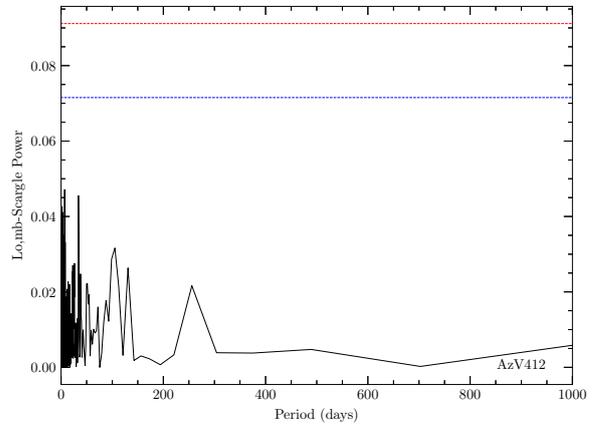

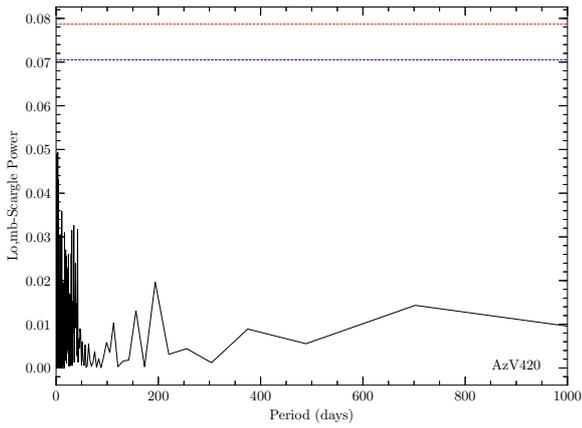
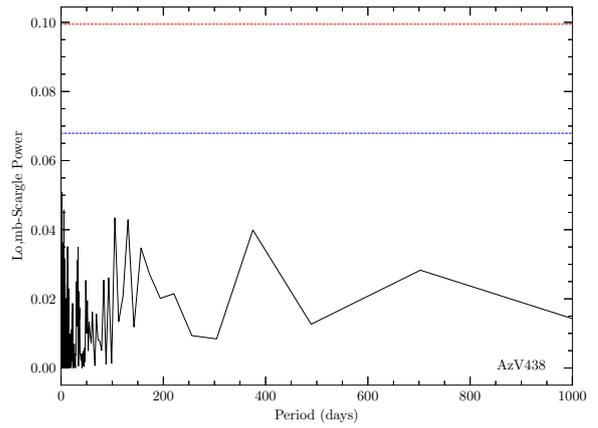

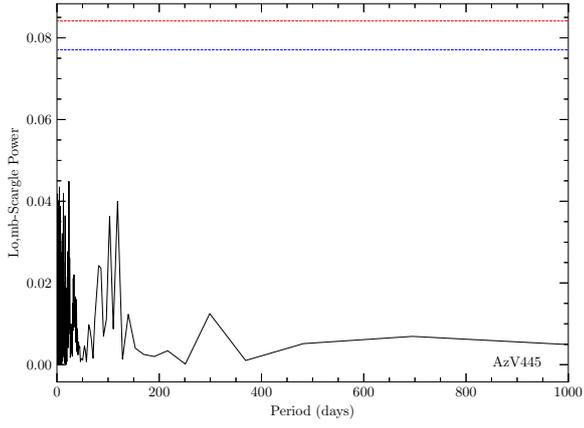
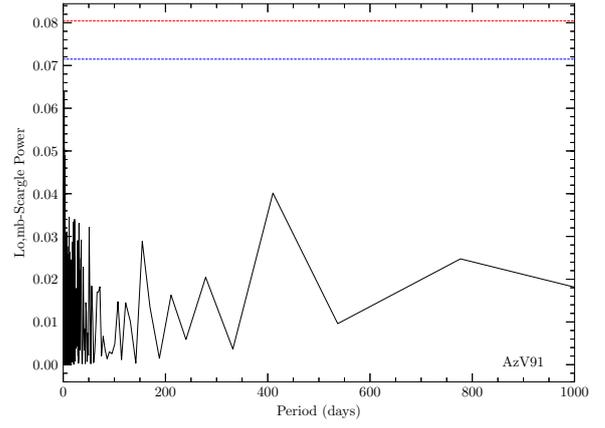

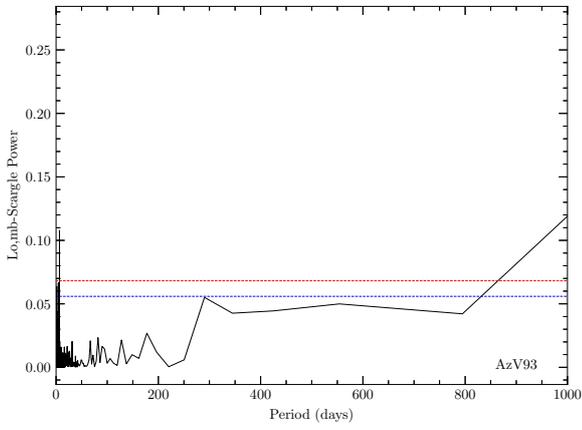
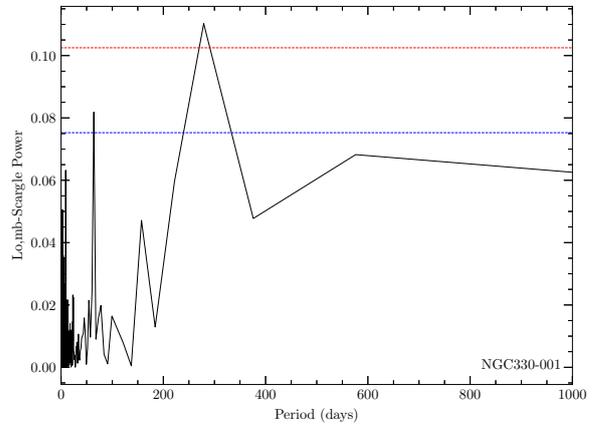

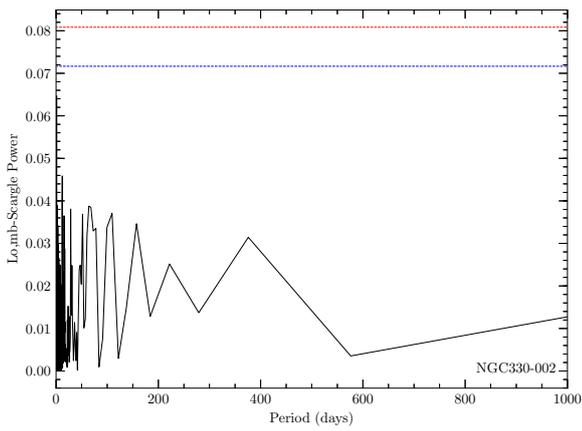
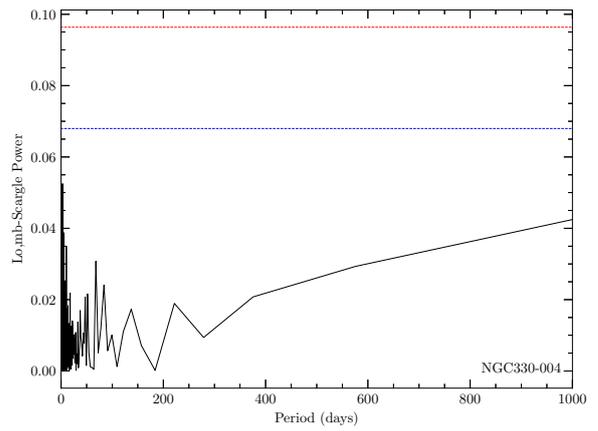

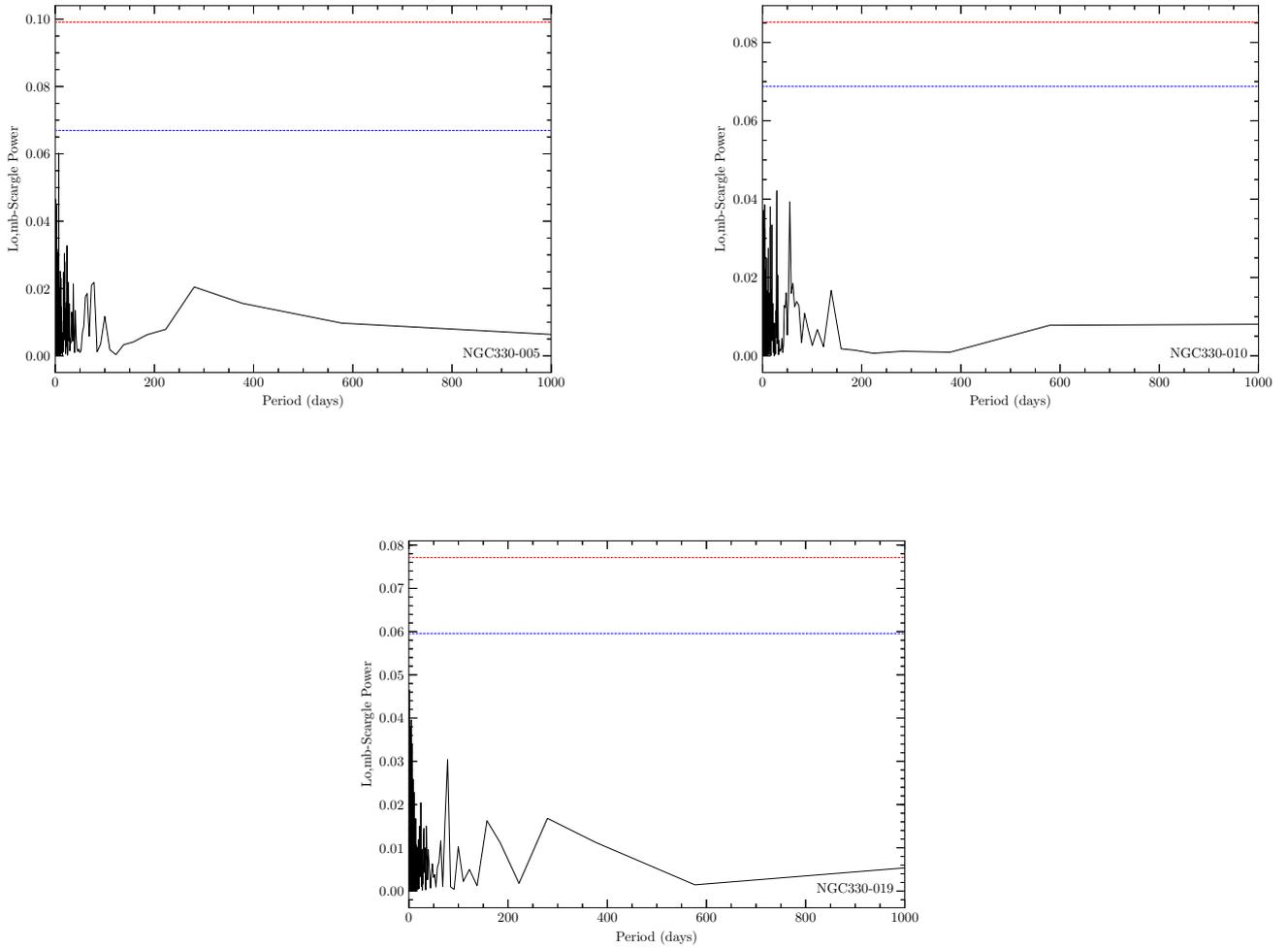

Fig. B1: Lomb Scargle periodgram of 64 Bsgs. The period in days is plotted against the power. The red and blue lines represent the 99% and 95% significance respectively calculated from the Monte Carlo bootstrap simulations.

**Appendix C: $\delta$m-$\delta$t two-dimensional histogram of Bsgs with OGLE II data**

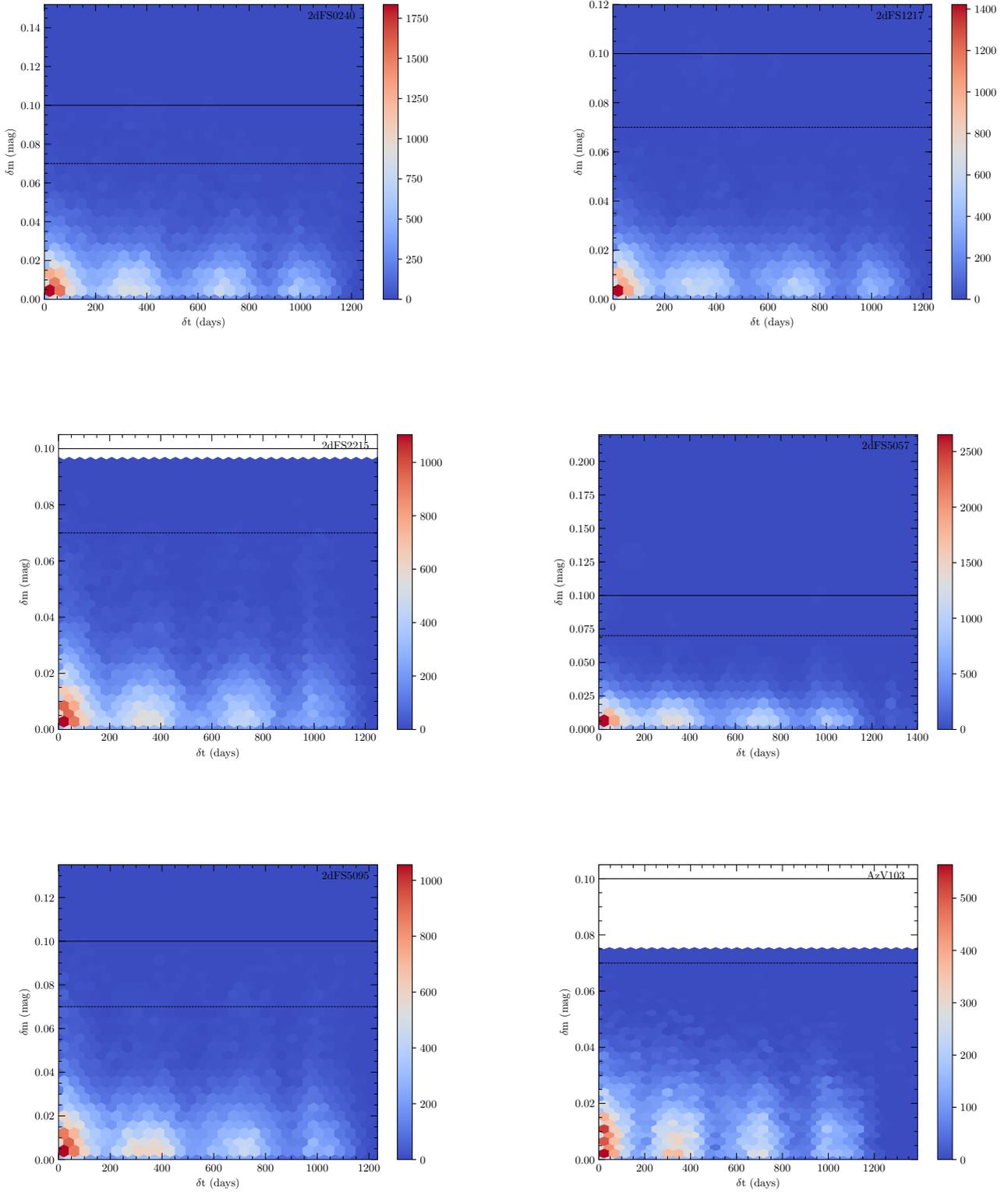

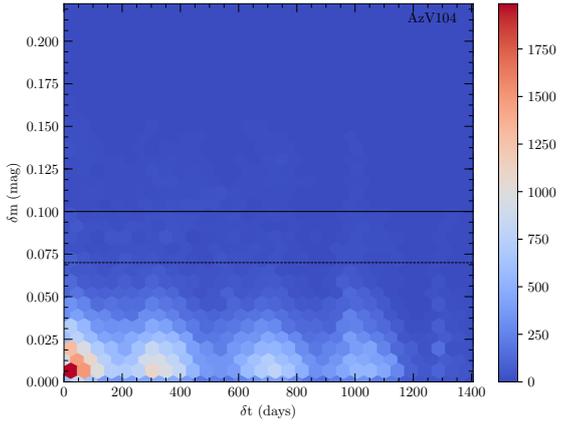
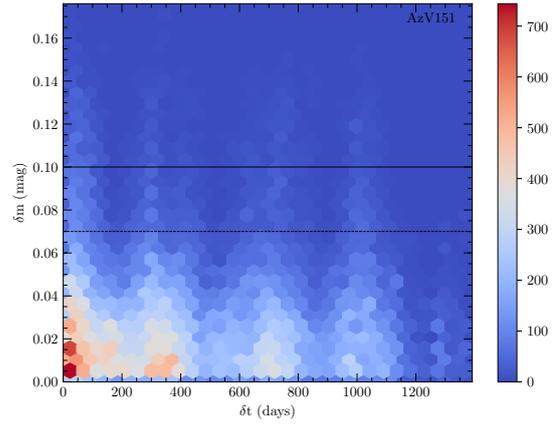

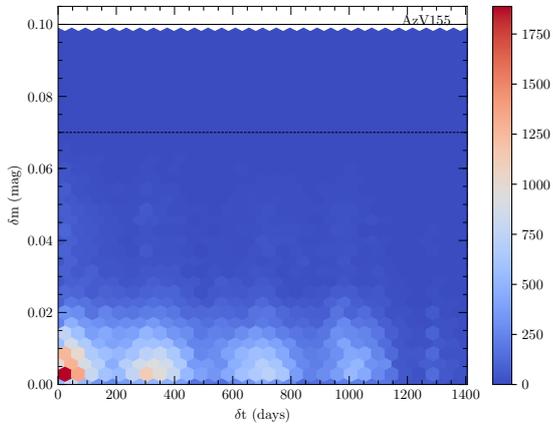
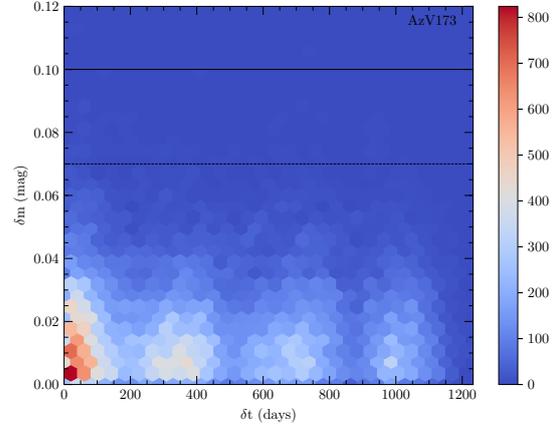

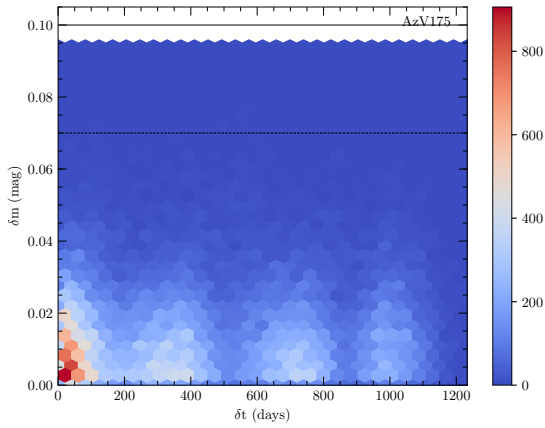
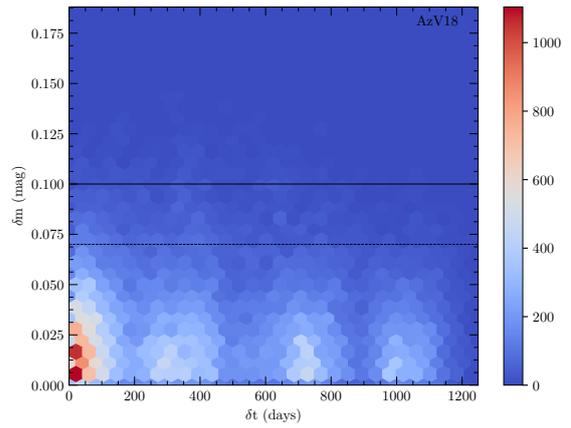

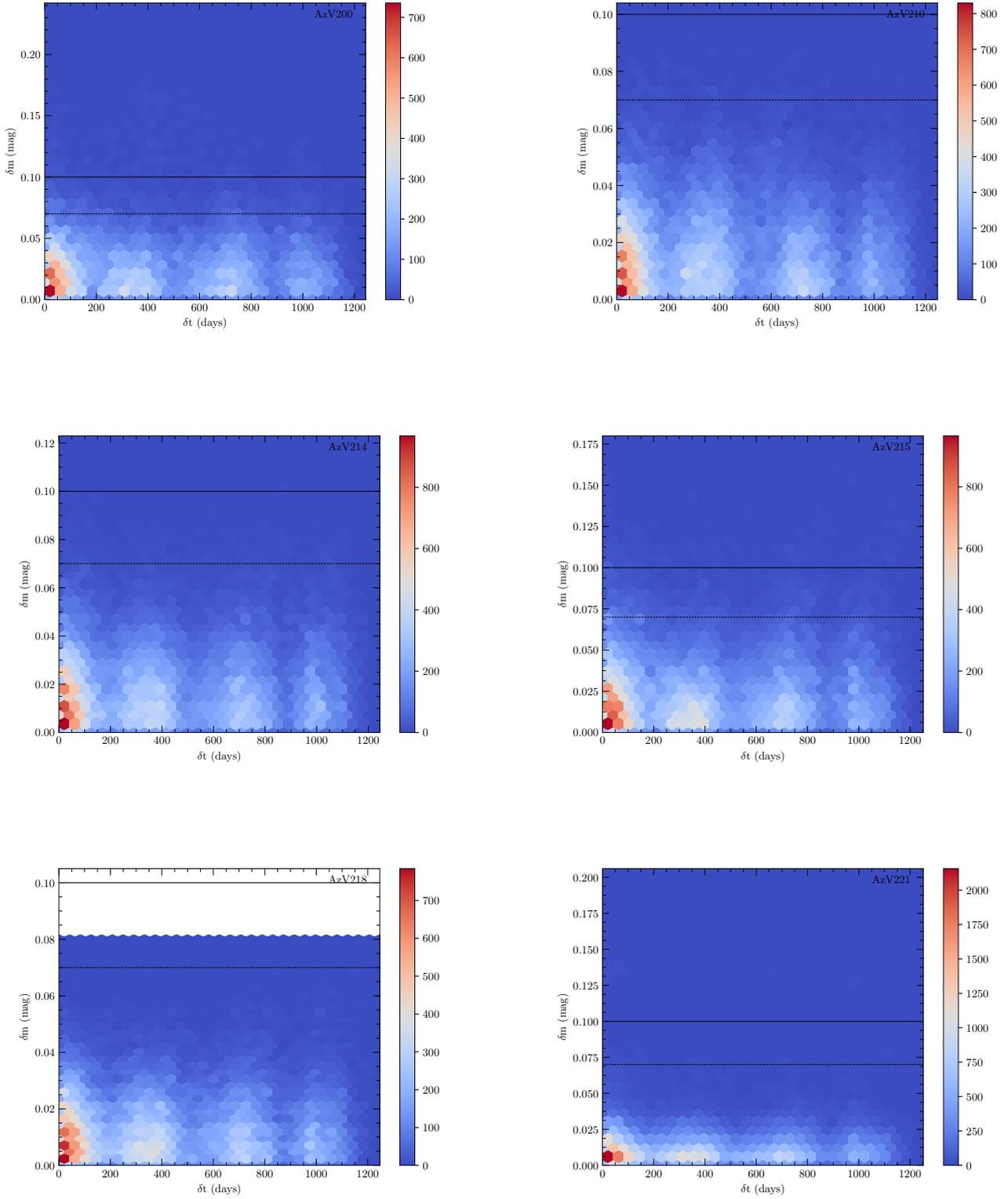

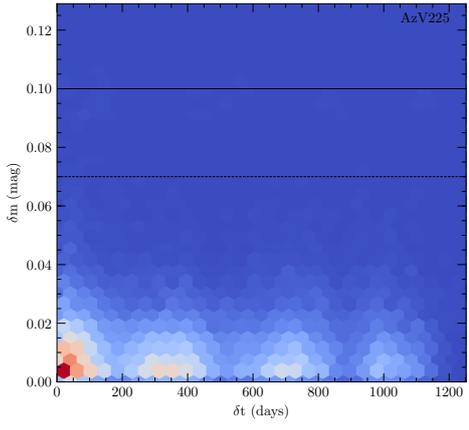
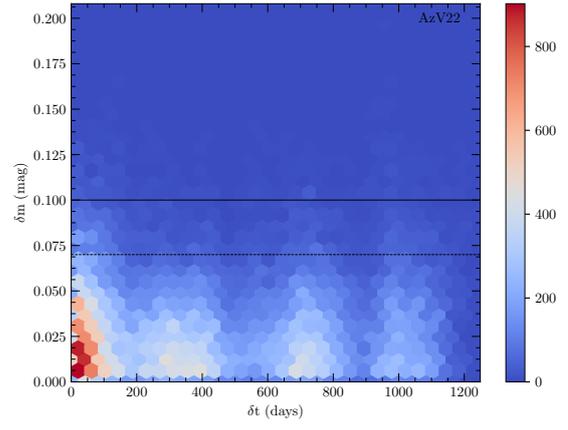

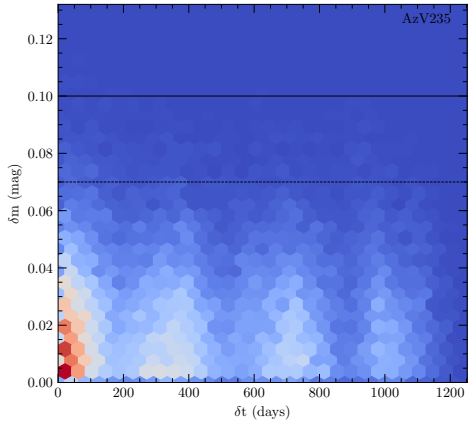
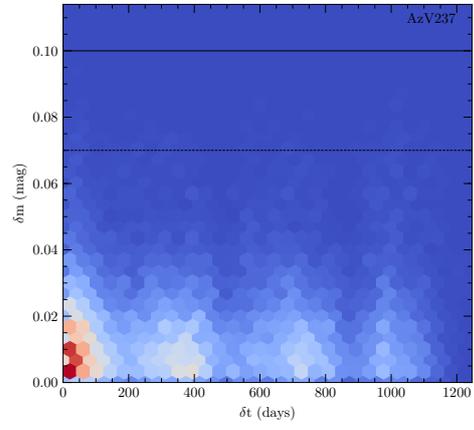

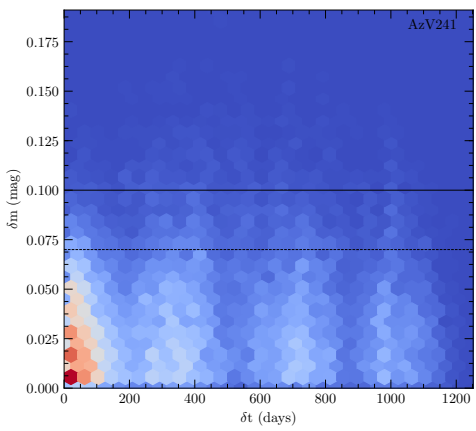
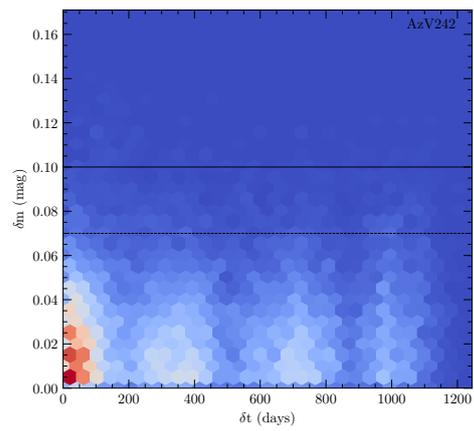

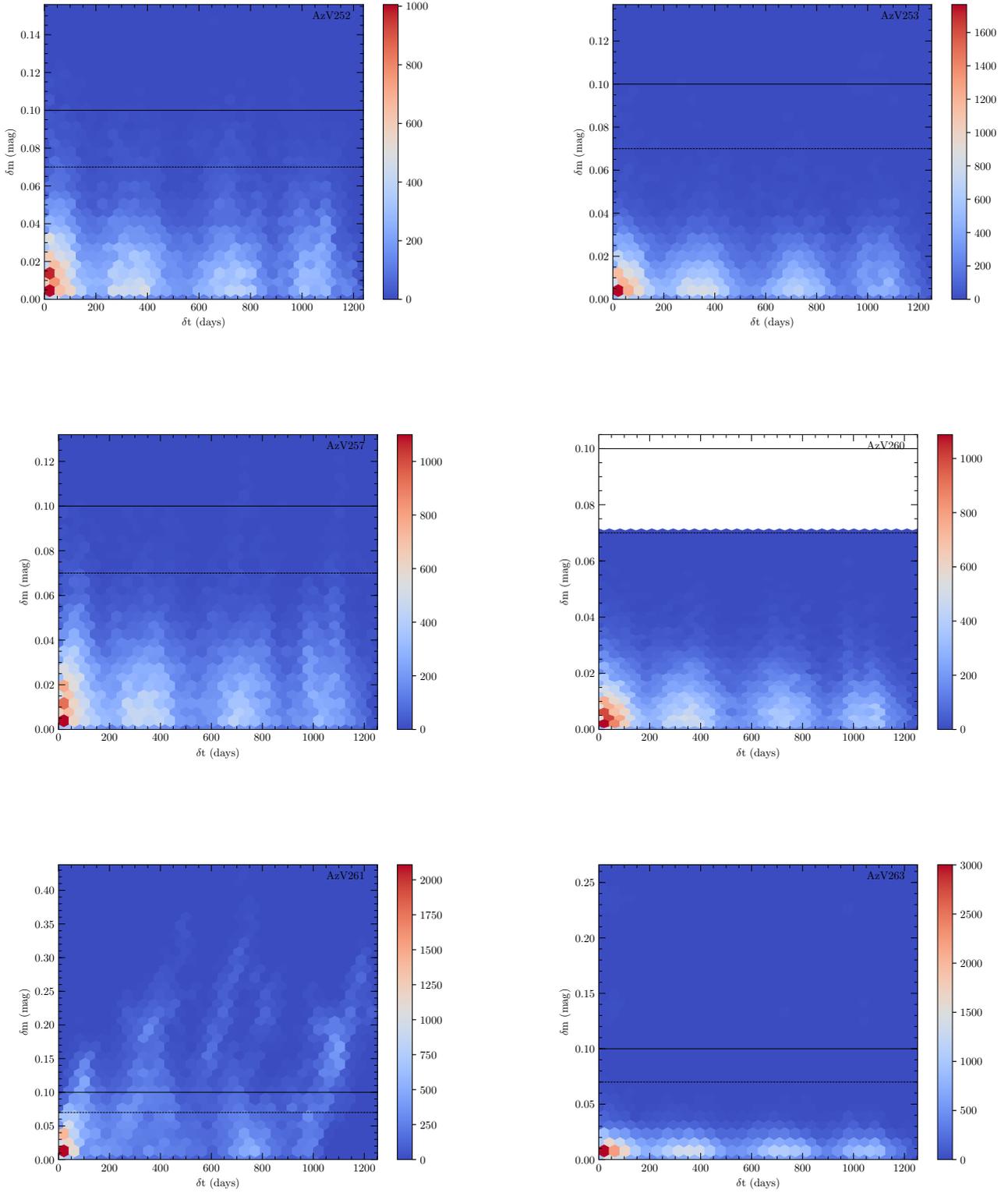

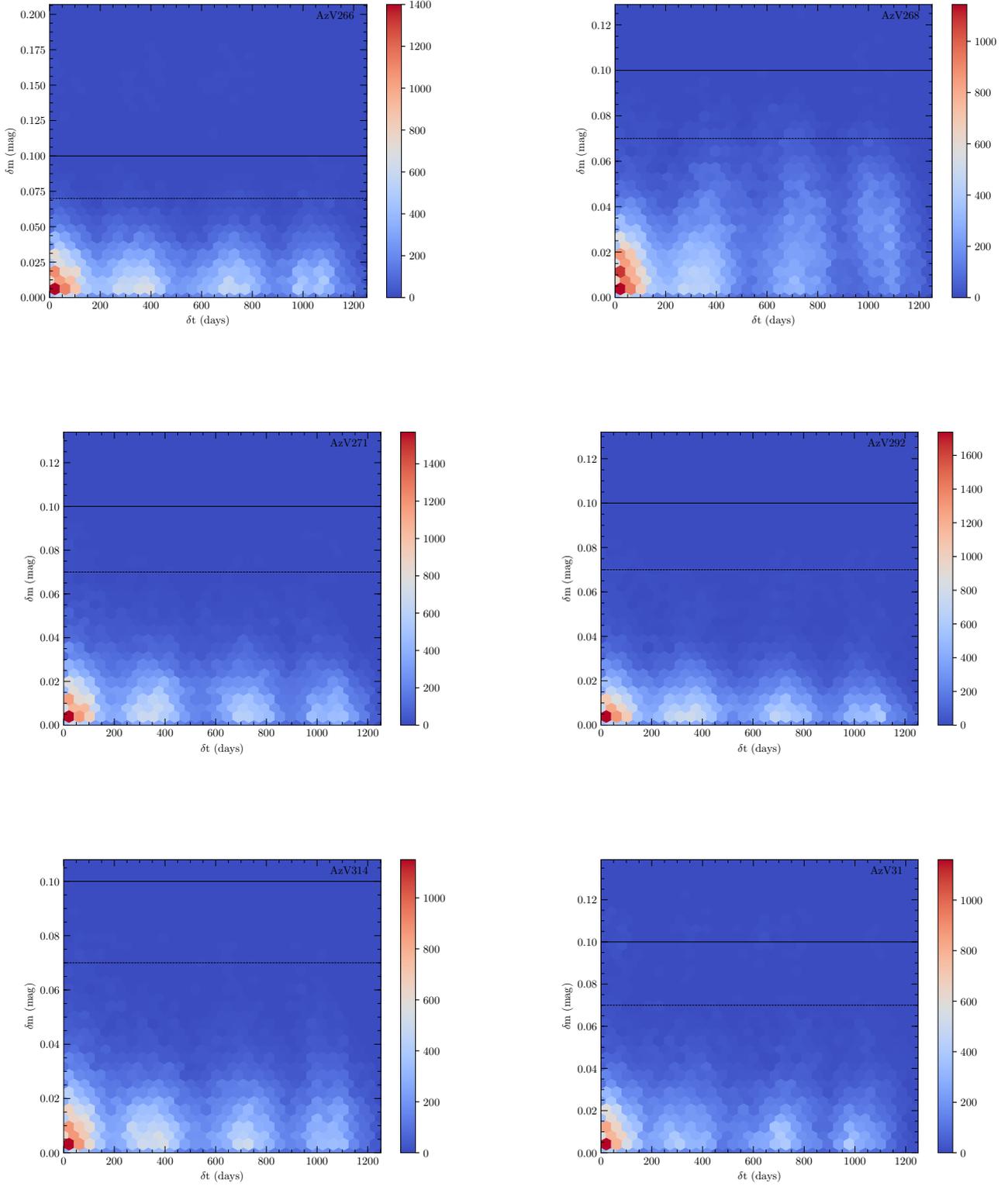

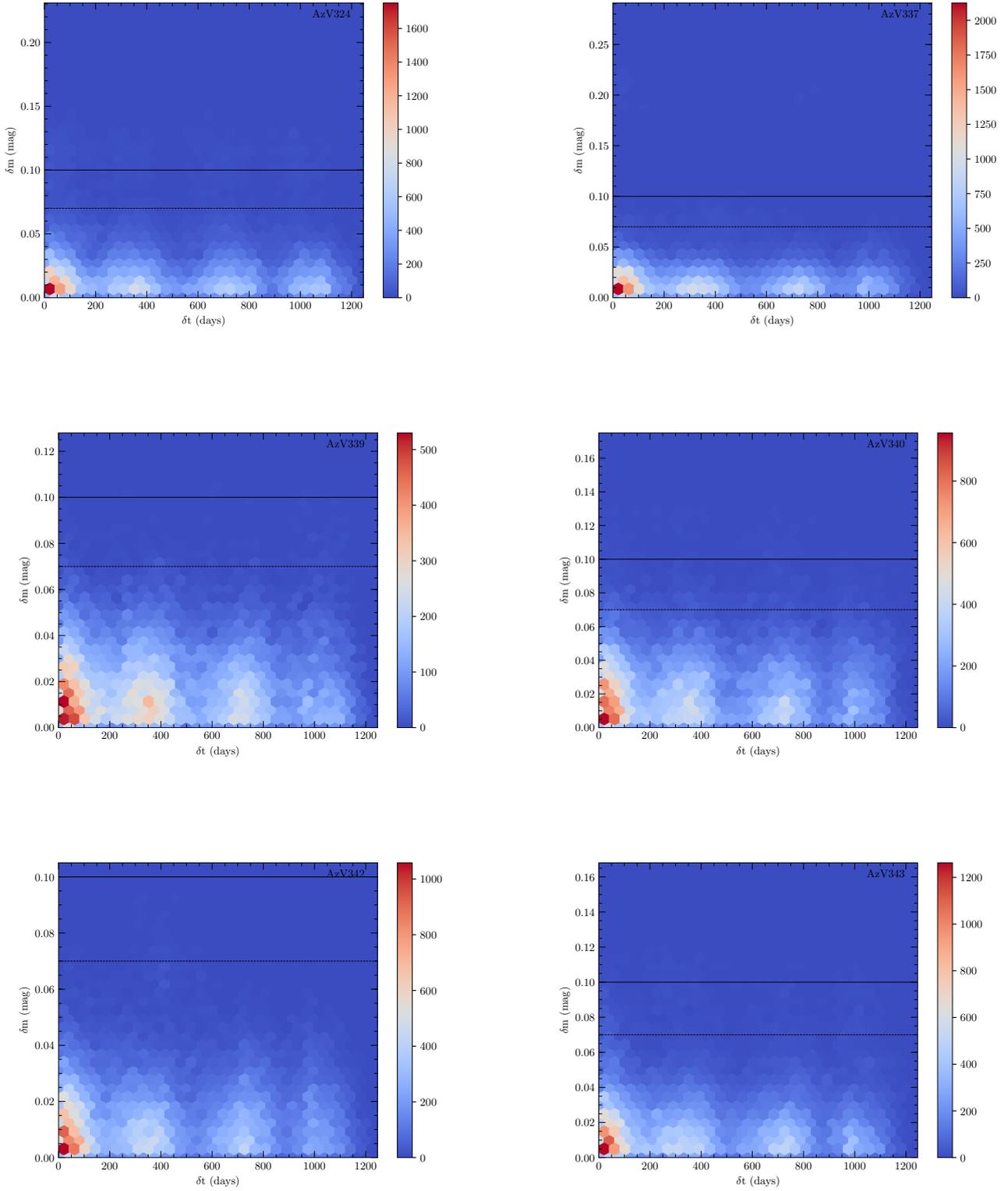

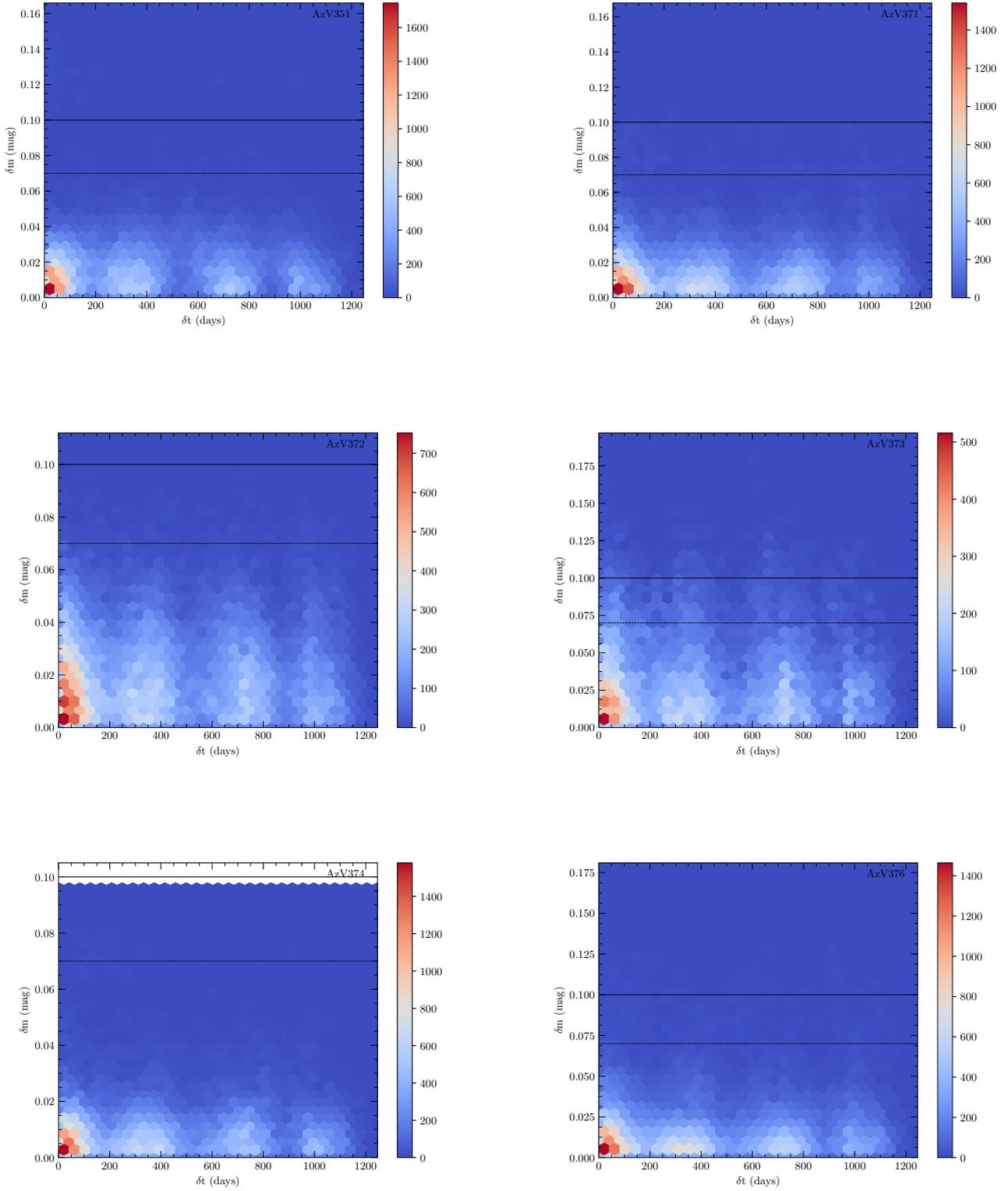

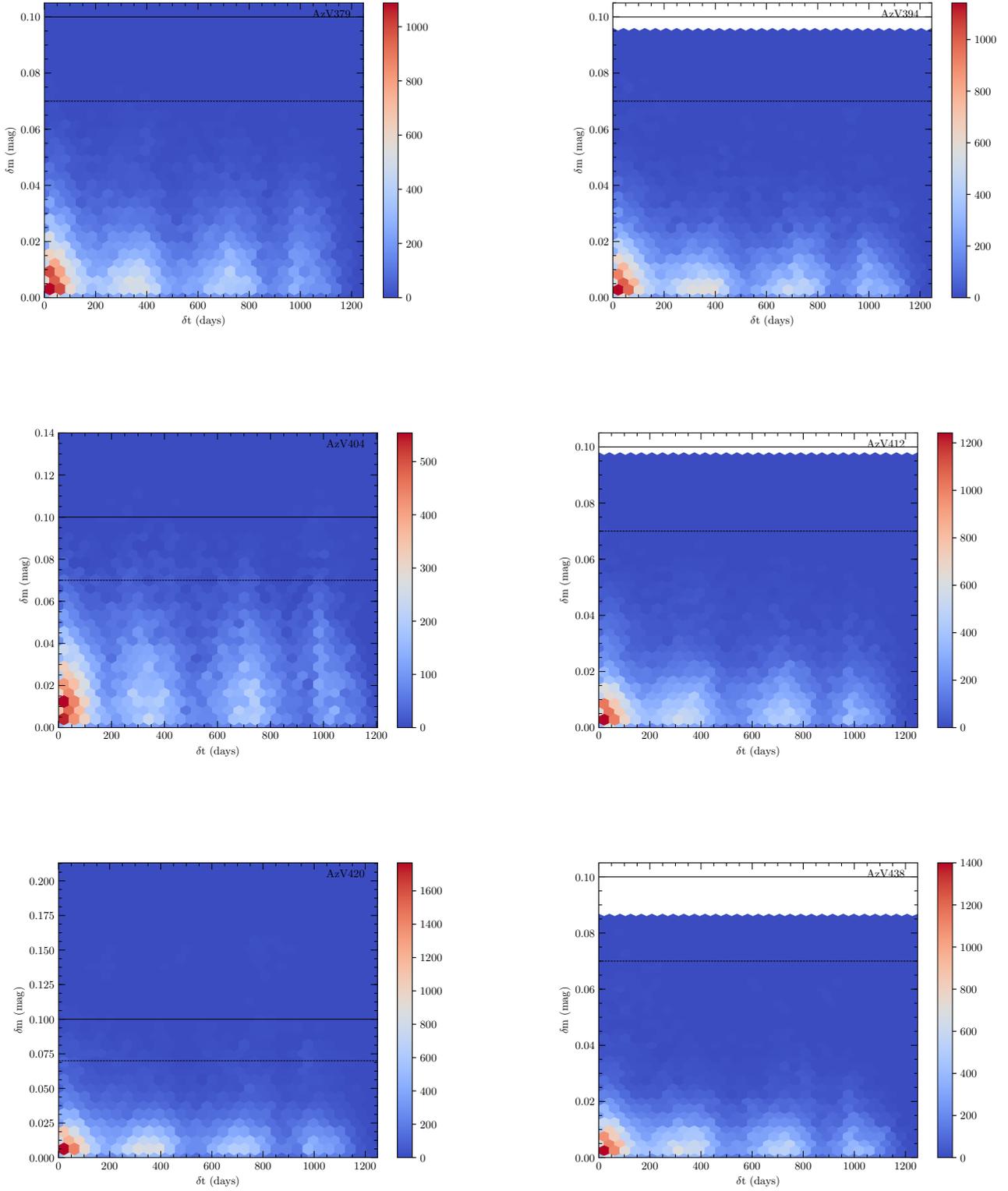

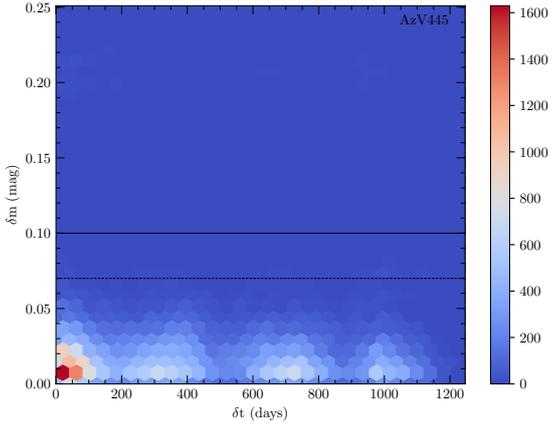
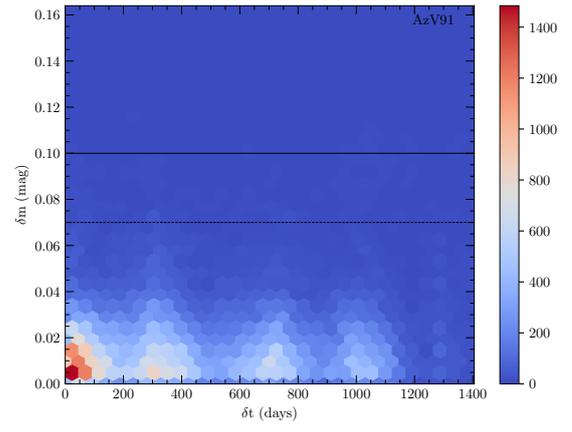

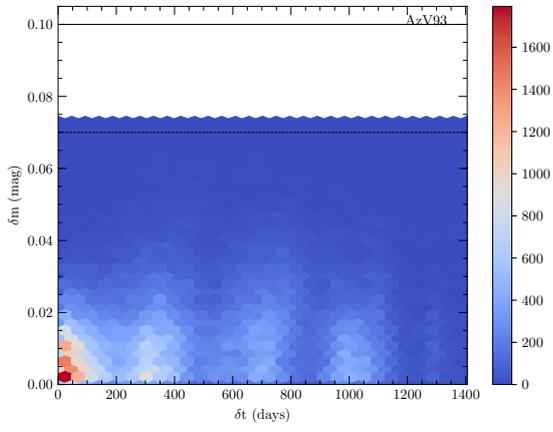
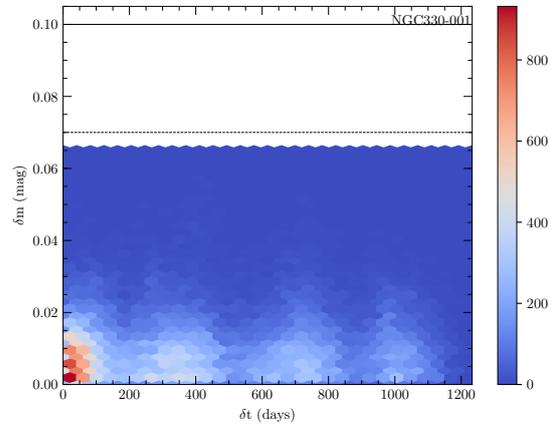

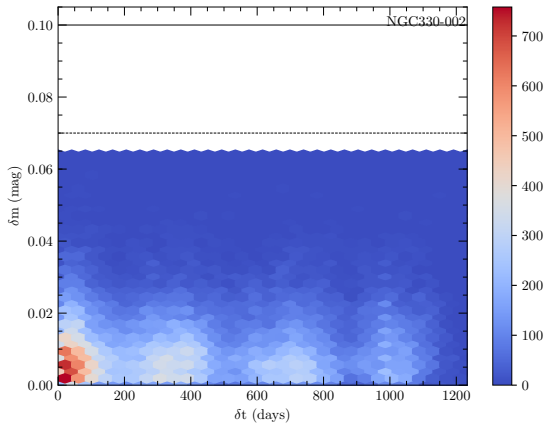
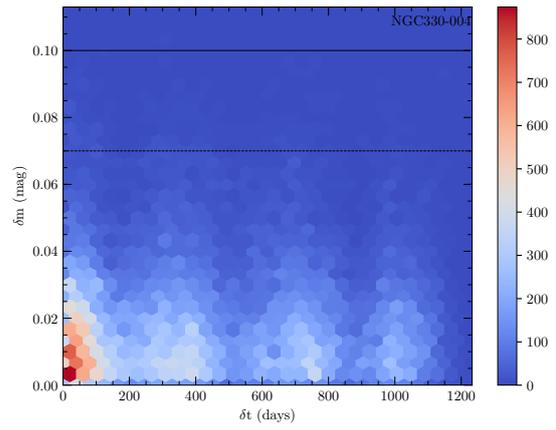

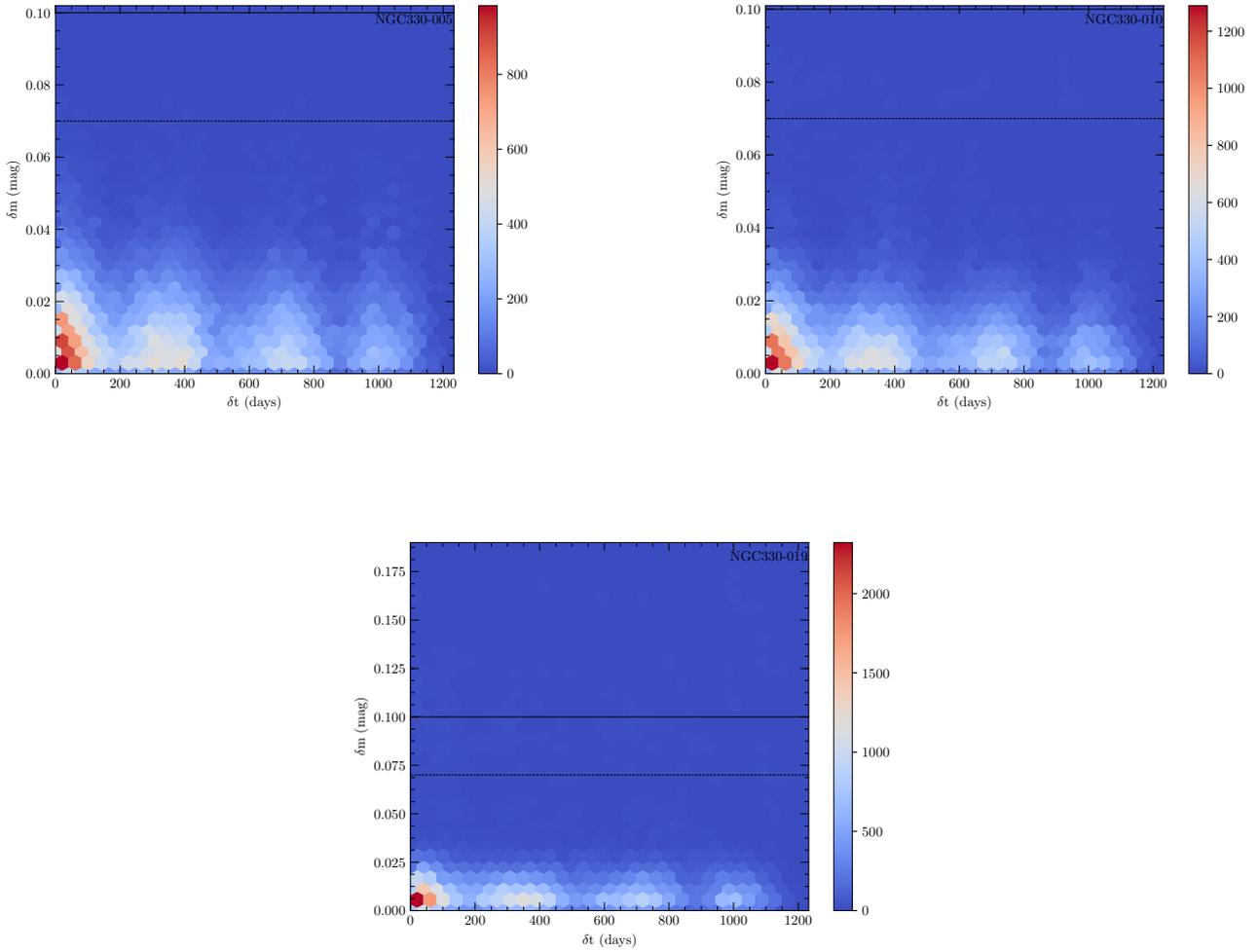

Fig. C1: Two-dimensional histogram of the δm-δt plot of 64 Bsgs. The δm in mag is plotted against the δt in days, and the colour is indicative of the density of the histogram following the colourbar.